\pdfoutput=1
\documentclass{JHEP3}

\usepackage{cite}
\usepackage{amsmath,amstext,amssymb}
\usepackage[pdftex]{graphics}
\usepackage{rotating}
\usepackage{graphicx,subfigure}
\usepackage[spanish,american]{babel}
 
\newcommand{\cA}{{\cal A}}  \newcommand{\cB}{{\cal B}}

\def\G{{\cal G}}

\def\ie{{i.\,e.\ }}
\def\eg{{e.\,g.\ }}
\def\dt{\tilde d}

\def\ft{\tilde f}
\def\vrho{\varrho}
\def\dt{\tilde d}

\def\w{\omega}

\def\im{\text{Im}\:}

\def\del{\partial}

\def\ii{{\mathrm i}}

\newcommand{\dd}{\mathrm{d}}

\newcommand{\vev}[1]{\left\langle{#1}\right\rangle}

\title{Quasinormal modes of massive charged flavor branes}

\author{Matthias Kaminski$^{a,b}$, Karl Landsteiner$^a$ and Francisco Pena-Benitez$^a$\\
   $^a$ Instituto de F\'{\i}sica Te\'orica CSIC/UAM, C-XVI Universidad Aut\'onoma de Madrid\\
  \hphantom{$^a$ }E-28049 Madrid, Spain\\
  $^b$ Department of Physics, Princeton University, Princeton, NJ 08544, USA. \\
  \hphantom{$^a$ }E-mail: \email{mkaminsk@princeton.edu, karl.landsteiner, fran.penna@uam.es}}

\author{Johanna Erdmenger, Constantin Greubel, Patrick Kerner \\
  Max-Planck-Institut f\"ur Physik (Werner-Heisenberg-Institut)\\
  F\"ohringer Ring 6, 80805 M\"unchen, Germany\\
  E-mail: \email{jke, greubel, pkerner@mppmu.mpg.de}
  }

\abstract{
We present an analysis and classification of vector and scalar fluctuations in a 
D3/D7 brane setup at finite termperature and baryon density. 
The system is dual to an $\mathcal{N}=2$ supersymmetric Yang-Mills theory with $SU(N_c)$
gauge group and $N_f$ hypermultiplets in the fundamental representation in the quenched
approximation. 
We improve significantly over previous results on the quasinormal mode spectrum of
D7 branes and stress their novel physical interpretation.
Amongst our findings is a new purely imaginary scalar mode that becomes tachyonic 
at sufficiently low temperature and baryon density. We establish the existence of a 
critical density above which the scalar mode stays in the stable regime for all temperatures.
In the vector sector we study the crossover from the hydrodynamic to the quasiparticle regime
and find that it moves to shorter wavelengths for lower temperatures. 
At zero baryon density the quasinormal modes move toward distinct discrete 
attractor frequencies that
depend on the momentum as we increase the temperature. At finite baryon density, however, the
trajectories show a turning behavior such that for low temperature the quasinormal mode
spectrum approaches the spectrum of the supersymmetric zero temperature normal modes.
We interpret this as resolution of the singular quasinormal mode spectrum that appears at
the limiting D7 brane embedding at vanishing baryon density. 
}
\date{\today}

\preprint{IFT--UAM/CSIC--09--43\\ MPP--2009--162\\ PUPT-2315} 


\begin{document}

%%%%%%%%%%%%%%%%%%%%%%%%%%%%%%%% I N T R O
\section{Introduction and summary}

One of the successes of the AdS/CFT correspondence \cite{Maldacena:1997re,Aharony:1999ti} and its
generalizations is its application to the plasma phase of non-Abelian 
gauge theories \cite{Policastro:2001yc}. This is of high interest because of its potential 
relevance for the description of the strongly-coupled quark-gluon
plasma as it is created in Heavy Ion Collision
at RHIC or in the near future at the LHC. 

The holographic modelling of the plasma phase invariably involves an 
asymptotically AdS black hole \cite{Witten:1998zw}. 
Of  particular interest are the quasinormal modes of such black holes as they are mapped to the
poles in the correlation function of the dual finite temperature field theory 
\cite{Birmingham:2001pj,Kovtun:2005ev}. One aspect of this is a relation between the quasinormal frequencies and the hydrodynamic
transport coefficients, \eg the shear viscosity. 

An important generalization of the AdS/CFT correspondence is the addition
of flavor degrees of freedom in the fundamental  representation of the
gauge group.  One convenient way of achieving this is via the addition of
D$7$-brane probes to the ten-dimensional supergravity background  
\cite{Karch:2002sh}. 
The meson spectrum can  then be studied via the fluctuations of the probe brane
\cite{Kruczenski:2003be, Erdmenger:2007cm}. 

For the configuration of a D$7$-brane probe added to the AdS Schwarzschild black
hole background, a first order phase transition occurs between D$7$-brane probes
either staying outside of the horizon or reaching down to it \cite{Babington:2003vm,
  Kirsch:2004km, Mateos:2007vn}. These two types of embeddings are called Minkowski 
  or black hole embeddings, respectively.
As discussed in \cite{Hoyos:2006gb}, the first case corresponds to stable mesons,
whereas in the second case the mesons are unstable. 
For stable mesons, the spectrum has been determined analytically 
at zero temperature in \cite{Kruczenski:2003be}.
In the second case, the
meson excitations may be identified with quasinormal modes and their finite width is
related to an infalling energy boundary condition at the black hole horizon. 
The spectral functions for this configuration were first studied in
\cite{Myers:2007we}. If the temperature is high compared 
to the quark mass the quasinormal frequencies lie deep inside the lower complex half plane and
the spectral function is smooth and without noticeable peaks. As the temperature is lowered
the quasinormal modes move towards the real axis, producing rather well defined quasiparticle
peaks in the spectral function. This behavior is particularly strong at finite baryon density.

Furthermore a tachyonic mode is present in the scalar
sector close to the first order phase transition discussed above.  
For null momentum, the spectral function of this
configuration has been studied in \cite{Mateos:2007yp} and the presence of the tachyonic mode
has been argued for by way of a WKB analysis. We find indeed a so far overlooked purely imaginary
mode in the scalar spectrum that crosses the real axis and therefore becomes unstable at a temperature
that is in excellent numerical agreement with \cite{Mateos:2007yp}. 

At finite baryon chemical potential, as obtained by considering a non-zero profile for the
time component of the gauge field on the D$7$-brane,  there is a rich phase
structure for the D$7$-brane embedded in the AdS Schwarzschild background
\cite{Kobayashi:2006sb, Mateos:2007vc, Nakamura:2006xk, Nakamura:2007nx,
  Karch:2007br}. In these works the strict supergravity limit is considered, i.e. $\alpha'\to 0$,
and the results are thus perturbative on the gravity side. 
Non-perturbative effects such as worldsheet instantons have been studied in~\cite{Faulkner:2008hm}. 
Taking these effects into account the first-order phase
transition mentioned above becomes third order at a critical chemical
potential \cite{Faulkner:2008hm}. In this paper however 
we will consider the strict supergravity limit only.   

The spectral functions for vector mesons at finite baryon density have been
studied in  \cite{Erdmenger:2008yj}, where a particular turning behavior of the
quasi-normal frequencies in the complex frequency plane was observed as
function of the ratio of quark mass over temperature: At low values of
$M_q/T$, the position of the quasinormal modes moves to smaller real parts
when $M_q/T$ increases, while for large values, it moves to larger real
parts. For large $M_q/T$, the spectrum approaches the form of the
supersymmetric case. The momentum dependence of the spectral function at finite
baryon density has been investigated in \cite{Myers:2008cj, Mas:2008jz}.   

\bigskip

\paragraph{Summary of results}
The purpose of the present paper is to present an in depth study of both scalar
and vector modes at either finite baryon density or finite momentum. 

\begin{figure}[htbp]
 \begin{center}
  \subfigure[]{\includegraphics[width=0.45\textwidth]{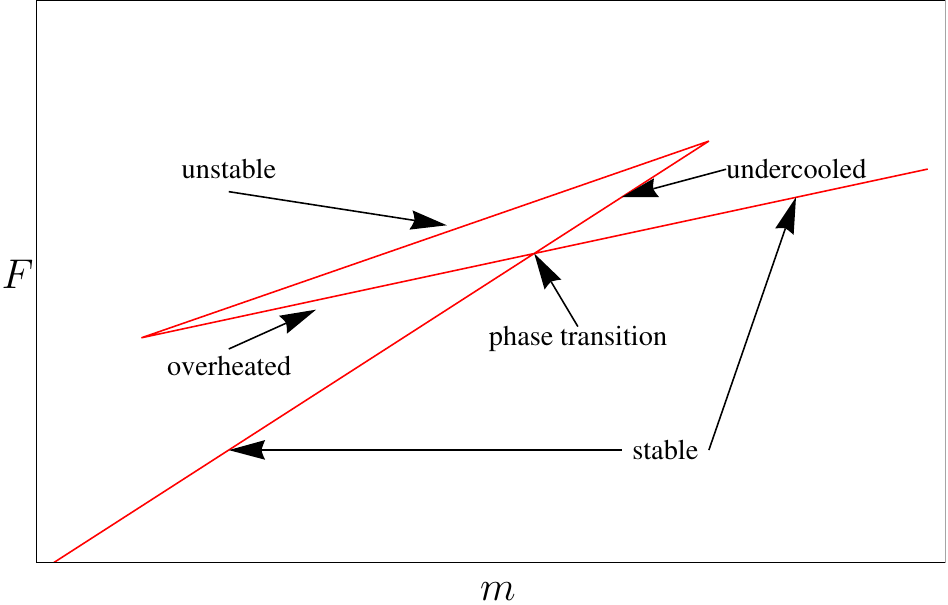}}
   \hfill
  \subfigure[]{\includegraphics[width=0.45\textwidth]{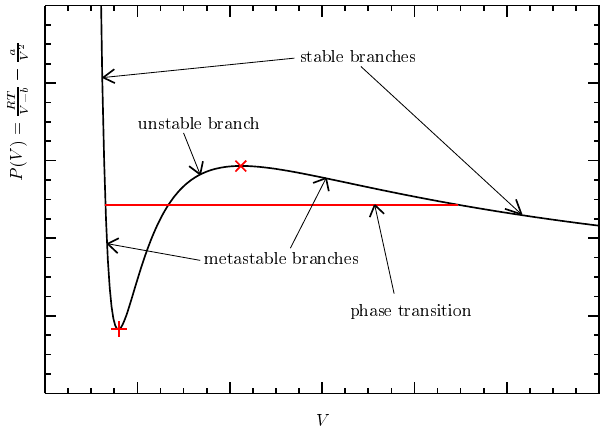}}
 \caption{\label{fig:vdw_phtr}
 (a) Sketch of the free energy $F$ of the flavor fields versus the quark
 mass over temperature ratio $m\propto M_q/T$ close to the first order phase 
 transition. (b) Pressure versus volume of the van-der-Waals gas.
 The red line marks the phase transition which is obtained by the Maxwell
 construction.}
 \end{center}
\end{figure}

At zero density and zero momentum we determine the first and second quasinormal frequency of
the scalar and vector fluctuations. We calculate their trajectories in the complex plane as
we increase the quark mass over temperature ratio. In general these modes move along a curve
where initially the real and imaginary parts can grow a little but then
move continuously towards lower values. As shown in \cite{Paredes:2008nf} the endpoints of
these curves asymptote to a single point on the
real frequency axis that is reached as the embedding reaches the limiting embedding.
This happens when
the brane just touches the black hole horizon\footnote{Note that this
  limiting embedding is often called critical embedding in the literature \eg \cite{Mateos:2007vn}. However in this
  paper we use the term `critical embedding' for the embedding at which the
  phase transition occurs (cf. figure~\ref{fig:vdw_phtr} a). }. There it has been argued that the spectrum of  
quasinormal modes becomes singular as they all coincide at one (real) energy value. 
It is however unclear if
this limiting embedding can really be reached since the embeddings become locally unstable at
rather high temperatures. Indeed in the scalar sector we find a mode that becomes tachyonic
shortly after the system has become thermodynamically metastable due to the presence of the first 
order phase transition. Once the tachyonic mode is in the spectrum the embedding is not even metastable
and so far it is not known what the true ground state would be or if such a state exists at all.

At finite momentum but zero density, we find that at a critical momentum the
system undergoes a crossover transition from the hydrodynamic behavior at long
wavelengths to a collisionless behavior at small wavelengths. We observe this
crossover in the longitudinal vector channel which has a hydrodynamic mode,
\ie a mode whose dispersion relation does not show a gap at zero momentum.
This mode describes baryon charge diffusion and is thus called diffusion mode. The
transition from the hydrodynamic to the collisionless regime is defined as the
point where the imaginary part of the diffusion mode becomes larger than the
imaginary part of the first quasinormal mode which is not a 
hydrodynamic mode. Additionally we observe that the first quasinormal mode of
the scalar and vector fluctuations with increasing momentum show an increasing
number of spirals in their trajectories parametrized by the quark mass over
temperature ratio. The number $n$ of these spirals apparently is correlated
with certain {\it "attractor" frequencies}, i.e. the real part of the quasinormal
modes asymptotes to $\w_n\in \mathbf{R}$ as the D$7$-brane embedding approaches
the critical embedding between the black hole and Minkowski embeddings.
Although the asymptotic values $\w_n$ lie deep in the unstable sector, the spirals mostly
lie in the physical sector of the theory. So they are physical features of the
(meta)stable theory. 

At finite quark density but zero momentum, we find the following: In the
vector mode sector, there are two distinct movements of the quasinormal
frequencies in dependence of the quark mass over temperature ratio. At small
densities there is a turning point as already observed in
\cite{Erdmenger:2008yj} and as described above. This turning point is no
longer present for large densities. 
The critical value for the normalized
density is $\dt_c=0.04$. The relation between the normalized density $\dt$ and
the baryon density $n_B$ is given by 
\begin{equation}
  \label{eq:1}
  \dt=\frac{2^{\frac{5}{2}}n_B}{N_f\sqrt{\lambda}T^3}\,,
\end{equation}
where $T$ is the temperature, $N_f$ the number of flavors and $\lambda$ the 't
Hooft coupling. In the scalar channel, the quasinormal frequencies show also
distinct movements in dependence of the quark mass over temperature ratio. In
this case we find three distinct behaviors separated by critical
densities. As a function of $M_q/T$, the quasinormal modes perform
  either a left turn, a right turn or no turn in the complex frequency
  plane. We will study these different behaviors in detail. 

For the scalar channel, 
we find in addition that the tachyonic mode mentioned
above is only present below a critical normalized density, $\dt<0.00315$.  
We identify the unstable region in which the tachyon is present with an
instability related to the first order phase transition found in the 
thermodynamics of the system. This behavior may be understood by
comparing it to 
the first order phase transition of the well-known van-der-Waals gas,
as described for instance in \cite{0034-4885-50-7-001}.
Figure~\ref{fig:vdw_phtr} (b) shows the first order phase transition
of the van-der-Waals gas, which follows the red line in the $p$-$V$-diagram as
obtained from the Maxwell construction. Additionally there are two metastable
branches: One of them reaches from the point where the phase transition
begins when increasing the volume to the minimum denoted by the red
plus sign in the figure. The second reaches from the maximum denoted
by the red cross to the point where the phase transition ends when
increasing the volume. 
The region between the two extrema is unstable since the pressure 
raises when the
volume is increased.

Our calculations show a very similar behavior of the system
considered here. We have calculated both the free energy as function
of $m\propto M_q/T$ close to the phase transition, 
\footnote{A similar calculation of the free energy was 
  performed in  \cite{Kobayashi:2006sb}.} 
displayed schematically in Figure~\ref{fig:vdw_phtr} (a), and the
quasinormal modes for the same range of $M_q/T$. Both calculations
show a similar structure of stable, metastable and unstable branches,
with the numerical values for the boundaries  of these branches
agreeing up to four digits in both calculations. In particular in
analogy to the van-der-Waals gas we find that in addition 
to the stable branches\footnote{Note that in \cite{Kobayashi:2006sb} it was
  observed that for a small region of the stable phase in which the mass
  parameter $m$ is just slightly larger than the critical one there is an
  charge instability. This instability is distinct from the instability
  discussed above. We do not observe this charge instability in our
  quasinormal mode analysis. This may be due to inadequacy of the our numerical
  methods at large mass over temperature ratios.} there are metastable branches close
to the phase transition, denoted as overheated and
undercooled\footnote{In the context of the
  D$3$/D$7$ system, the terms `overheated' and `undercooled'
were introduced in \cite{Paredes:2008nf}.}
in  figure~\ref{fig:vdw_phtr} (a). Furthermore we find an unstable branch which
connects these two metastable branches.  The metastable phases
are stable against fluctuations, while on the unstable branch a tachyonic
mode appears in the quasinormal spectrum. 
In particular we emphasize that at zero density the limiting
embedding which has a conical singularity at the black hole horizon, 
and is often discussed in the literature, 
lies clearly deep inside the unstable region. Thus any
observation found by using this embedding should be considered with great care
since it does not correspond to a physical state. 
Furthermore we should stress that on the field theory side there exists a
stable ground state for any combination of the mass $m$ and chemical potential $\mu$.

\smallskip

We also present a qualitative analysis of the quasinormal spectrum at
either
finite momentum or finite density which uses
the fact that the equation of motion for the fluctuations can be transformed
into a Schr\"odinger equation. The Schr\"odinger potential analysis
was introduced for instance in
\cite{Hoyos:2006gb,Myers:2007we,Paredes:2008nf,Myers:2008cj}. 

The results of this analysis are in qualitative agreement with the
results obtained by stating the quasinormal modes listed above: In particular, 
at finite density we find a general feature in the Schr\"odinger potential: At
low quark mass over temperature ratio there are only unbound scattering
states which correspond to quasinormal modes. By 
increasing the quark mass over temperature ratio, a barrier forms in the
Schr\"odinger potential \cite{Myers:2008cj} which leads to a local minimum.
In \cite{Paredes:2008nf} it was found for Minkowski embeddings that the
Schr\"odinger potential is a box whose extent in the AdS radial
direction coincides with the region filled by the D$7$-brane probe. Here
we observe that for black hole embeddings at finite density, a barrier
forms in the Schr\"odinger potential at the same radial position as
the IR boundary of the box in the Minkowski case. 
This barrier separates the bulk of the AdS space
from the horizon of the black hole. As the barrier increases when the ratio of
quark mass over temperature $m\propto M_q/T$ raises, less energy can leak into
the black hole. Thus `bound' states which correspond to `normal' modes are 
formed. We study the appearance and the behavior of this barrier in detail.

The paper is organized as follows: In section \ref{sec:holohydro} we give a
general introduction to the relation between quasinormal modes and
hydrodynamics. In section \ref{sec:setup} we introduce the D$3$/D$7$-brane setup.
In the sections \ref{sec:noDnoK}, \ref{sec:finiteK} and \ref{sec:finiteD}
we determine the quasinormal modes of the scalar and vector fluctuations
first at vanishing momentum and density and later at finite momentum or finite
density. In every section we also calculate the corresponding Schr\"odinger
potentials which we use to describe the qualitative behavior of the quasinormal
modes. At the end of each section we summarize the physical results which we
found.

%%%%%%%%%%%%%%%%%%%%%%%%%%%%%%%% H O LO H Y D R O 
\section{Quasinormal modes and Holography} \label{sec:holohydro}
In this section we recall the definition of  quasinormal modes of black holes and the
role they play in determining the response of a holographic field theory close to equilibrium. 

Quasinormal modes of a black hole are distinct perturbations of the black
hole solution. Roughly they can be understood as resonances of the black hole.
However since the energy of the perturbation can leak into the black hole, these
fluctuations are not normal modes and thus have been dubbed quasinormal. Their
corresponding frequencies consist of a real and an imaginary part. As for the
damped oscillator, the real part of the frequency essentially determines the energy of the
fluctuations, while the imaginary part is responsible for the damping. In AdS spacetimes the
quasinormal modes satisfy the following boundary conditions. At the horizon they
are purely ingoing, whereas at the conformal AdS boundary they have an asymptotic
behavior that corresponds to a normalizable mode. In this paper we determine
the quasinormal spectrum of the D$7$-branes. The corresponding modes can be grouped in terms of
their transformation properties under spatial rotations of $SO(3)$. We
consider scalar modes given by perturbations of the brane embedding as 
well as vector modes given by perturbations of the gauge field on the brane.
As is well established by now, the quasinormal frequencies of the dual gravity theory
can be identified with the poles of correlation functions in dual thermal
gauge theories \cite{Birmingham:2001pj,Kovtun:2005ev}.

A system which is close to equilibrium can be described by linear response theory.
There the effect of an external perturbation is given by the retarded two point
function folded against the source of the perturbation. By a Cauchy integration
in the complex frequency plane the response can be written as
a sum over the contributions of the different quasinormal modes (see e.g. \cite{Amado:2008ji}). 
Writing the time dependence as $\exp(-i\omega t)$ we note that a relaxation towards 
equilibrium can only happen if all the quasinormal modes lie in the lower complex half plane.
Following \cite{Amado:2007yr,Amado:2008ji} the response caused by an external perturbation $\tilde\jmath(\omega,k)$
can therefore be written as a sum over quasinormal frequencies
\begin{equation}\label{eq:linear_response_QNM}
\vev{\Phi(t,k)} = i \theta(t) \sum_{n} R_n \,\tilde\jmath(\omega_n,k) \,e^{ -i \Omega_n t - \Gamma_n t} ~,
\end{equation}
where the quasinormal frequencies are $\omega_n = \Omega_n-i\Gamma_n$ and their residues are $R_n$.
If a mode comes to lie in the upper half plane it results in an exponentially growing mode 
and therefore represents an instability of the system. As we will see in the following, 
such an instability does indeed occur in the scalar sector of the D$7$-brane fluctuations.

Of particular interest is the hydrodynamic limit that considers only fluctuations with small frequency and large
wavelength. This hydrodynamic expansion can be seen as an
effective field theory where the degrees of freedom with large
frequency and small wavelength are integrated out. In this effective field
theory only the poles of correlation functions closest to the origin are important. More precisely,
the modes with a dispersion relation obeying $\lim_{\vec{k}\rightarrow 0} \omega(\vec{k}) =0$
represent the hydrodynamic regime. Hydrodynamic transport coefficients such as
the shear viscosity or the diffusion constants can be read off form these
poles. Finally let us mention that recently it has been shown that the determinant of wave operators
in some asymptotically black hole backgrounds can be written in terms of the quasinormal modes \cite{Denef:2009kn}.

%%%%%%%%%%%%%%%%%%%%%%%%%%%%%%%% S E T U P
\section{Holographic Setup} \label{sec:setup}

We are interested in a large $N_c$ thermal field theory, at finite baryon 
chemical potential including fundamental and adjoint matter. Our matter content
is that of the $\mathcal{N}=4$ super Yang-Mills theory and a number $N_f$
of $\mathcal{N}=2$ super Yang-Mills fundamental hypermultiplets. We consider
the quenched approximation with $N_c\rightarrow \infty$ and $N_f\ll N_c$ fixed. In this
limit the theory stays conformal at leading order in $N_c$.

The dual gravity setup is given by a stack of $N_c$ D3-branes and $N_f$ probe D7-branes. The metric
generated by the D3-branes is a non-extremal AdS black hole background
placing the field theory at finite temperature. The D7-branes allow for strings
which have one end on a D3- and the other end on the D7-brane. These
3-7 strings correspond to quarks in the fundamental representation. The length
of these strings, i.e. the separation between the stacks of D3- and D7-branes
corresponds to the quark mass $M_q$ in the field theory. On the world volume of
the D7-branes we introduce a background gauge field $A$ which generates
a chemical potential $\mu$ in the field theory at the boundary. Depending
on the problem it is convenient to work in different coordinate systems. 
We now introduce the $\rho$-coordinates and the $z$-coordinates.
%______________________________________________
\subsection{The $\rho$-coordinate system} \label{sec:rhoCoords}

We find it convenient to use the $\rho$-coordinates of
\cite{Kobayashi:2006sb} to write the AdS black hole background in Minkowski
signature as 
\begin{equation}
  \label{eq:AdSmetric}
  ds^2=\frac{\vrho^2}{2R^2}\left(-\frac{f^2}{\ft}\dd
    t^2+\ft
    \dd\vec{x}^2\right)+\left(\frac{R}{\vrho}\right)^2(\dd\vrho^2+\vrho^2\dd\Omega_5^2)\,,
\end{equation}
with $\dd\Omega_5^2$ the metric of the unit 5-sphere and
\begin{equation}\label{eq:AdSmetric_f}
  f(\vrho)=1-\frac{\vrho_H^4}{\vrho^4},\quad
  \ft(\vrho)=1+\frac{\vrho_H^4}{\vrho^4}\,, 
\end{equation}
where $R$ is the AdS radius, with
\begin{equation}
  R^4=4\pi g_s N_c\,{\alpha'}^2 = 2\lambda\,{\alpha'}^2\,.
\end{equation}
This type of radial coordinate is better suited for calculation of the quasinormal
modes with the shooting method, where one places a cutoff at some large value of $\rho$.

The temperature of the black hole given by \eqref{eq:AdSmetric} may be
determined by demanding regularity of the Euclidean section. It is given by
\begin{equation}
  T=\frac{\vrho_H}{\pi R^2}\,. 
\end{equation}
In the following we may use the dimensionless coordinate
$\rho=\vrho/\vrho_H$, which covers the range from the event horizon at
$\rho=1$ to the boundary of the AdS space at $\rho\to\infty$.

To write down the DBI action for the
D$7$-branes, we introduce spherical coordinates $\{r,\Omega_3\}$ in the
4567-directions and polar coordinates $\{L,\phi\}$ in the 89-directions
\cite{Kobayashi:2006sb}. The angle between these two spaces is denoted by
$\Theta$ ($0\le\Theta\le\pi/2$). The six-dimensional space in the
$456789$-directions is given by
\begin{equation}
  \begin{split}
    \dd\varrho^2+\varrho^2\dd\Omega_5^2=&\,\dd r^2+r^2\dd\Omega_3^2+\dd L^2+L^2\dd\phi^2\\
    =&\,\dd\varrho^2+\varrho^2(\dd\Theta^2+\cos^2\Theta\dd\phi^2+\sin^2\Theta\dd\Omega_3^2)\,,
  \end{split} 
\end{equation}
where $r=\varrho\sin\Theta$, $\varrho^2=r^2+L^2$ and $L=\varrho\cos\Theta$.
Due to the symmetry, the embedding of the D$7$-branes only depends on the
radial coordinate $\rho$. Defining $\chi=\cos\Theta$, we parametrize the
embedding by $\chi=\chi(\rho)$ and choose $\phi=0$ using the $O(2)$
symmetry in the 89-direction.  The induced metric $G$ on the D$7$-brane
probes is then
\begin{equation}
  \label{eq:inducedmetric}
    ds^2(G)=\frac{\vrho^2}{2R^2}\left(-\frac{f^2}{\ft}\dd
      t^2+\ft\dd\vec{x}^2\right)+\frac{R^2}{\vrho^2}
    \frac{1-\chi^2+\vrho^2(\del_\vrho\chi)^2}{1-\chi^2}\dd\vrho^2 
    +R^2(1-\chi^2)\dd\Omega_3^2\,.
\end{equation}
The square root of the determinant of $G$ is given by
\begin{equation}
  \sqrt{-G}=\frac{\sqrt{h_3}}{4}\varrho^3f\ft(1-\chi^2)\sqrt{1-\chi^2+\varrho^2(\del_\varrho\chi)^2}\,,
\end{equation}
where $h_3$ is the determinant of the 3-sphere metric.

%______________________________________________
\subsection{Brane setup and background fields in $z$-coordinates}
Here we introduce the coordinates also used in~\cite{Hoyos:2006gb}.
This set of coordinates maps the compact interval $[0,1]$ to the
distance between the conformal boundary and the black hole horizon.
Such a compact radial coordinate is particularly well suited for the 
calculation of quasinormal modes using the relaxation method as
explained in appendix \ref{sec:appRelax}. 
We work in a non-extremal black-hole background generated by the 
stack of D3-branes giving the effective metric
\begin{equation}\label{eq:metricz}
\frac{\mathrm ds^2}{R^2}=\frac{1}{z^2}\left [-f(z)\mathrm dt^2+\frac{\mathrm dz^2}{f(z)}+\mathrm d\vec x^2\right ]+\mathrm d\Omega_5^2,
\end{equation}
with $f(z)=1-z^4$. Note that the horizon is located at $z=1$, the AdS-boundary at $z=0$.
Frequencies and momenta measured in \eqref{eq:metricz} are related to physical
  frequencies and momenta by $(\omega_{ph},k_{ph}) = \pi T (\omega,k)$, where $T$ is the temperature. 

The D7-brane action reads
\begin{equation}\label{eq:D7Action}
S=\int\mathrm d^8\xi\sqrt{-\mathrm{det}(P[G]+2\pi\alpha'F)} \, ,
\end{equation}
where $P[G]$ is the metric's pull-back on the $N_f$ D7-branes. Here the induced metric
is given by
\begin{equation}
\frac{\mathrm ds_{D7}^2}{R^2}=-\frac{f(z)}{z^2}\mathrm dt^2 + \left(\frac{1}{z^2f(z)} - \Theta'(z)\right)\mathrm dz^2 + \frac{1}{z^2}\mathrm d\vec x^2 + \sin^2\Theta(z)\, \mathrm d\Omega_3^2 \, ,
\end{equation}
where $\Theta(z)$ describes the D7-brane embedding in the AdS-Schwarzschild
background and the embedding coordinates are $x^a=(t,\vec
x,z,\alpha_1,\alpha_2,\alpha_3)$.  To second order in the field strength the D7-brane
Lagrangian can be written in the following way
\begin{equation}
\mathcal{L}=\sqrt{-P[G]}\left[1 + \pi^2\alpha'^2F_{ab}F^{ab}\right]\, .
\end{equation} 
With the zero order term, we obtain the background equation of motion
\begin{eqnarray}
0 &=& 3\cos\Theta(z)[-1+z^2(-1+z^4)\Theta'(z)^2] \\
&&- z\sin\Theta(z)[(3+z^4)\Theta'(z)+2z^2(1-z^4)(2-z^4)\Theta'(z)^3-z(-1+z^4)\Theta''(z)]\,\nonumber .
\end{eqnarray}
The brane embedding can be found by integrating this equation from the horizon out to the boundary. As initial
conditions one chooses $\chi_0 = \cos(\Theta(1))$ and demands regularity on the horizon.
The quark mass can be read off from the behaviour at radial AdS infinity located at 
$z=0$ \cite{Hoyos:2006gb}
\begin{equation}
M_q = \frac 1 2 m \sqrt{\lambda} T,\quad m=\chi'(z=0)\, .
\end{equation}

\begin{figure}[htbp]
 \begin{center}
 \includegraphics[width=0.45\textwidth]{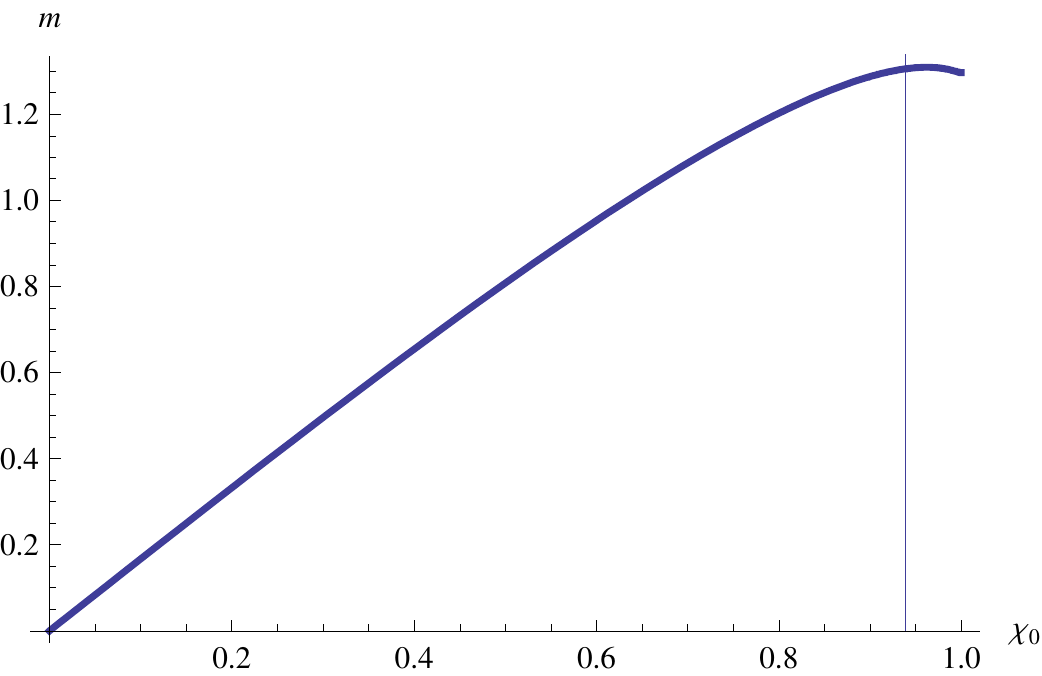}  
 \caption{\label{fig:massplot}
Plot of the dimensionless mass parameter $m$ vs the cosine $\chi_0=\cos\Theta_0$ of the embedding angle $\Theta(z)$ at the horizon, i.e. $\Theta_0=\Theta(z=1)$. 
The mass is
not a single valued function of $\chi_0$ on $[0,1]$! It takes a maximum value of $m=1.31$ at $\chi_0=0.962$.
The horizontal line indicates the first order phase transition at $\chi_0=0.939$.}
 \end{center}
\end{figure}
\begin{figure}[htbp]
 \begin{center}
 \includegraphics[width=0.45\textwidth]{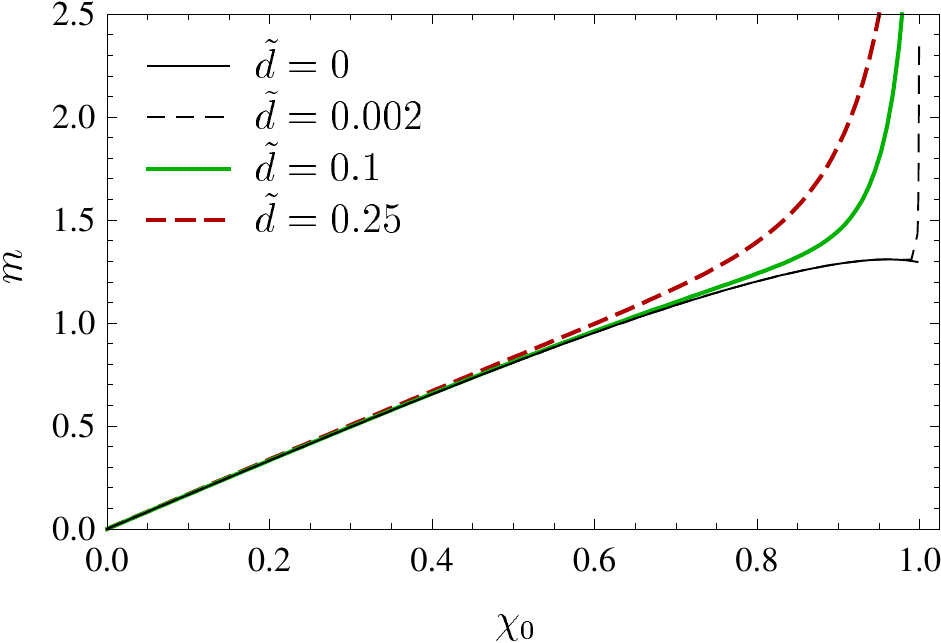}  
\caption{\label{fig:massplotD}
              At finite charge density: The dependence of the dimensionless
              quark mass~$m=2 M_q/\sqrt{\lambda} T$
		on the horizon value~$\chi_0=\lim_{\rho\to1}\chi$ of the embedding
		as shown in~\cite{Erdmenger:2007ja}.}
 \end{center}
\end{figure}

We have plotted the mass parameter $m$ as a function of $\chi_0=\cos\Theta(\vrho_H)$ in figure 
\ref{fig:massplot}. The mass is not a single valued function of $\chi_0$. Although $m$ is the physical parameter of the brane embedding we prefer to quote $\chi_0$ values instead of $m$ since much of our investigations will take place in the regime where $m$ ceases to be single valued.
The change in the sign of $\partial m / \partial \chi_0$ is also suggestive of an instability. Indeed the maximum of $m=1.31$ is reached at $\chi_0=0.962$, which is precisely the value from where an unstable mode appears in the scalar 
sector of the quasinormal mode spectrum.

Looking at figure \ref{fig:massplotD} it is obvious that at finite charge density the
unstable region disappears at a critical density near $\tilde d_c=0.00315$ and 
for larger densities the quark mass is a monotonously increasing function of the
embedding parameter $\chi_0$, i.e. $\chi_0\sim M_q/T$ for $\tilde d_c\ge 0.00315$.
This critical value will be confirmed in our analysis of the quasinormal mode spectrum
below.

The second order term produces the Maxwell equation
\begin{equation}
\partial_a(\sqrt{-P[G]}F^{ab})=0,
\end{equation}
where $F_{ab}=\partial_aA_b-\partial_bA_a$. We can choose the $A_z=0$ gauge and expand $A_a$ in plane wave modes. Moreover if we write the equations in a gauge invariant way using the electric 
fields in longitudinal $E_L=k_{ph} A_0+\omega_{ph} A_ 1$ and transverse direction $\vec E_T=\omega_{ph} \vec A_T$ the equations of motion are
\begin{eqnarray}
E_L''(z) + \left[ C(z) + \frac{f'(z)\omega^2}{f(z)(\omega^2-f(z)k^2)}\right] E'_L(z) + B(z)(\omega^2-f(z)k^2)E_L(z) &=& 0\, ,\\
E_T''(z) + \left[ C(z) + \frac{f'(z)}{f(z)}\right] E'_T(z) + B(z)(\omega^2-f(z)k^2)E_T(z) &=& 0\, ,
\end{eqnarray}
with
\begin{align*}
B(z)&=\frac{1}{f^2(z)}+\frac{z^2\Theta'(z)^2}{f(z)}\,,\\
C(z)&=-\frac{1}{z} + 2z(-2+z^4)\Theta'(z)^2\, .  
\end{align*}

\bigskip
Our results are summarized in the following three sections.

%%%%%%%%%%%%%%%%%%%%%%%%%%%%%%%% V A N I S H I N G  MOMENTUM & DENSITY
\section{Vanishing momentum and density} \label{sec:noDnoK}
Our analysis produced a considerable amount of data and we will not show
all of it because of some redundancy in the results. Technical details on 
our numerical methods
are also deferred to the appendices~\ref{sec:appShooting} and~\ref{sec:appRelax}.
The idea for this and the following two sections is, to have example figures for each
case of interest and a listing of all the effects we observe. We also provide
a qualitative analysis by studying the correspondent Schr\"odinger equations. The discussion subsection 
in each of the cases is then devoted to the physically most interesting effects, i.e. the tachyon, diffusion mode, turning point.

%______________________________________________
\subsection{Transverse vectors}

The transverse equation of motion can be written in this simplified form
\begin{equation} \label{eq:transVecEom}
E_T''(z) + A_1(z) E'_T(z) + B(z)(\omega^2-f(z)k^2)E_T(z) = 0\, ,
\end{equation}
where $A_1(z)=C(z) + \frac{f'(z)}{f(z)}$. Close to the boundary of AdS $(z\to 0^+)$, the differential equation reduces to
\begin{equation}
E_T''(z) -\frac{1}{z} E'_T(z)  = 0\, ,
\end{equation}
which has the solution $E_T(z)=\cA+\cB z^2$. According the dictionary of the correspondence, $\cA$ should be zero in order to study the quasinormal states. Close to the horizon $(z\to 1^-)$, the differential equation is given by
\begin{equation}
E_T''(z) +\frac{1}{z-1} E'_T(z) + \frac{\omega^2}{16(z-1)^2}E_T(z) = 0\, ,
\end{equation}
with the solution $E_T(z)=\cA'(1-z)^{i\omega/4}+\cB'(1-z)^{-i\omega/4}$. The infalling boundary condition is fulfilled by
choosing $\cA'=0$.

At this point we can perform the next transformation $E_T(z)=(1-z)^{-i\omega/4}y(z)$, in order to split the infalling singular part from the regular part of the function. 
In consequence the function $y(z)$ must satisfy the boundary conditions $y(0)=0$ and $y(1)=1$ and the differential equation turns out to be
\begin{equation}
y_t''(z)+[\alpha_1+i\omega\gamma_1]y'_t(z)+[\alpha_0+i\omega\beta_1+\omega^2\beta_2]y_t(z)=0\, , 
\end{equation}
with $\alpha_1=A_1$, $\gamma_1=\frac{1}{2(1-z)}$,  $\alpha_0=-k^2f(z)B(z)$,  $\beta_1=\frac{1+A_1(1-z)}{4(1-z)^2}$ and $\beta_2=-\frac{1}{16(1-z)^2}+B(z)$.

Results for the quasinormal modes of the transverse vectors in this case 
are shown in figure~\ref{fig:vectorK0}.

%______________________________________________
\subsection{Longitudinal vectors}

The equation of motion for longitudinal vectors is given by
\begin{equation}\label{eq:longVecEom}
E_L''(z) + \left[\frac{A_1(z)(\omega^2-f(z)k^2)+C_0(z)}{\omega^2-f(z)k^2} \right]E'_L(z) + B(z)(\omega^2-f(z)k^2)E_L(z) = 0,
\end{equation}
with $C_0(z)=k^2f(z)$. The asymptotic behavior of this equation is the same for the transverse e.o.m, then if we do the same transformation that above, we obtain this equation:
\begin{equation}
y_l''(z)+\left[\frac{\alpha'_1+\alpha'_2\omega^2}{\omega^2-k^2f(z)}+i\omega\gamma'_1\right]y'_l(z)+\frac{\alpha'_0+\beta'_2\omega^2+\beta'_3\omega^4+i(\omega\beta'_1+\omega^3\beta'_4)}{\omega^2-k^2f(z)}y_l(z)=0,
\end{equation}
with 
\begin{align*}
&\alpha'_1 = C_0(z)-A_1(z)k^2f(z)\,, \qquad \qquad \qquad \qquad \qquad
&&\alpha'_2 = A_1(z)\,,\\
&\gamma'_1 = \frac{1}{2(1-z)}\,,
&&\alpha'_0 = k^4f^2(z)B(z)\,,\\
&\beta'_1 = \frac{C_0(1-z)+k^2(A_1(z-1)-1)f(z)}{4(z-1)^2}\,,
&&\beta'_3 = B(z)-\frac{1}{16(1-z)^2}\,,\\ 
&\beta'_2 = \frac{k^2 \left(1-32 (-1+z)^2 B(z)\right) f(z)}{16 (-1+z)^2}\,,
&&\beta'_4 = \frac{1+A1(z)(1-z)}{4 (-1+z)^2}\,.
\end{align*}

\begin{figure}[!h]
\begin{center}
\includegraphics[scale=0.7]{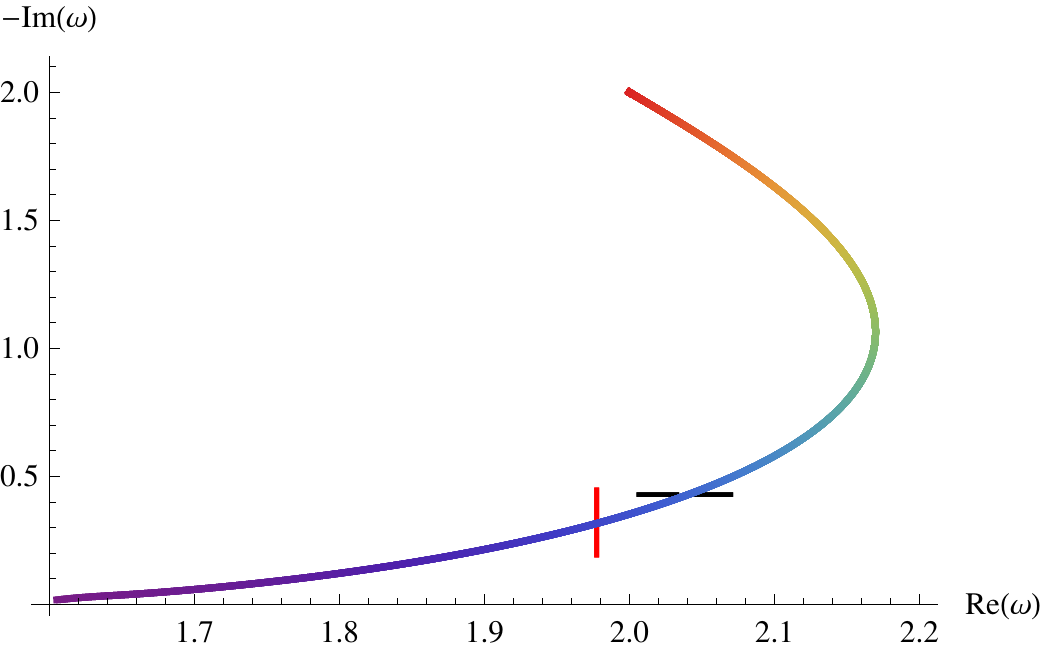}
\includegraphics[scale=0.7]{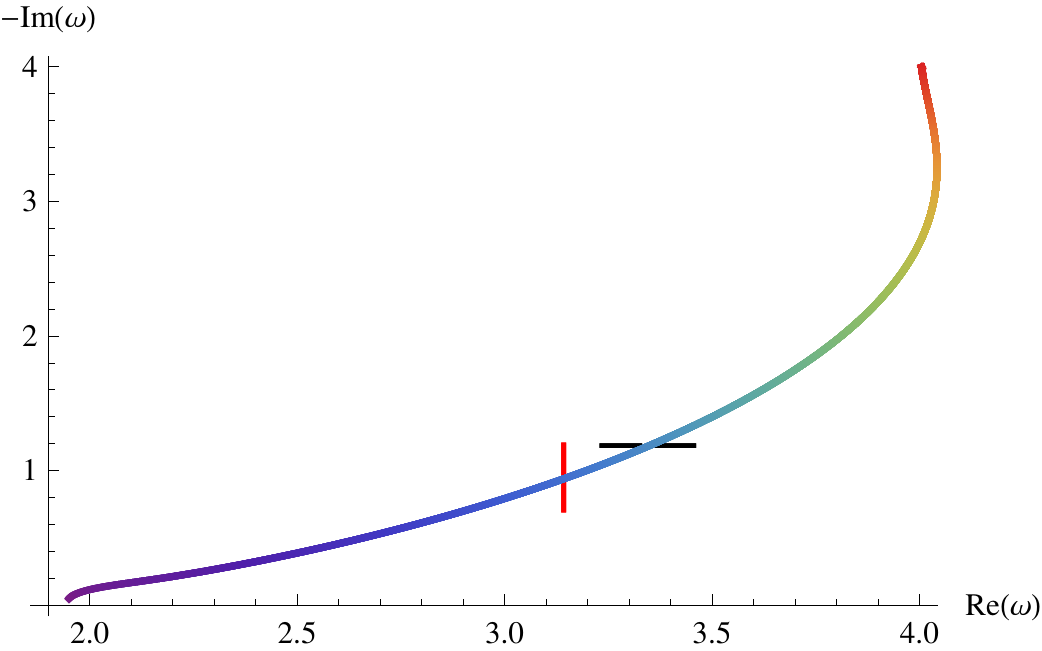}
\caption{\label{fig:vectorK0}
Location of the first~(left) and second~(right) quasinormal modes in the complex
frequency plane for the {\it vector} fluctuations at
vanishing momentum ($k=0$) as a function of the embedding $\chi_0$. 
Red color indicates small quark mass, or high temperature, while
the temperature decreases towards blue colors.
The horizontal (black) dash indicates the frequency at the first order phase transition where the 
angle is $\chi_0=0.939$. The vertical (red) dash indicates the frequecy at which the embeddings become
locally unstable at $\chi_0=0.962$. The modes are followed down to embeddings with $\chi_0=0.999875$.}

\end{center}
\end{figure}

In the case with $k=0$ the differential equations for transverse and longitudinal fluctuations are the same, in consequence their quasinormal spectra coincide,
see figure~\ref{fig:vectorK0}.

\paragraph{Results for vectors}
Figure~\ref{fig:vectorK0} shows the first and second QNM in the
complex frequency plane. Starting with zero quark mass, i.e. at high
temperature~(red color), the imaginary part monotonously decreases with 
decreasing temperature. This means the corresponding mode becomes 
more and more stable. In contrast to that the real part of the quasinormal
frequency first grows until it reaches a maximum and then decreases
as well with decreasing temperature. This maximum in the real part of
the QNM lies above the meson melting transition~(indicated by a short
horizontal dash). The melting transition takes place at a critical angle $\chi_0=0.939$ 
from which on the Minkowski embeddings are thermodynamically
favored, not the black hole embeddings. We have chosen to remain
in the so-called {\it undercooled} phase keeping the black hole 
embeddings even beyond the transition. This phase is accessible 
since the meson-melting is a first order transition. So the undercooled phase is metastable. However we will see in the following section that at a smaller temperature
below the melting transition, i.e. 
a larger angle $\chi_{0}^{\text{tachyon}}\approx 0.962$, this undercooled 
phase is destabilized by the scalar fluctuation becoming tachyonic. In the figures this is indecated by
a red vertical dash.
   
Exemplary numerical values for the vector QNM frequencies 
at $k=0$ are given in table~\ref{tab:vectorK0}. 
We find a remarkable agreement between the values obtained with
two different methods: the relaxation method and the shooting method.
The results are in good agreement for all parameter regions and in all the
cases we treat in this work. Therefore we exclusively show results produced 
with the relaxation method from now on.
\begin{table}
\begin{center}
\begin{tabular}{|c|c|c||c|c|}
 \hline
 &\multicolumn{2}{c||}{relaxation}&\multicolumn{2}{c|}{shooting}\\
 \hline
 $\chi_0$
 &$\text{Re} \omega$& $\text{Im} \omega$&$\text{Re} \omega$& $\text{Im} \omega$\\
 \hline\hline
 1st QNM & & & &\\
 \hline
 $0$ &2.0000 & -2.0000 & 2.0000 & -2.0000 \\
 \hline
  $0.48$ & 2.1075 & -1.5973 & 2.1075 & -1.5972 \\
 \hline
  $0.92$ & 2.0656  & -0.4853 & 2.0657 & -0.4852 \\
 \hline\hline
 2nd QNM & & & & \\
 \hline
 $0$ &4.0054 &-3.9976 & 3.9995& -4.0004\\
 \hline
  $0.48$ &4.0417 & -3.3366 & 3.9324& -3.3386\\
 \hline
  $0.92$ & 3.4397 & -1.3093 & 3.4397& -1.3093\\
 \hline  
\end{tabular}
\caption{\label{tab:vectorK0}
Exemplary values for the first and second vector QNM frequencies
at $k=0$ for different values of $\chi_0$ parametrizing 
the D7-embedding. See figure \ref{fig:massplot} for the relation between $\chi_0$ and the 
quark mass $M_q$. The first pair of values in each row is obtained 
from the relaxation method, the second pair stems from requiring
the shooting solution to vanish at the AdS boundary. We find a
remarkable agreement.
}
\end{center}
\end{table}
\begin{figure}
\begin{center}  
\includegraphics[width=0.6\linewidth]{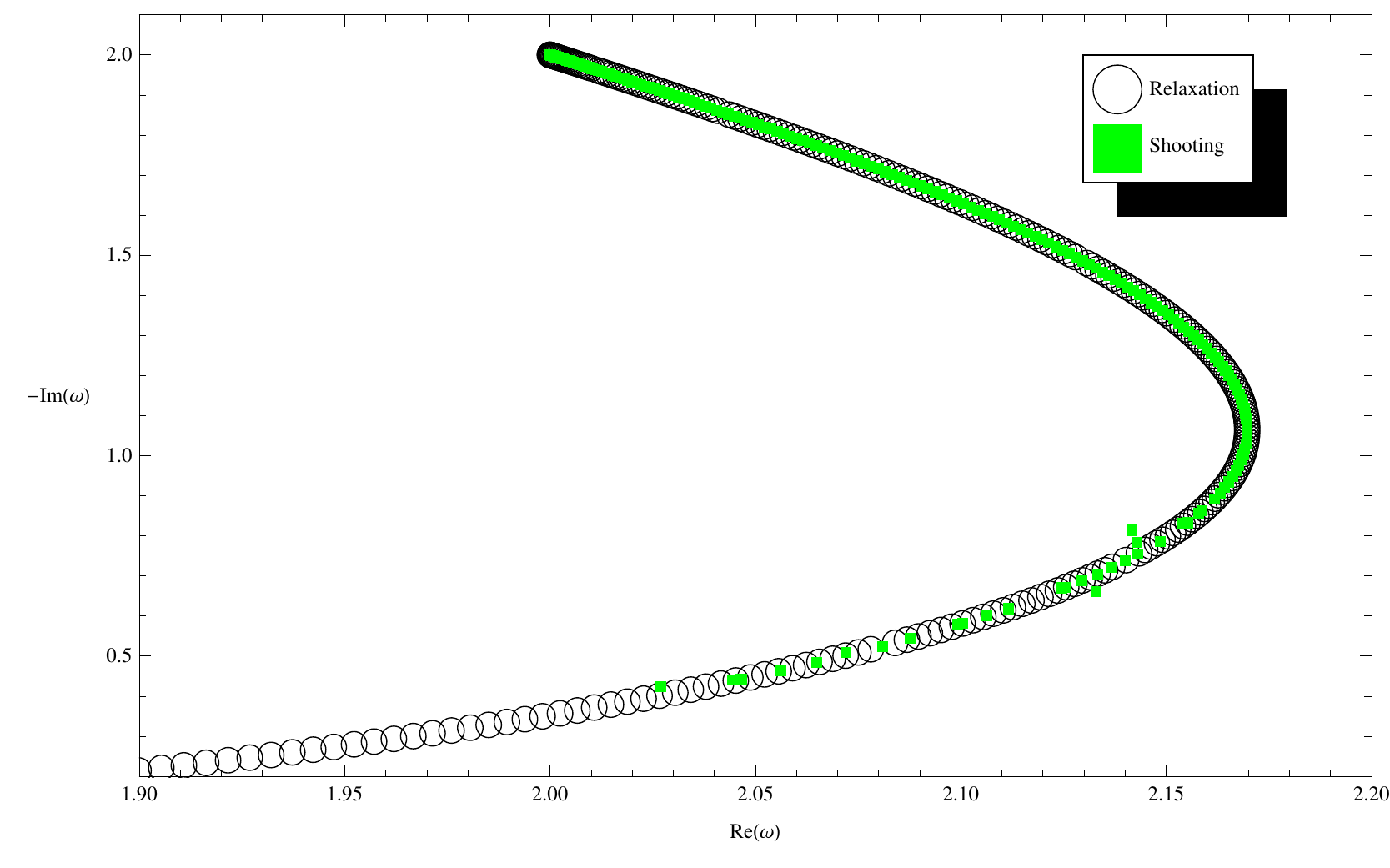}
\caption{
Shoot and relax: Comparison of the shooting method result~(green squares) with 
the relaxation method results~(black circles) for the location of the first 
transverse vector quasinormal mode at vanishing momentum $k=0$, density $\tilde d=0$.  
Along the curves the temperature is varied. Our two methods agree very well.  
}
\end{center}
\end{figure}  

%_______________________________________________
\subsection{Scalar}
In the DBI-action~\eqref{eq:D7Action} we let the $\Theta$-angle fluctuate 
and we split this fluctuation $\delta\Theta$ into a product of its singular
and regular parts $\delta\Theta(z)=(1-z)^{-i\omega/4}zy(z)$. With this change the infalling boundary condition at the horizon is translated into $y(1)=1$\footnote{Notice that the equation for $y(z)$ is linear and that we can
scale $y$ by and arbitrary constant, therefore we can always choose the boundary condition at the horizon to be $y=1$.} and the Dirichlet condition at the boundary implies $y(0)=0$.
The equation of motion for scalar fluctuations then reads
\begin{equation}\label{eq:scalarEom}
y''(z) + \left[ a_1(z) + ic_1(z)\omega\right] y'(z) + \left[a_0(z) + ib_1(z)\omega + b_2(z)\omega^2\right]y(z),
\end{equation}
with 
\begin{align*}
&a_1(z) = A_1(z) + \frac{2}{z}\,, \qquad \qquad \qquad \qquad \qquad
&&a_0(z) = \frac{A_1(z)}{z} + A_0(z) - B(z)^2k^2f(z)\,,\\
& c_1(z) = \frac{1}{2(1-z)}\,,
&&b_2(z) = B(z)^2-\frac{1}{16(1-z)^2}\,,\\
&b_1(z) = -\frac{(1-A_1(z)(1-z))z-2}{4z(1-z)^2} \, .
\end{align*}

\begin{figure}[!h]
\begin{center}
\includegraphics[scale=0.7]{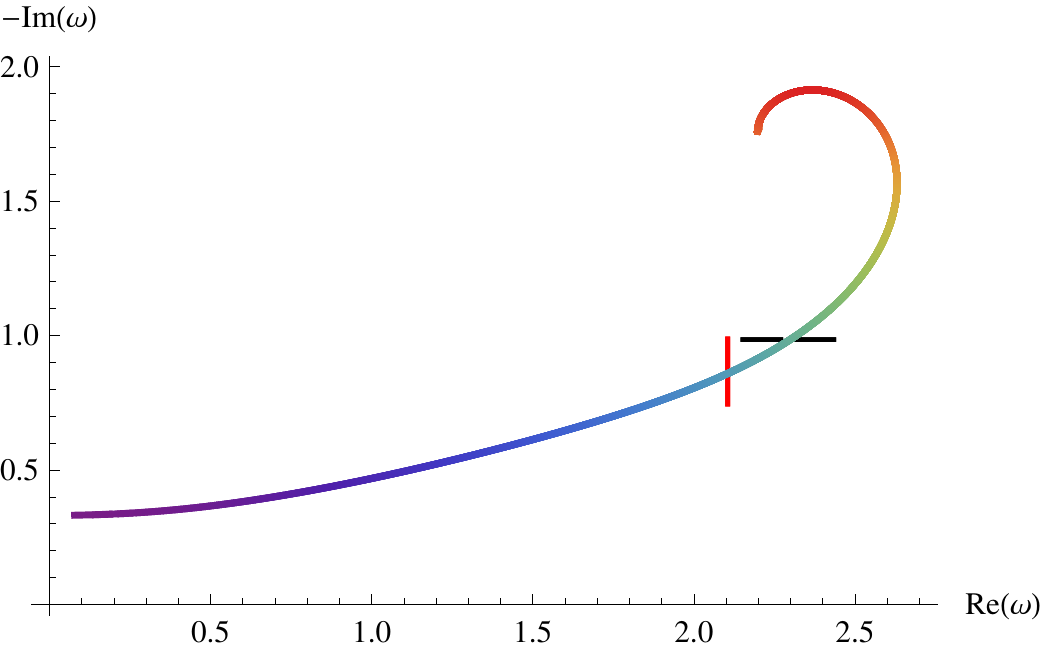}
\includegraphics[scale=0.7]{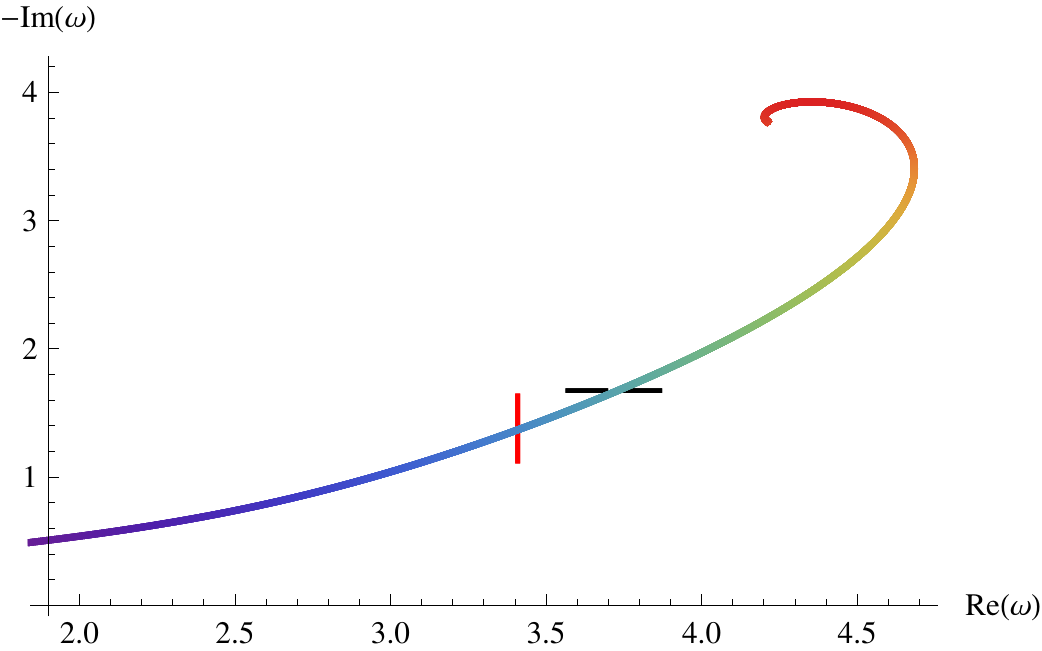}
\caption{\label{fig:scalarK0}
Location of the first and second quasinormal modes in the complex
frequency plane for the {\it scalar} fluctuations at
vanishing momentum ($k=0$) as a function of the embedding $\chi_0$. 
Red color indicates small quark mass, or high temperature, while
the temperature decreases towards blue colors.
The horizontal (black) dash indicates the frequency at the phase transition where the 
angle is $\chi_0=0.939$ whereas teh vertical (red) dash indicates the onset of the instability at
$\chi_0=0.962$.}

\end{center}
\end{figure}

\paragraph{Results for the scalar} 
The first and second scalar QNM at vanishing momentum can be found
in figure~\ref{fig:scalarK0}. The basic behavior is similar to that of the vector modes.
Increasing the quark mass from zero the real part of the QNM frequency
again shows a turning behavior moving first to larger values, then to smaller
values of $\text{Re}(\omega)$. However, in contrast to the vectors, the 
scalar QNM frequency also shows a turning behavior in the imaginary
part $\text{Im} (\omega)$. This means that increasing the quark mass, i.e. 
decreasing the temperature, the corresponding modes first decay faster, then beyond the turning point they decay
slower and slower as 
the mode approaches the real axis. Moreover the scalar QNMs do not
asymptote to the real axis as fast as the vector QNMs do. Instead the
scalar QNM frequencies even at large masses still have a considerable 
imaginary part of roughly $1/2$. All the values for the scalar modes are excellent 
agreement with the ones obtained previously in \cite{Hoyos:2006gb}.

\begin{figure}[!h]
\begin{center}
\includegraphics[scale=0.7]{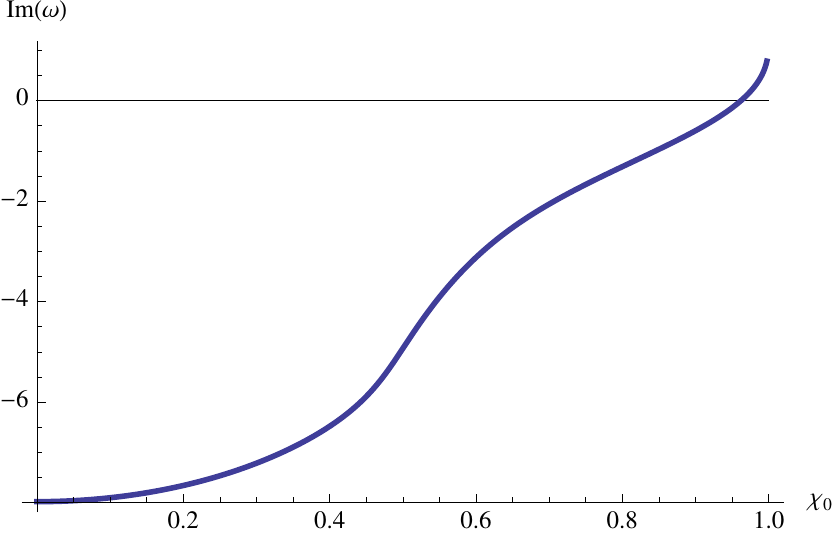}
\includegraphics[scale=0.7]{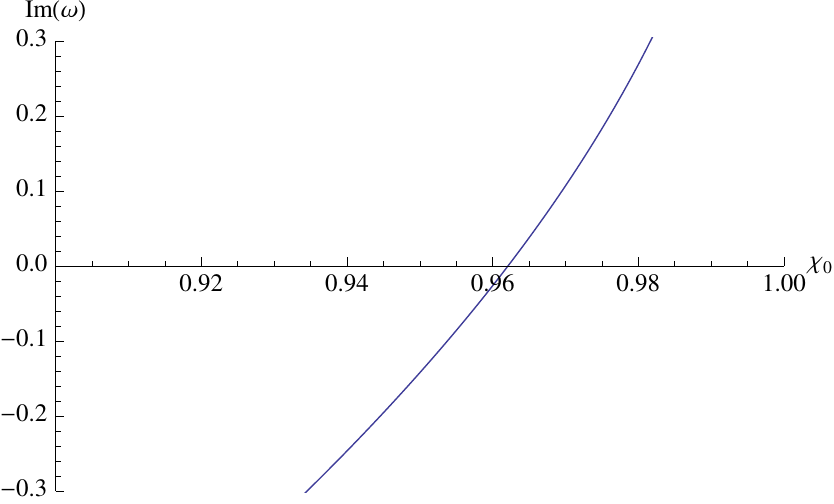}
\caption{\label{fig:tachyonK0D0}
Left: The plot  shows a purely imaginary quasinormal mode at $k=0$ as a function of the embedding. Right: Zooming into the region where the scalar
mode crosses the real axis becoming tachyonic approximately at $\chi_0=0.96221$
 }
\end{center}
\end{figure}

The short dash in the figures again shows the
location of the known meson melting transition where the initial angle is
$\chi_0=0.939$. Increasing the mass further while staying in the 
black hole phase, we observe a scalar QNM to become tachyonic. This point is marked by a vertcal dash in
the figures.
Figure~\ref{fig:tachyonK0D0} shows the appearance of an unstable mode explicitely. This particular
mode is special since it has vanishing real part but it starts with an extra-ordinary 
large imaginary part of the QNM frequency at zero quark mass. Increasing
the quark mass this purely damped mode moves closer
to the real axis until it crosses to become unstable at $\chi_0=0.962$ corresponding to the
maximal mass for black hole embeddings of $m\approx 1.31$. 
This particular mode has not been observed in previous studies because at 
vanishing quark mass it is located very deep in the complex frequency plane 
near $\text{Im}\w\approx -8$, while for example the first scalar QNM has $\text{Im}\w=-2$ 
at vanishing quark mass. In principle there could be an infinite tower of such purely
imaginary modes, each crossing the real axis at the larger and larger quark mass.
However, the accuracy of our numerics proved insufficient to establish additional modes beyond this lowest
one. In any case once this mode has crossed the real axis the D7 brane embedding is locally unstable
and can not be taken as a (metastable) ground state. This raises the question of what is the
true ground state in this regime. It might be that there is another type of D7-brane embedding
that is reached somehow by condensation of the scalar mode. Another possibility is that there is
simply no locally stable embedding beyond that point.

%_______________________________________________
\subsection{Schr\"odinger potential analysis}
\label{sec:schr-potent-noDnoK}
In this section we present a qualitative analysis of the quasinormal spectrum
using the fact that the equations of motion for the fluctuations can be
rewritten in the form of the Schr\"odinger equation (see Appendix
\ref{sec:appSchroedingerPot} for more details)
\begin{equation}
  \label{eq:schroedingereq}
  -\partial_{R*}^2\psi+V_S\psi=E\psi\,,
\end{equation}
where $R*$ is a tortoise-like coordinate. The Schr\"odinger potential $V_S$
determines the energy spectrum $E$ which is related to the quasinormal
spectrum by $E=\w^2$. 

\begin{figure}
  \centering
  \subfigure[]{\includegraphics[width=0.45\textwidth]{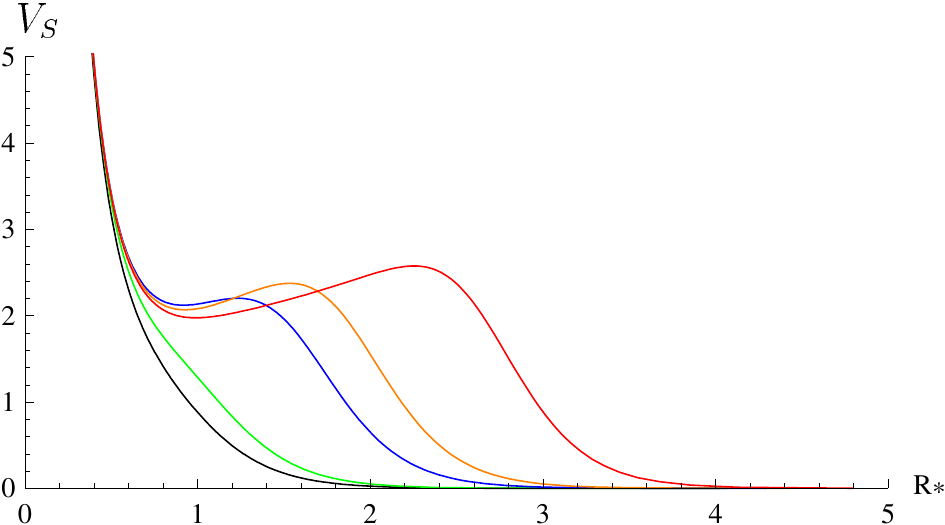}}
  \hfill
  \subfigure[]{\includegraphics[width=0.45\textwidth]{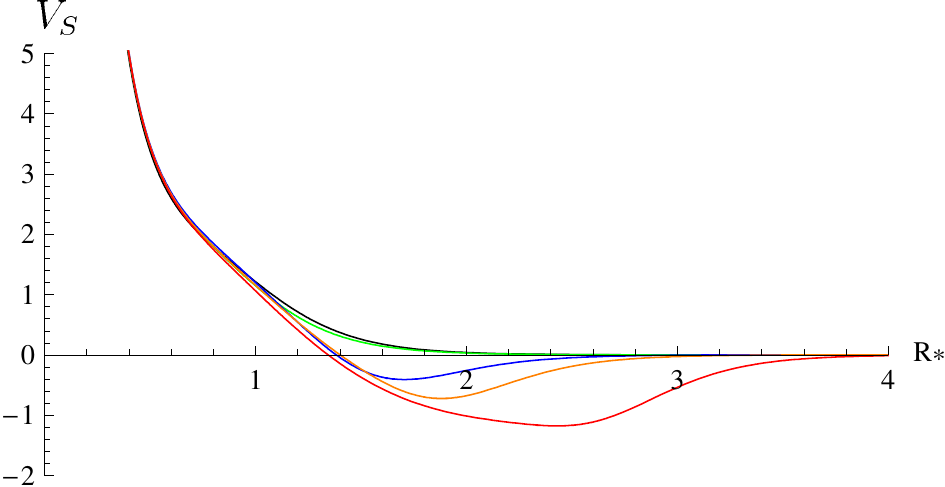}}
  \caption{Schr\"odinger potential of the vector (a)  and scalar (b)
    fluctuations for different values of $\chi_0$. The different colors
    correspond to $\chi_0=0$ (black), $0.5$ (green), $0.9$ (blue), $0.95$
    (orange), $0.99$ (red).}
  \label{fig:potdt0q0}
\end{figure}

At zero baryon density and zero momentum, the potentials for the
vector and scalar modes are already discussed in \cite{Paredes:2008nf} and
\cite{Myers:2007we}, respectively. Nevertheless we include the discussion here
for completeness. In fig.~\ref{fig:potdt0q0} we present the Schr\"odinger
potential for the vector and scalar fluctuations at different quark masses
parametrized by  $\chi_0$. In these plots we observe an infinite wall in the
potential at $R*=0$ which corresponds to the AdS boundary. In addition to this
wall the potential for the vector modes develops a step-shape as we increase
the quark mass. In \cite{Paredes:2008nf} it is shown that the imaginary part
of the quasinormal frequency decreases as the step gets longer which is
consistent with our result found in fig.~\ref{fig:vectorK0}.  

 For the scalar modes a negative well arises in the Schr\"odinger potential.
 This well becomes deeper and wider as we increase the quark mass and
 therefore support a `bound' state with $E<0$ which corresponds to a tachyonic
 quasinormal frequency $\im\w>0$. This well and the `bound' state are studied
 in \cite{Myers:2007we}. The Schr\"odinger analysis clearly shows the
 existence of a tachyonic mode which we already found in
 fig.~\ref{fig:tachyonK0D0}. 

%______________________________________________
\subsection{Discussion: Tachyon and De-singularization}

In this section we discuss why the so-called {\it undercooled phase} shows unphysical meson
spectra which do not approach the known ones in the supersymmetric limit. Further we 
discuss that finite density cures this behavior by de-singularizing the
geometry, \ie by smoothing out the limiting embedding.  

As mentioned above the scalar fluctuation becomes tachyonic once the
quark mass parameter has reached its maximum as a function of $\chi_0$. It is not to be expected that the region beyond
that point contains physically 
relevant or meaningful signatures. This region contains the limiting embedding
which only touches the horizon and geometrically separates Minkowski from black hole embeddings.
Here meson spectra had been studied earlier~\cite{Myers:2007we,Paredes:2008nf}. These meson spectra display a singular behavior in
the sense that all the quasinormal modes (first, second, \dots) approach one single
'attractor' energy (or frequency) near the limiting embedding. The geometric reason for 
this is the scaling symmetry for the embeddings in the near-critical region~\cite{Mateos:2007vn}. 
That scaling symmetry implies that near the critical quark mass (or temperature) there 
exists no preferred scale on the brane. In this sense there is no scale which could determine the 
distance between resonances in the brane fluctuations, i.e. between the distinct 
meson mass resonances, or quasinormal modes equivalently.

At finite fixed baryon density however, this particular scaling symmetry is broken~\cite{Kobayashi:2006sb}.
Therefore the chemical potential introduced together with that density does set the scale
for the separation between the distinct quasinormal modes. This is somewhat analogous to 
the behavior of the thermal quasinormal mode spectrum when the temperature of the dual black hole 
background is lowered towards zero. In that case the quasinormal modes are known to
approach one single value, i.e. the spectrum becomes singular in our sense. 
There the scale which determines the distance between distinct 
excitations is clearly the temperature since it is the only scale in the theory. In the 
zero temperature limit this scale vanishes and the scaling symmetry is restored.            
 
As observed earlier~\cite{Erdmenger:2007ja} the meson spectrum at finite density approaches
the supersymmetric one at low temperatures. In this sense the theory is de-singularized 
by finite baryon density. We will see this explicitly in the quasinormal mode spectra
at finite density in section~\ref{sec:finiteD}.

%%%%%%%%%%%%%%%%%%%%%%%%%%%%%%%% F I N I T E    M O M E N T U M
\section{Finite momentum but vanishing density} \label{sec:finiteK}
We now turn to the case of non-zero 
spatial momentum.
%______________________________________________
\subsection{Transverse vectors}\label{sec:transVecK}
We obtain results at finite momentum from the numerical solution
of equation~\eqref{eq:transVecEom} with distinct non-zero values of $k$.
\begin{figure}[!h]
 \begin{center}
$$
\begin{array}{cc}
\includegraphics[scale=0.8]{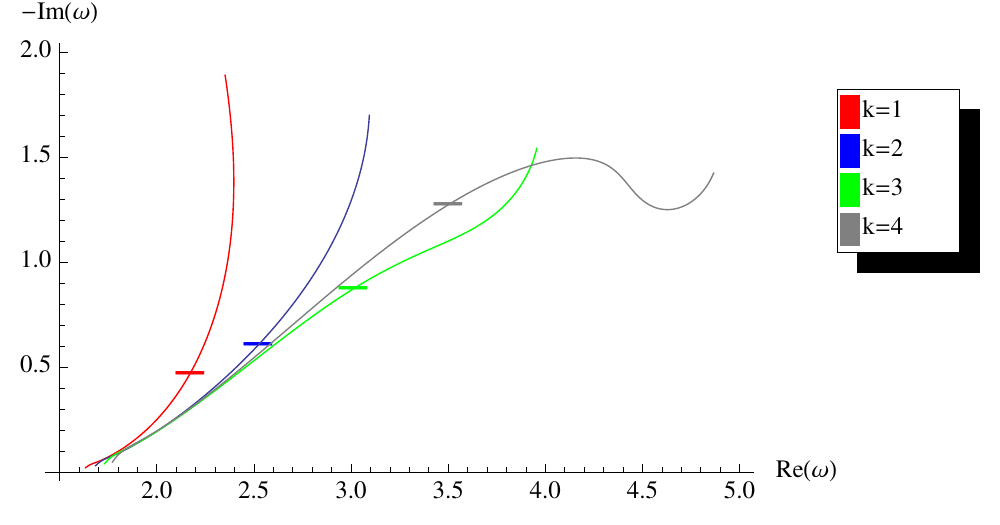} & \includegraphics[scale=0.8]{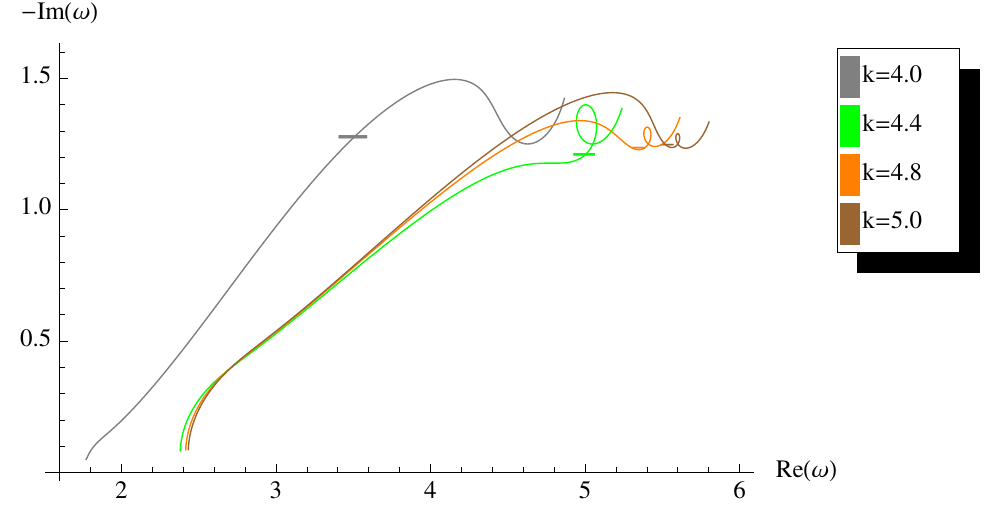}\\
\includegraphics[scale=0.8]{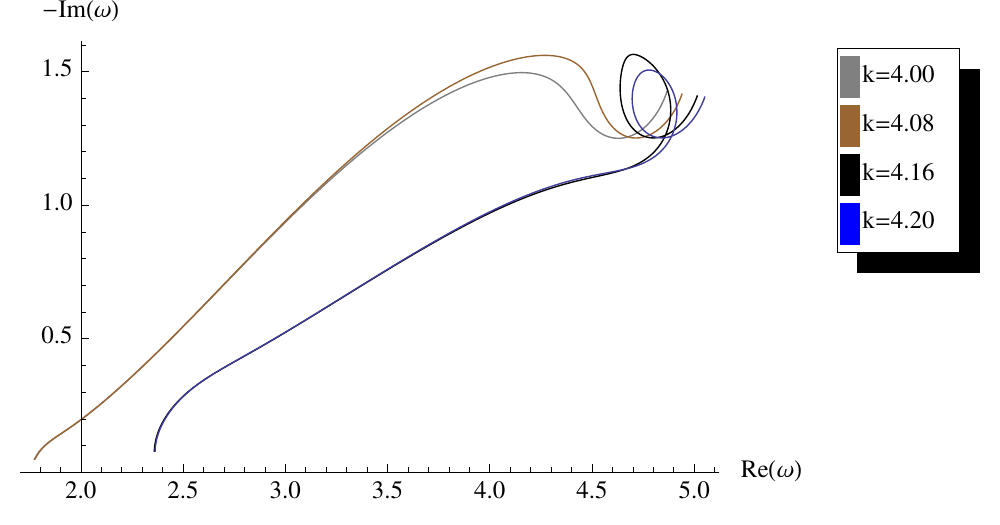} & \includegraphics[scale=0.8]{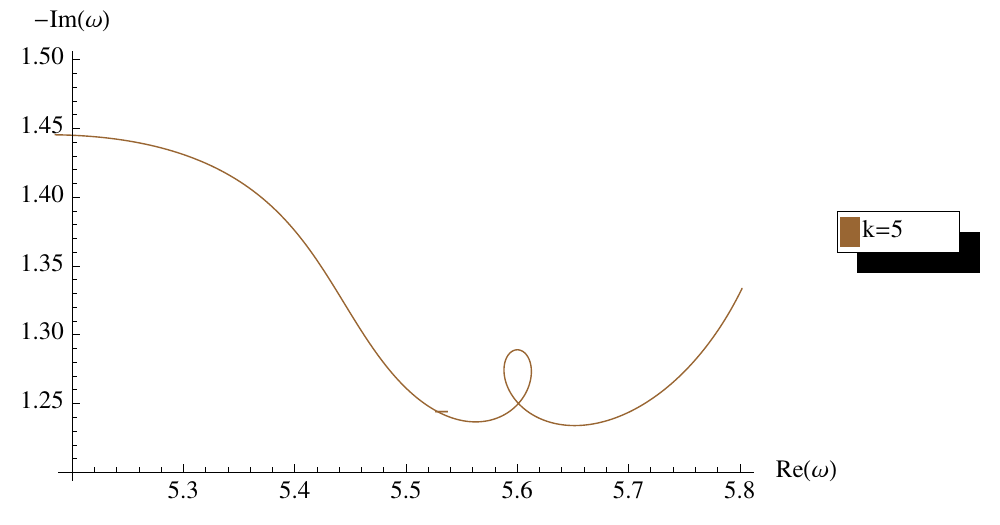}
\end{array}
$$
\caption{\label{fig:transVecK}
First quasinormal mode for transverse vector fluctuations at
different spatial momenta $k$. Horizontal short dashes across the curves
show the location of the meson melting transition. At $k\approx 4.16$
a spiral structure appears and the curves also asymptote to a distinct 
small temperature value from that $k$ on. Along the curves the quark
mass parameter $\chi_0=\cos\Theta_0$ is varied. See figure~\ref{fig:massplot} for its
relation to the quark mass and temperature. }
\end{center}
\end{figure}

Figure~\ref{fig:transVecK} shows the first of the transverse vector QNMs at different
values of the spatial momentum $k$. The behavior in the region 
$k=0, \dots, 4$ is very similar to the $k=0$ case. Although its trajectory
in the complex frequency plane becomes more wavy at larger $k$, the 
first scalar QNM still starts at quite large real and imaginary parts in order to 
approach the real axis and smaller real parts when temperature is decreased.
Distinct curves for different values of $k$ within numerical accuracy approach a 
single limiting value $\omega_0$ at small temperatures.
It is interesting to note that the turning point in 
$\text{Re}(\omega)$ mentioned
in the previous section for $k=0$ disappears when $k$ reaches values 
between $k=1$ and $k=2$.

A quite substantial qualitative change appears at $k=4.16$, where the
trajectory of the mode in the frequency plane develops one loop. Also
the trajectories at higher $k$ have this looping behavior. At the same time
these curves with one loop do asymptote to a single small temperature 
frequency value $\omega_1$ as well. But this limiting value is distinct from the limiting value which is approached by the low $k$ curves without the loop, i.e. 
$\omega_1\not =\omega_0$. This fact 
suggests that the loop-behavior and the distinct limiting value are somehow
related. For the longitudinal vector fluctuations (see figure~\ref{fig:longVecK}) 
we will explicitly see that this relation generalizes
to all fluctuations and to higher loops in this way: All the first QNM trajectories 
for scalar and vector fluctuations with $k<k_n$ have $n$ loops and they asymptote to
a small temperature limit frequency $\omega_n$ (within numerical 
accuracy) with $\omega_{n+1}>\omega_{n}$. Note that the loop behavior
appears in a physical, thermodynamically preferred phase, 
i.e. {\it before} the meson melting transition and also way before 
the tachyon appears. 

The loops are absent in the second quasinormal mode as seen from figure~\ref{fig:transVecK2ndQNM}.
Nevertheless the second quasinormal mode also asymptotes to distinct
low temperature limits $\omega_n$ above distinct certain momenta $k_n$.

\begin{figure}[!h]
 \begin{center}
$$
\begin{array}{cc}
\includegraphics[scale=0.7]{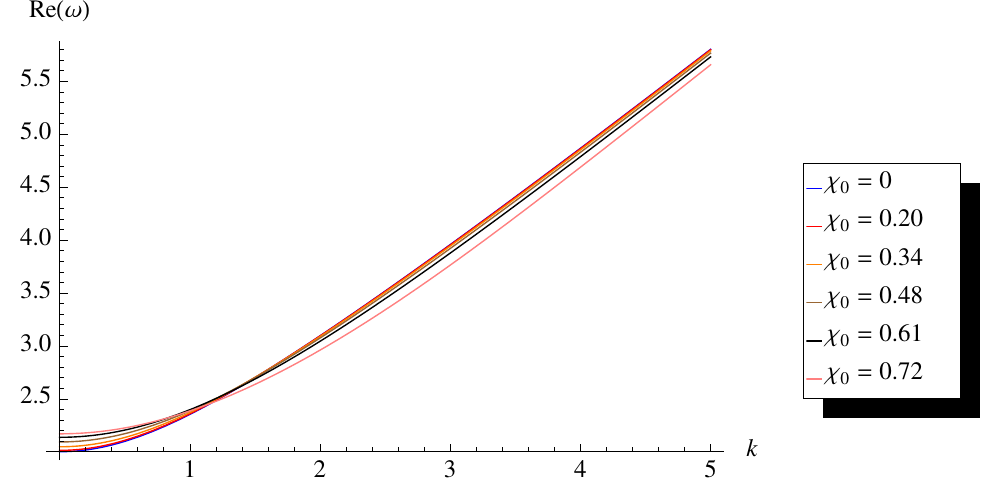} & \includegraphics[scale=0.7]{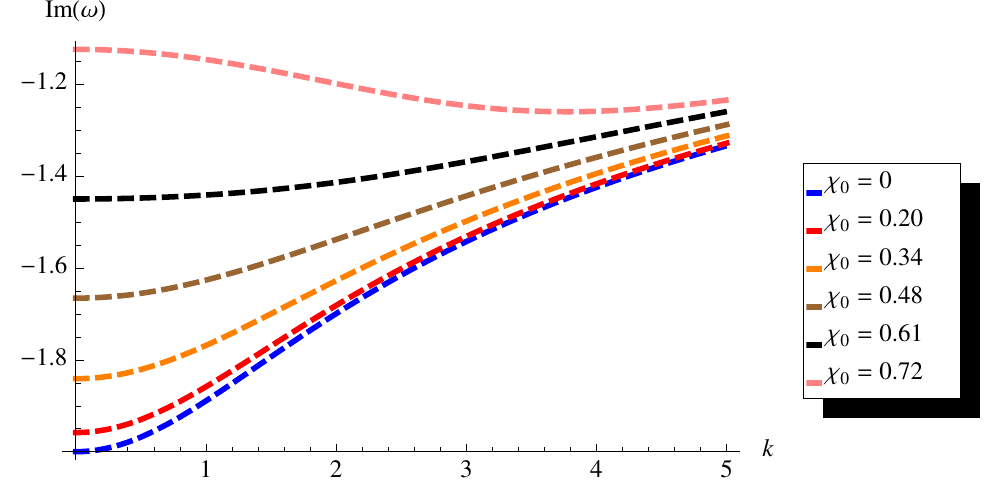}
\end{array}
$$
\caption{\label{fig:dispTransVec}
Dispersion relation for the first transverse quasinormal mode at distinct
quark masses, or equivalently temperatures, parametrized by the embedding 
parameter at the horizon, $\chi_0$. See figure~\ref{fig:massplot} for the
relation between $\chi_0$, the temperature and quark mass $M_q$.}
\end{center}
\end{figure}

Figure \ref{fig:dispTransVec} captures the dispersion relation of
the first transverse vector QNM at different values
of $\chi_0$.

%______________________________________________
\subsection{Longitudinal vectors}
We obtain results at finite momentum from the numerical solution
of equation~\eqref{eq:longVecEom} with distinct non-zero values of $k$.
\begin{figure}[!h]
 \begin{center}
$$
\begin{array}{cc}
\includegraphics[scale=0.7]{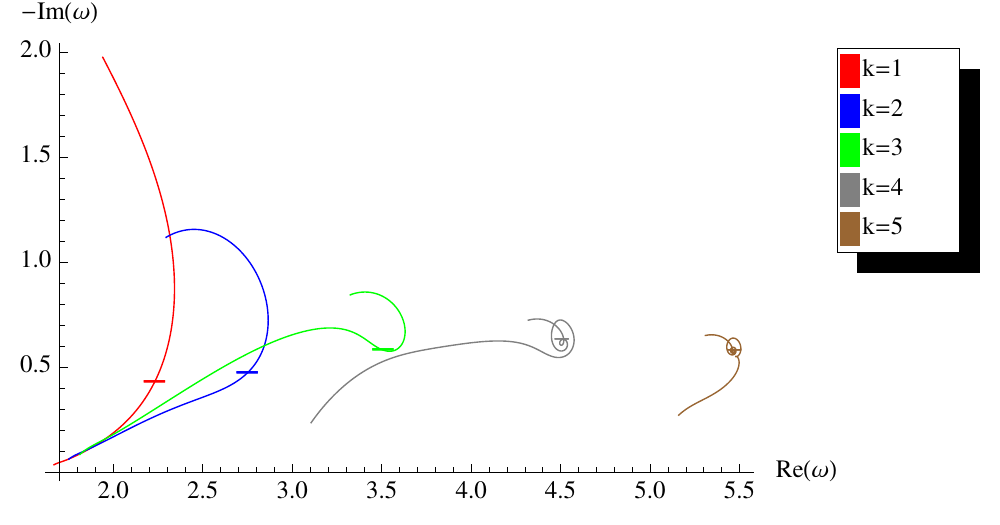} & \includegraphics[scale=0.7]{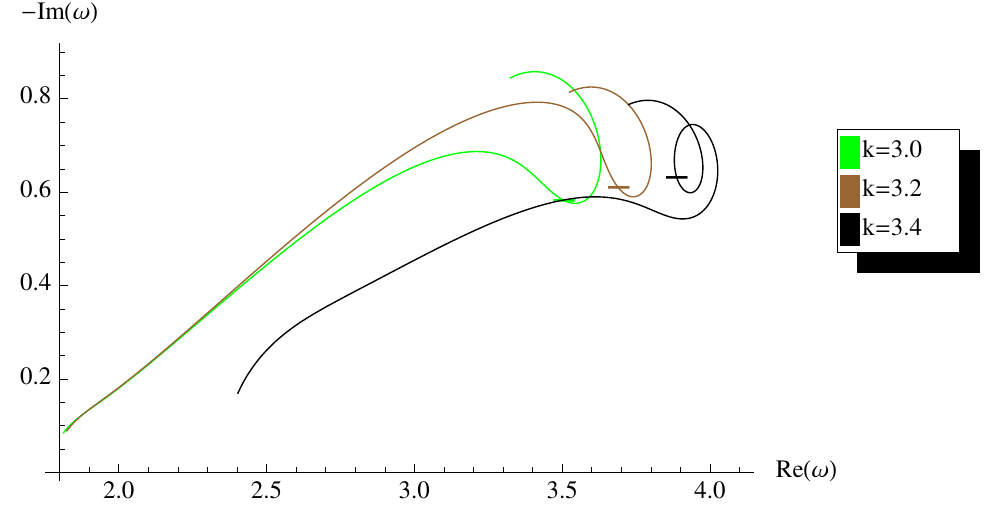}\\
\includegraphics[scale=0.7]{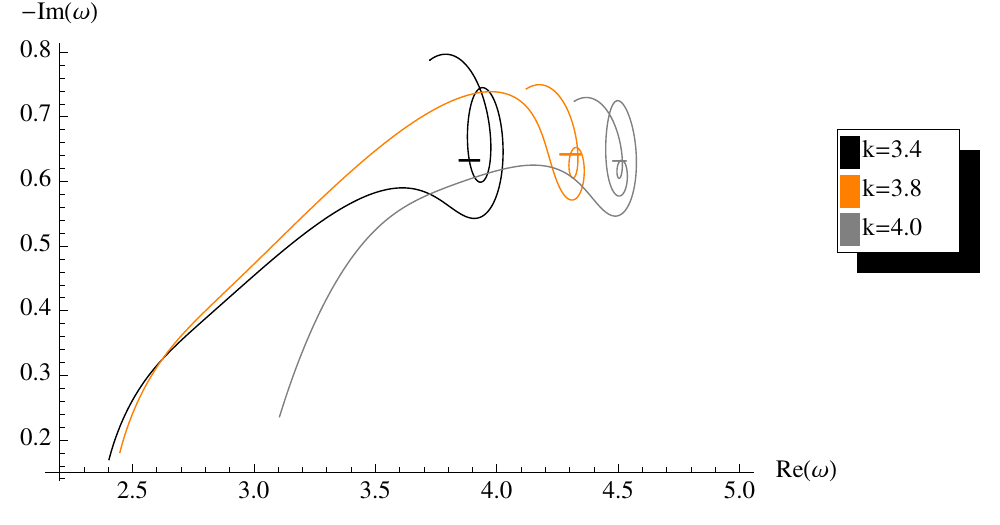} & \includegraphics[scale=0.7]{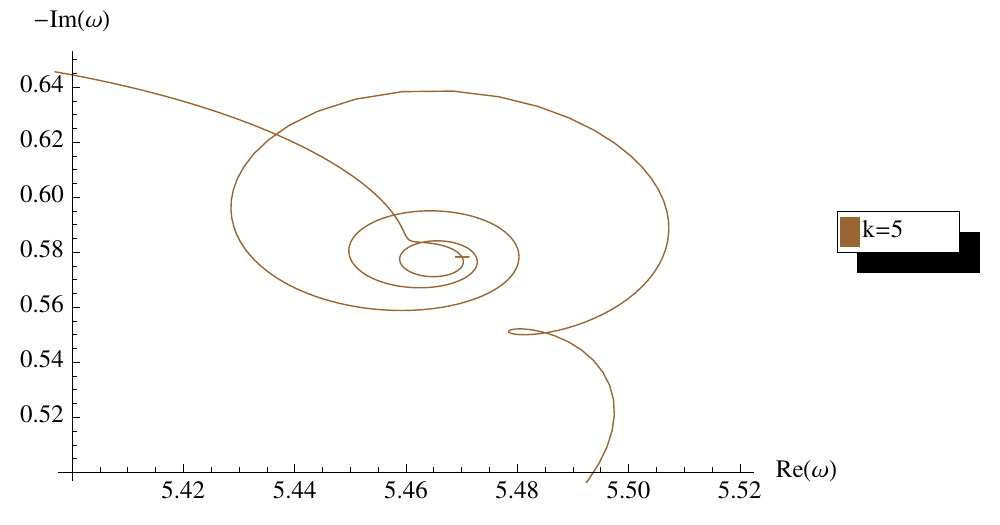} 
\end{array}
$$
\caption{\label{fig:longVecK}
The first quasinormal mode for longitudinal vector fluctuations at distinct
values of k. Along the curves the quark
mass parameter $\chi_0=\cos\Theta_0$ is varied. See figure~\ref{fig:massplot} for its
relation to the quark mass and temperature.}
\end{center}
\end{figure}

The behavior of the longitudinal vector QNMs is qualitatively similar to that of the 
transverse ones discussed in section~\ref{sec:transVecK}. The mentioned loops in
the frequency plane trajectory do appear at smaller values $k\approx 3.4$ in the 
longitudinal channel than they do in the transverse one. However, while in the
transverse vector case the
loops appeared before the meson melting transition, in the longitudinal case
the transition takes place before the first loop is terminated as can be seen 
from the figure~\ref{fig:longVecK}. Just as for the transverse vectors, also the
second QNM of the longitudinal vectors does not have any loops in its
complex frequency plane trajectory, as figure~\ref{fig:longVecK2ndQNM} shows.

The dispersion relations for the first and second QNM of the longitudinal
vector fluctuation are depicted in figure~\ref{fig:dispLongVec} and
figure~\ref{fig:dispLongVec2ndQNM} in appendix~\ref{app:secondQNMs}, respectively.
\begin{figure}[!h]
 \begin{center}
$$
\begin{array}{cc}
\includegraphics[scale=0.7]{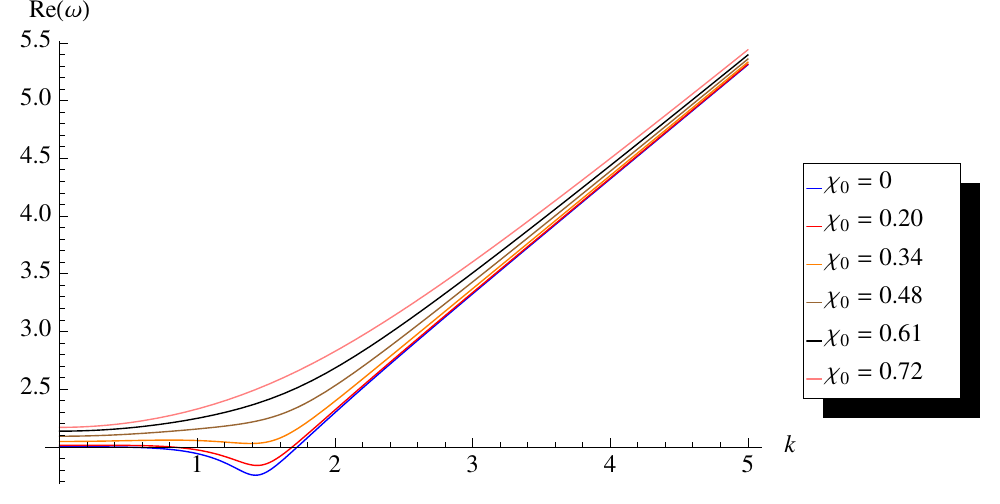} & \includegraphics[scale=0.7]{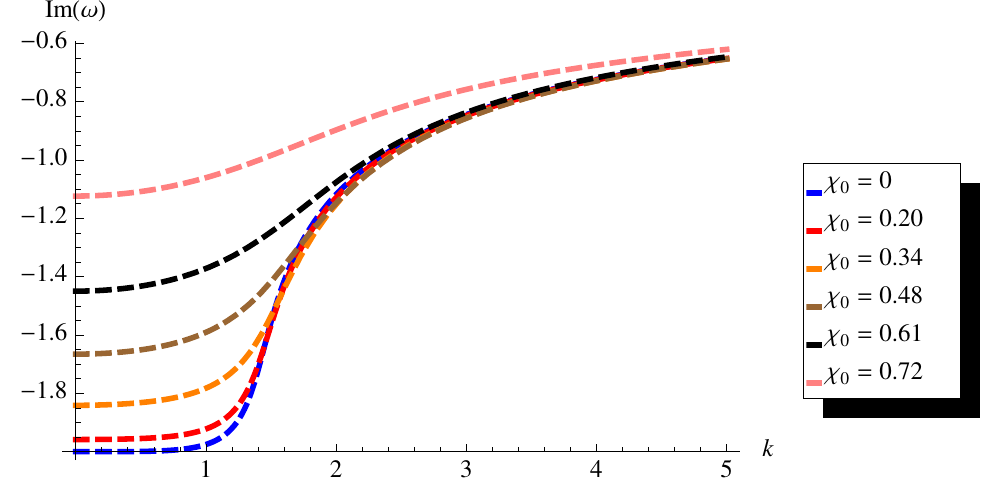}
\end{array}
$$
\caption{\label{fig:dispLongVec}
Dispersion relation for the first quasinormal longitudinal vector mode fluctuation at 
distinct values of the quark mass parameter $\chi_0$ (see figure~\ref{fig:massplot} for 
its relation to the quark mass at vanishing density).}
\end{center}
\end{figure}

%______________________________________________
\subsection{Scalar}
We obtain results at finite momentum from the numerical solution
of equation~\eqref{eq:scalarEom} with distinct non-zero values of $k$.

\begin{figure}[!h]
 \begin{center}
$$
\begin{array}{cc}
\includegraphics[scale=0.7]{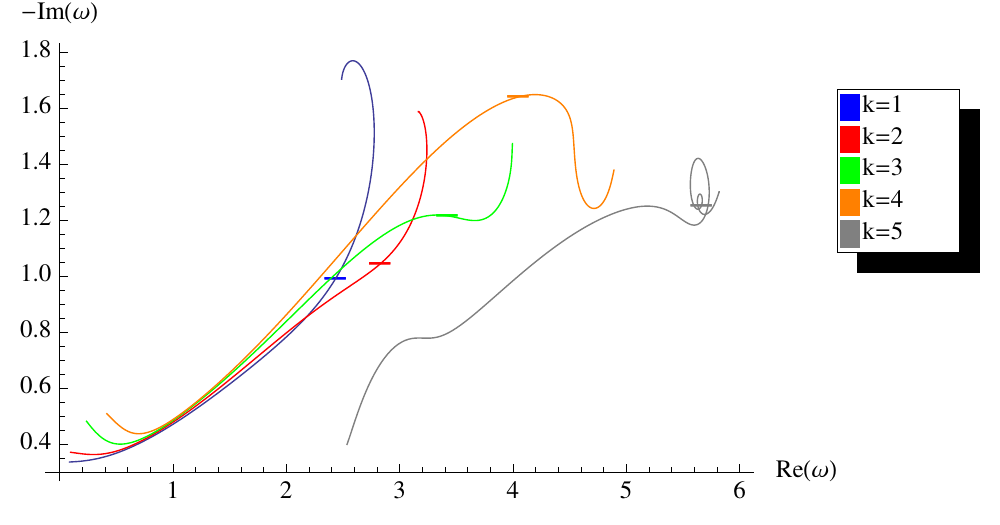} & \includegraphics[scale=0.7]{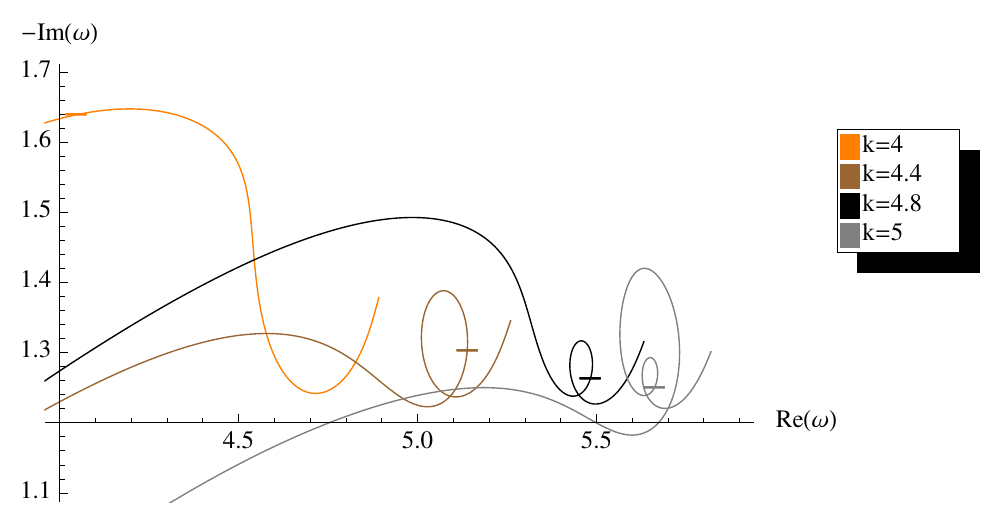}
\end{array}
$$
\caption{\label{fig:scalarK}
First scalar quasinormal mode at distinct momenta $k$. Along the curves the quark
mass parameter $\chi_0=\cos\Theta_0$ is varied. See figure~\ref{fig:massplot} for its
relation to the quark mass and temperature.}
\end{center}
\end{figure}

\begin{figure}[!h]
 \begin{center}
$$
\begin{array}{cc}
\includegraphics[scale=0.7]{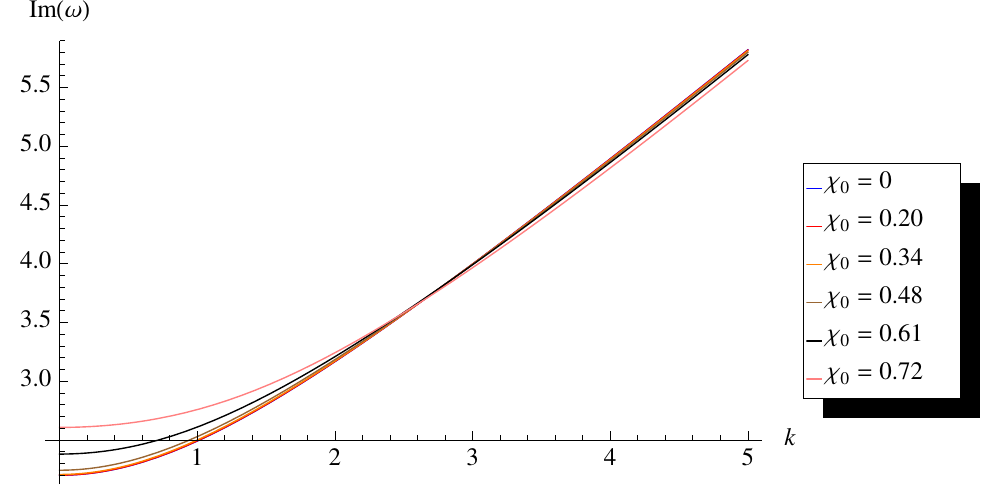} & \includegraphics[scale=0.7]{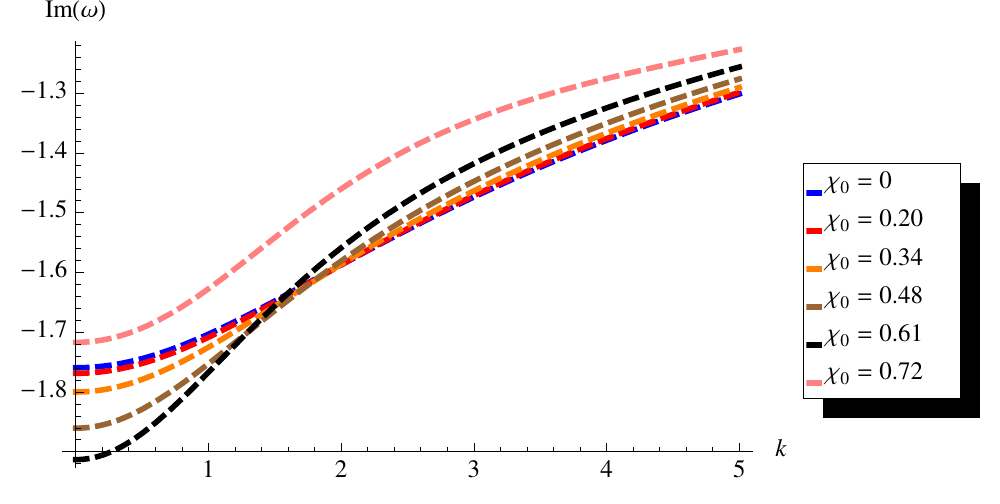}
\end{array}
$$
\caption{\label{fig:dispScalarK}
Dispersion relation for the first scalar quasinormal mode at distinct
values for the quark mass parameter $\chi_0$ (see figure~\ref{fig:massplot} for 
its relation to the quark mass at vanishing density). The real
part of the QNM frequency is shown on the left, the imaginary part on the right.}
\end{center}
\end{figure}

The general behavior of the scalar QNMs is qualitatively similar to that
of the longitudinal vector QNMs. Figure~\ref{fig:scalarK} shows the first
of the scalar QNMs at momenta between $k=1$ and $k=5$. Also in this
case the overall behavior is that the real and imaginary parts decrease
as temperature is decreased along the curves. At small $k$, e.g. $k=1$,
there is a turning point present in the real as well as in the imaginary part.
These turning points again disappear between $k=1$ and $k=3$. Just 
like for the first of the longitudinal vector QNMs multiple loops form 
successively for larger values of $k$. The meson melting transition appears
before the first of the loops has terminated. Again the number of loops seems to
be directly related to the low temperature value $\omega_n$ to which the 
curves for all $k>k_n$ asymptote. Dispersion relations for the first scalar
QNM are shown in figure~\ref{fig:dispScalarK}. The corresponding figures for 
the second scalar QNM are figure~\ref{fig:scalarK2ndQNM} 
and~\ref{fig:dispScalarK2ndQNM} in appendix~\ref{app:secondQNMs}. Just like for the vectors there are 
also no loops in the second QNM for the scalars.

%_______________________________________________
\subsection{Schr\"odinger potential analysis}
%----------------------------------------
Just as in the previous section at zero density and momentum, we
here compute the effective potential for the scalar and vector fluctuation
equations~\eqref{eq:scalarEom} 
and~\eqref{eq:longVecEom},~\eqref{eq:transVecEom} (see Appendix
\ref{sec:appSchroedingerPot} for more details). 

\begin{figure}
\includegraphics[width=0.9\textwidth]{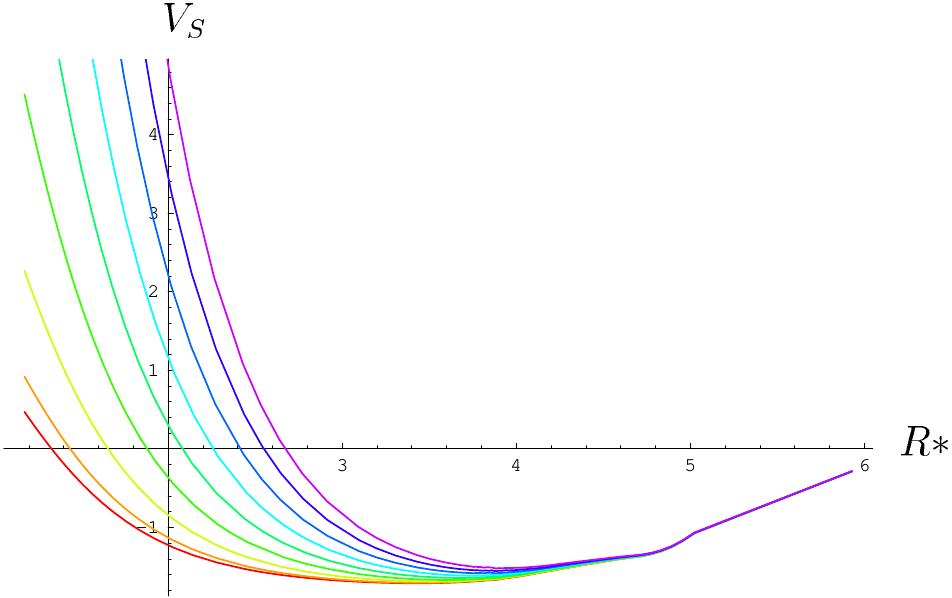}
\caption{ \label{fig:scalarSchroePotFiniteK}
The {\it scalar} Schr\"odinger potential $V_s$ (zooming in on the minimum of
the potential) versus the radial coordinate $R*$ defined in section \ref{sec:schr-potent-noDnoK}
at increasing momenta $k=0,\, 2,\, 4,\,6,\,8,\,10,\,12,\,14,\,16$ 
from bottom to top curve with the quark mass parameter $\chi_0=0.9999$ (see figure~\ref{fig:massplot} for 
its relation to the quark mass at vanishing density), $\tilde d=0$. 
The dip supporting the tachyon narrows.
}
\end{figure}

\begin{figure}
\includegraphics[width=0.9\textwidth]{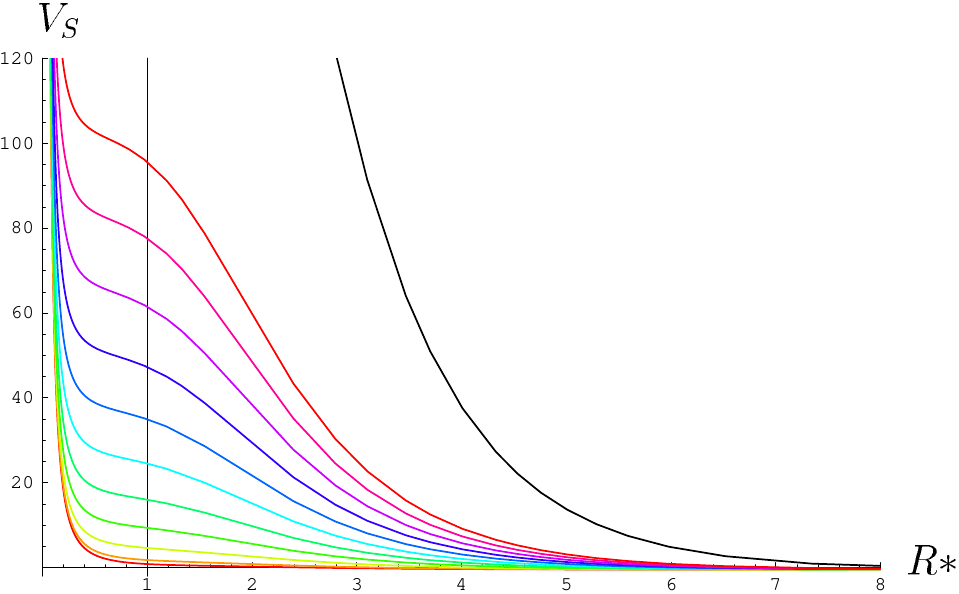}
\caption{ \label{fig:scalarSchroePotFiniteKStep}
The {\it scalar} Schr\"odinger potential $V_s$ (zooming out to larger values
of the potential) versus 
the radial coordinate $R*$ 
at increasing momenta $k=0,\, 2,\, 4,\,6,\,8,\,10,\,12,\,14,\,16,\,18,\,20,\,40$ 
from bottom to top curve with quark mass parameter 
$\chi_0=0.9999$ (see figure~\ref{fig:massplot} for 
its relation to the quark mass at vanishing density), $\tilde d=0$. 
A step forms towards the boundary, similar  
as that for the vectors. The grid line at $R*=1$ serves to guide the eye only.
}
\end{figure}

We begin by examining the Schr\"odinger potential for the scalar fluctuations
in figure~\ref{fig:scalarSchroePotFiniteK}. The lowest (red) curve shows the
potential at vanishing momentum and density at $\chi_0=0.9999$. That is 
near the limiting embedding, far beyond the thermodynamic transition to
Minkowski embeddings and far beyond the appearance of the tachyon in the
spectrum. The results at smaller $\chi_0$ are qualitatively the same, but we plot
large $\chi_0$ in order to investigate the tachyon and  the reason for 
the different "attractor" frequencies in that large mass regime.
The scalar potential clearly exhibits a wide negative dip in which the
tachyonic scalar mode resides, compare figure~\ref{fig:tachyonK0D0}. 
As the momentum is increased the potential is lifted and the negative
dip is narrowed. In this way the lowest possible excitation is pushed towards
more positive energy values becoming non-tachyonic at large momenta. 
However the theory is clearly unstable against condensation of the scalar
fluctuations already at $\chi_0\ge 0.962$ and $k=0$.
In figure~\ref{fig:scalarSchroePotFiniteKStep} we zoom out to larger values of
the potential. For increasing momentum, a step forms near the boundary. When comparing
to figure~\ref{fig:potdt0q0} (a) we see that the scalar
potential at finite momentum is similar to the vector case at zero momentum.
In figure~\ref{fig:scalarSchroePotFiniteKStep} the step becomes higher and
longer for increasing momentum, while its plateau becomes shorter, i.e.  
most of the plateau is located near the boundary. Therefore conceptually the 
analysis of~\cite{Paredes:2008nf} as discussed in section
\ref{sec:schr-potent-noDnoK} applies as in the vector case: When  
the step becomes longer, the imaginary part of the quasinormal frequencies
decreases. This is consistent with our observations in figure~\ref{fig:scalarK} (compare for example the 
initial points where $\chi_0=0$ for increasing momenta from curve to curve). 
The increasing real part of the quasinormal frequency observed in
figure~\ref{fig:scalarK} is due to the fact that the potential step rises
closer to the horizon at larger momentum. This narrows the potential 
and rises the excitation energies.

\begin{figure}
\includegraphics[width=0.9\textwidth]{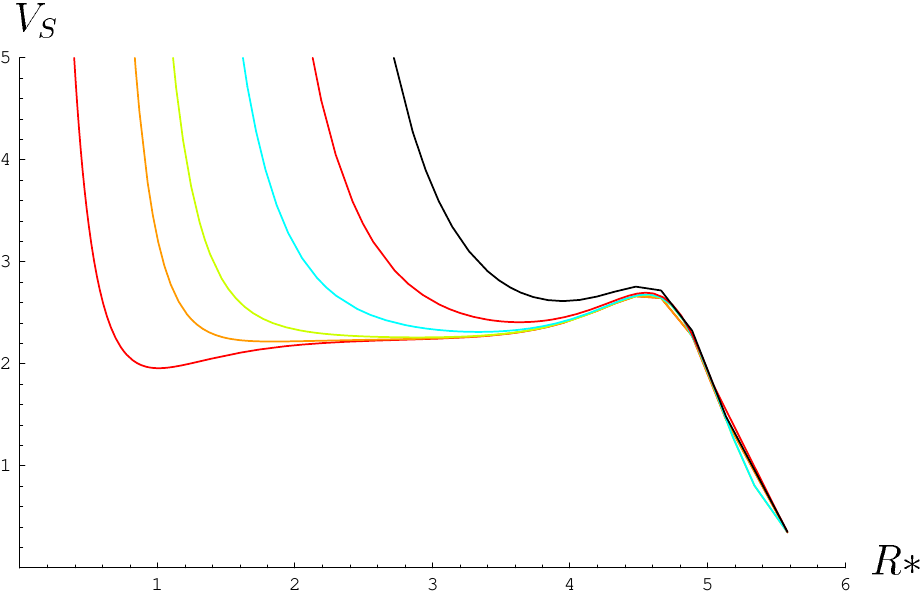}
\caption{ \label{fig:transVectorSchroePotFiniteK}
The {\it transverse vector} Schr\"odinger potential $V_s$ versus 
the radial coordinate $R*$ 
at increasing momenta $k=0,\, 2,\, 4,\,10,\,20,\,40$ 
from bottom to top curve with quark mass parameter $\chi_0=0.9999$
(see figure~\ref{fig:massplot} for 
its relation to the quark mass at vanishing density), $\tilde d=0$. The potential step
becomes shorter since the potential rises farther and farther away
from the boundary $R*=0$.
}
\end{figure}

Turning now to the transverse vector fluctuations, we observe a step
potential in figure~\ref{fig:transVectorSchroePotFiniteK}. The larger the momentum,
the earlier the potential rises towards infinity when approaching the boundary 
at $R*=0$. So effectively the boundary moves towards the horizon
and the length of the plateau of the potential step becomes shorter.
This is different from the scalar modes discussed above. While the imaginary
part of the quasinormal frequency decreases for increasing momentum just as in
the scalar case, the Schr\"odinger potential shows a different behavior: For
increasing momentum, the Schr\"odinger potential approaches the shape of a
wall. Where as in the scalar case, the formation of the step is responsible
for lowering the imaginary part of the quasinormal frequency, here we expect that
the formation of the wall is responsible for a similar decrease of the
imaginary part of the quasinormal frequency.

%______________________________________________
\subsection{Discussion: Breakdown of hydrodynamics, 'attractors' and the tachyon}
In this discussion we focus on three distinct physical implications of the 
quasinormal modes described previously in this section. The longitudinal vector modes 
tell us when the hydrodynamic approximation breaks down,
while the novel purely imaginary scalar mode renders the whole theory
unstable as it becomes tachyonic. Both scalar and vector modes asymptote
to 'attractor' frequencies. This behavior is probably related to a spiraling behavior
of the quasinormal mode's trajectories with changing quark mass or temperature.
\subsubsection{Hydrodynamics to Collisionless Crossover}
\label{sec:hydroToCollisionless}
\begin{figure}[!h]
 \begin{center}
$$
\begin{array}{cc}
\includegraphics[scale=0.7]{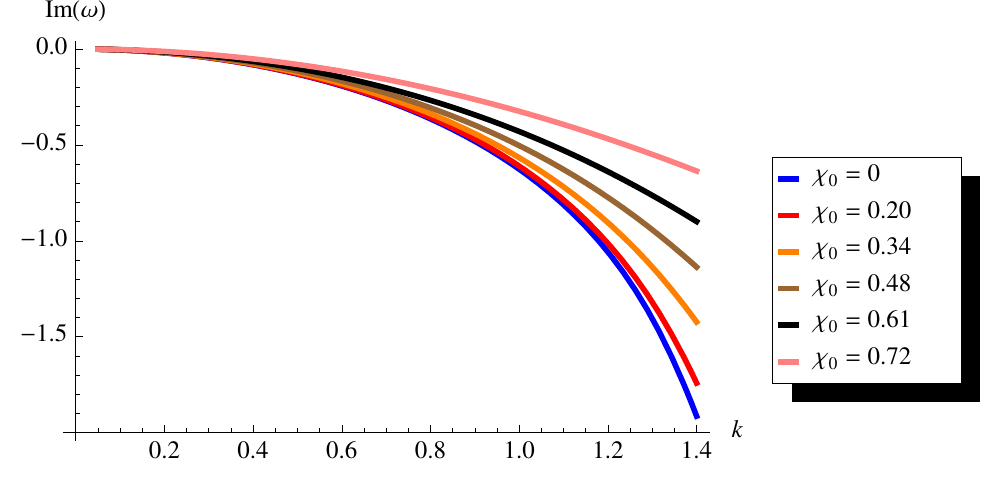} &  \includegraphics[scale=0.7]{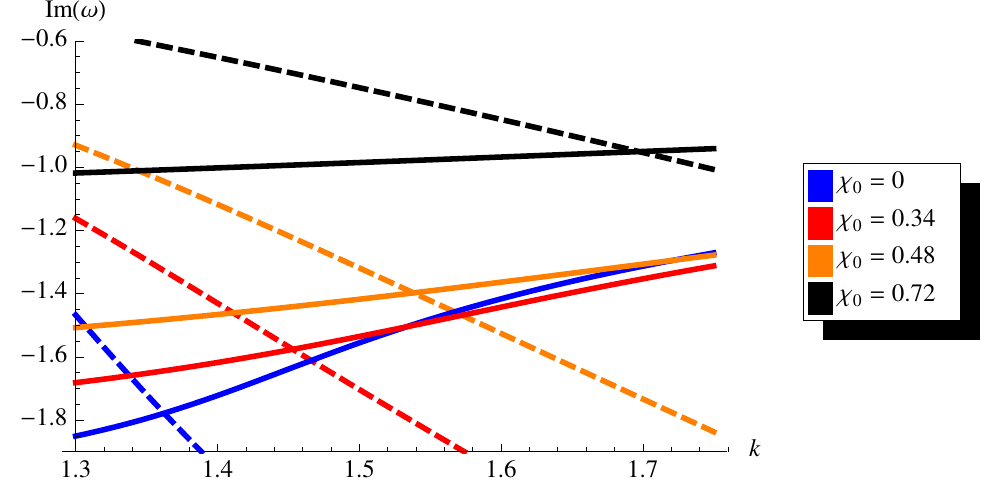}
\end{array}
$$
\end{center}
\caption{Left: The dispersion relation for the diffusion mode at distinct
values for the mass/temperature parameter $\chi_0$ (see figure~\ref{fig:massplot} for 
its relation to the quark mass at vanishing density). 
Right: Intersection of the diffusive mode with the imaginary part of the first longitudinal quasinormal mode. The intersection point moves to larger values of $k$, but to smaller
values of $\text{Im}(\omega)$ as the mass/temperature parameter
$\chi_0$ is increased.}
\label{fig:vecDiffDispersion}
\end{figure}

Contrary to the transverse vector and scalar channel the longitudinal vector channel also
has a hydrodynamic quasinormal mode, i.e. a mode whose dispersion relation does not show a gap at zero momentum, $\lim_{k\rightarrow} \omega(k) =0$ (see figure
\ref{fig:vecDiffDispersion}, left).
This mode represents diffusion of baryon charge. It is a mode whose frequency is purely imaginary and therefore results in a purely damped time evolution without any oscillation. At small momentum the dispersion relation is well approximated by the diffusion kernel $\omega = -i D k^2$, as seen in figure~\ref{fig:vecDiffDispersion} on the left. 
Fitting our numerical data
to this we can extract the diffusion constant $D$. Since it has been calculated before in
the literature in \cite{Myers:2007we} we do not restate this result. We checked however that our values are consistent with the results there. 

On general grounds one expects that a many body system shows a crossover from
hydrodynamic behavior at long wavelengths to a coherent or collisionless behaviour at small wavelengths.
In the holographic context this has first been discussed in an $AdS_4$  example in \cite{Herzog:2007ij} by calculating spectral functions. A more direct way to see this
crossover can be obtained by studying the quasinormal mode spectrum. At small
momentum the hydrodynamic mode should be the dominant one, i.e. the one with the 
largest imaginary part. At small wavelength we expect the dominant modes to
have  frequencies whose imaginary part is much smaller than their real parts. 
This means that at large wavelength the response would be simply an
exponential decay whereas at short wavelength the response would be a slowly
decaying oscillation. 
In terms of the quasinormal mode spectrum this implies that the purely imaginary diffusion mode as a function of
momentum has to cross the imaginary part of the dispersion relation for the
lower non hydrodynamic modes. Indeed this is what happens for the R-charge and 
momentum diffusion in the strongly coupled $\mathcal{N}=4$ theory as discussed in \cite{Amado:2007yr,Amado:2008ji}. 
We therefore define the crossover from the hydrodynamic regime to the collisionless regime
through the momentum at which the imaginary part of the lowest
non-hydrodynamic mode crosses the purely imaginary diffusion mode. From that
wavelength on it is the lowest gaped quasinormal mode which dominates the late
time response\footnote{There is a small puzzle related to that. If the
  prolongation of the hydrodynamic mode to large momenta is constantly
  increasing, the front velocity computed from it seems to violate causality.
  As is well-known already the diffusion kernel violates causality because of
  the $k^2$ behaviour and the extension to larger momenta shows even higher 
exponents in the dependence on $k$. So how manages the theory to preserve causality? The resolution has been presented in
\cite{Amado:2008ji}: the residues of the diffusive quasinormal mode vanish for large momenta and therefore this 
mode ceases to exist in
the dangerous limit of large $k$. On the other hand one can study the hydrodynamics by fixing a real frequency and then
looking for complex roots in the momentum $k$ as it has been done in
\cite{Amado:2007pv}. The quasinormal frequencies or the complex momenta
respectively are roots of infinite order polynomials (or at least of extremely
high order polynomials in a truncated approximation of the holographic Green
function). Therefore it is not possible to simple infer the behaviour of the
complex momentum modes from the quasinormal modes. Indeed as shown in
\cite{Amado:2007pv} the front velocity of hydrodynamic mode as calculated from
the complex momentum roots behaves perfectly causal and approaches $1$ to very
good numerical accuracy.}.  
It should be mentioned that there is at least one other way of how this
crossover can be established in terms of quasinormal modes. It could also be
that the purely imaginary diffusion mode pairs up with another purely
imaginary but non-hydrodynamic mode which allows them to develop real parts as
well and to move off the imaginary axis. This seems to be the preferred
mechanism for $AdS_4$ black holes \cite{Miranda:2008vb} and it also appears on
probe D-branes representing defects in the four-dimensional strongly coupled
CFT \cite{Myers:2008me}. As will be shown in a companion publication this is
also what happens if finite baryon density is introduced
\cite{collabsantiago}. This crossover has recently also been investigated for
AdS black holes in various dimension in the time domain in
\cite{Morgan:2009pn}.  

We have numerically determined the crossover point defined above for different embeddings.
As seen from the right side of figure~\ref{fig:vecDiffDispersion}, the
crossover moves to higher momenta as the embedding angle~$\chi_0$ is
increased. 
This means that the brane responds to baryon charge fluctuations in a purely
absorptive way for smaller and smaller wavelengths as the quark mass is
increased, or equivalently as the temperature is decreased. Of course the
actual rate of absorption given by the absolute value of the imaginary part of
the frequencies decreases with decreasing temperature. Nevertheless it is
somewhat surprising that the crossover towards the collisionless regime takes
place at smaller wavelengths for lower temperatures. 

\subsubsection{"Attractor" frequencies}
Here we briefly discuss the appearance of spiraling QNM trajectories
and their relation to the "attractor" frequencies found in section~\ref{sec:finiteK}.

First we should note that in most cases the spirals in the QNM trajectories
occur before the scalar in the spectrum becomes tachyonic. Thus the spiral
is a physical signature on the stable or metastable branch of the theory.
However the "attractor" frequencies to which the trajectories asymptote
at large quark mass parameters $\chi_0\to 1$ are located deep in the
unstable phase of the theory.
Therefore the "attractor" frequency is no signature of the physical stable sector
of the theory. Nevertheless it would be interesting to understand the
apparent direct relation of these "attractor" frequencies and the number of
spirals in the QNM trajectories since the spirals are physical, as noted before.

Note that a spiraling behavior for changing the temperature has been observed in 
the quark condensate in this system for near-limiting brane embeddings in~\cite{Mateos:2007vn}. There the spirals are 
due to oscillations of the parameters of the embedding, \ie the quark mass and
  the quark condensate. In particular the asymptotic value $m$ 
oscillates. This behavior has only been observed in the near-limiting embeddings. In contrast 
to that our loops in the QNM-trajectories appear way above the
critical embedding 
already. Nevertheless, as stated before there is an apparent connection between the number of 
loops in our QNM-trajectories and the near-limiting 'attractor' frequency. In this way
we could argue that at finite momentum we see the near-limiting embedding oscillations reflected 
already in the non-critical region in spiraling QNM-trajectories. In other words
both the QNM-loops and the spiraling quark condensate might have the same origin, namely 
the oscillating embedding parameters which are related to the aforementioned scaling
symmetry of the near-limiting embedding~\cite{Mateos:2007vn}.

Unfortunately the Schr\"odinger potentials at finite momentum but vanishing
density close to $\chi_0=1$ in figure~\ref{fig:scalarSchroePotFiniteK}
and~\ref{fig:transVectorSchroePotFiniteK} do not show any distinct
feature hinting neither on discrete special frequencies $\w_n$ 
nor on the jumps between them at critical momenta.

\subsubsection{Tachyon: A new hydrodynamic mode}
We briefly discuss here the behavior of the scalar mode becoming
tachyonic as explained in the previous section. This mode turns into 
a hydrodynamic mode in a special case.

As expected this scalar mode becomes tachyonic at higher and
higher values for the quark mass parameter $\chi_0$ as the momentum
of the excitation is increased. In the parameter space spanned by charge
density, temperature and quark mass there is one interesting special point: that is the 
location $\chi_0^{crit}(\tilde d)$ where the scalar mode becomes tachyonic. Just at this special
quark mass/temperature value, this scalar mode develops a hydrodynamic dispersion
relation, i.e. $\lim_{k\to 0}\omega\to 0$. In other words the scalar mode which shows 
the instability of the system turns into a hydrodynamic mode just at the critical point.
This could signal that there is a transition to a new phase. For instance
  this transition might be similar to the glass transition in 
  supercooled liquids discussed \eg in \cite{glasstrans}.
Lastly there might not exist a new stable ground state since we
might be scanning a regime where no stable brane embedding
exists besides the thermodynamically preferred Minkowski embedding.

%%%%%%%%%%%%%%%%%%%%%%%%%%%%%%%% F I N I T E    D E N S I T Y
\section{Finite density but vanishing momentum} \label{sec:finiteD}
\def\Im{{\mathrm{Im}\,}}
\def\Re{{\mathrm{Re}\,}}
\def\L{{\mathcal{L}}}
\def\pd{{\partial}}
\def\wn{{\omega}}

In this section we turn back to zero momentum, but switch on a finite baryon density and chemical potential.
The non-normalizable mode of the zero component of the gauge field living on
the D$7$-brane gives the chemical potential in the field theory. Working in
the $\rho$ coordinates we define 
\begin{equation}
\lim_{\rho\rightarrow\infty}A_0(\rho)=\mu\;.
\end{equation}
In order to study vector mesons in this background we consider fluctuations of
the gauge field about this background field.
The equation of motion for the zero component of the gauge field thus reads~\cite{Kobayashi:2006sb}

\begin{equation}\label{eq:eom_vector_finDnoK_gf}
\del_\rho A_0=2\tilde d\frac{f\sqrt{1-\chi^2+\rho^2{\chi'}^2}}{\sqrt{\tilde f
    (1-\chi^2)\left[\rho^6\tilde f^3(1-\chi^2)^3+8\tilde d^2\right]}}\;. 
\end{equation}
Since the asymptotic behavior for the embedding function $\chi$ is known we can also give the asymptotics for the gauge field at infinity

\begin{equation}
A_0=\mu-\frac{1}{\rho^2}\frac{\tilde d}{2\pi\alpha'}+\ldots\;,
\end{equation}
where $\mu$ is, as already mentioned, the chemical potential and $\tilde d$ parametrizes the baryon number density $n_B$ by

\begin{equation}\label{eq:dtdef}
\tilde d=\frac{2^{\frac{5}{2}} n_B}{N_f \sqrt{\lambda} T^3}\;.
\end{equation}

We divided this chapter into four sections in the same manner as the former
ones, except that we do not have to handle longitudinal vectors here since
$k=0$. But we will investigate the transverse vectors and the scalar
modes. 

Then we will describe our results, especially the tachyon behavior and the
turning points of the meson mass. Furthermore we present a Schr\"odinger
potential analysis to understand the above results in a different way. We
investigate the turning behavior of the quasinormal frequencies further by an analytical calculation close to
the horizon at large frequencies.

%______________________________________________
\subsection{Transverse vectors}
\label{sec:finiteD_trans_vector}

To obtain the equation of motion for the transverse vector modes, we have to
vary the DBI action with respect to the gauge field fluctuations about the
background and retain these terms up to second order~\cite{Erdmenger:2007ja}. 
The equation of motion for the fluctuations of the gauge field reads

\begin{equation}\label{eq:eom_vector_finDnoK}
0=E''+\del_\rho\log\left(\frac{f}{\tilde f}\sqrt{1-\chi^2}\sqrt{\frac{\rho^6\tilde f^3 (1-\chi^2)^3+8\tilde d^2}{\tilde f(1-\chi^2+\rho^2{\chi'}^2)}}\right)E'+2\w^2\frac{\tilde f}{f^2}\frac{1-\chi^2+\rho^2{\chi'}^2}{\rho^4(1-\chi^2)}E\;.
\end{equation}
The connection between the gauge invariant field $E$ and the gauge field is simply $E=\omega_{ph} A_i$, where we are free to choose any Minkowski spatial direction $i=1,2,3$. Furthermore the parameter $\w$ is a dimensionless frequency defined as $\w=\omega_{ph}/\pi T$ with the physical frequency $\omega_{ph}$. A more detailed derivation of this equation of motion can be found in~\cite{Erdmenger:2007ja}.

The solution to this equation of motion can be obtained by numerical methods, described in appendix~\ref{sec:appShooting}. Then we can compute the correlator in the complex plane where the quasinormal modes appear as poles. We aim for an explanation of the turning behavior of the meson mass found in~\cite{Erdmenger:2007ja}:\\
At large quark mass the meson mass increases proportional to the quark mass $M_q$ as expected from the formula found in the supersymmetric case for vector
and scalar meson masses (for vanishing angular momentum $l=0$ on the $S^3$)
\cite{Kruczenski:2003be}
\begin{equation}\label{eq:susyMassFormula}
M_{meson}=\frac{2 M_q}{\sqrt{\lambda} T R^2}\sqrt{2(n+1)(n+2)}\, , \quad n\in \mathbb{N}\, .
\end{equation}

Whereas for small quark mass there exists a region where the meson mass is decreasing when the quark mass is raised.

\paragraph{Results for transverse vectors}

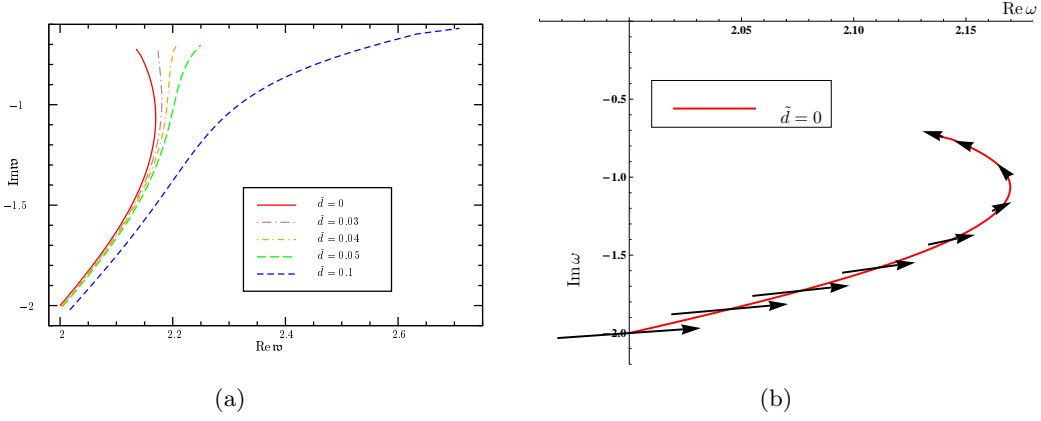
\begin{figure}[htbp]
  \begin{center}
   \subfigure[]{
     \includegraphics[width=6.5cm]{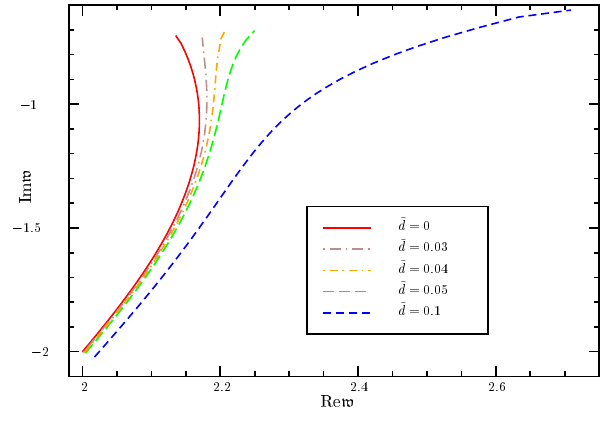}}
   \hspace{3mm}
   \subfigure[]{
   \includegraphics[width=6.5cm]{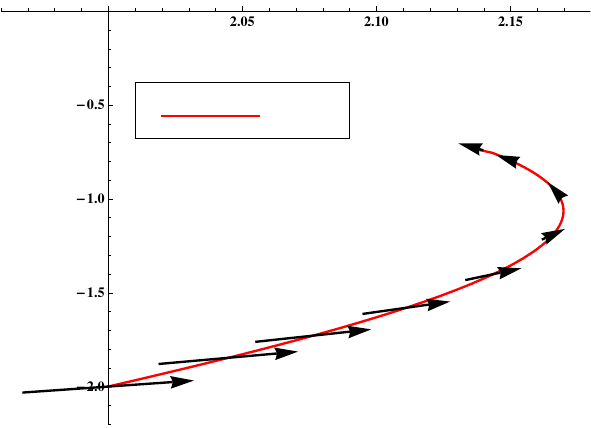}}
    \begin{picture}(10,10)
      \put(-15,133){\scalebox{0.6}{$\Re\w$}}
      \put(-98,92){\scalebox{0.6}{$\tilde d=0$}}
       \put(-180,30){\scalebox{0.6}{
        \begin{sideways}
        $\Im\w$
        \end{sideways}
       }}
    \end{picture}
    \caption{(a) First quasinormal modes for several densities $\tilde d$. For
      smaller densities a turning point in the real part occurs, which
      vanishes as the densities rises. (b) The first quasinormal mode for vanishing density. The arrows display the residue of the pole at the specific point, particularly the direction its phase and the length its absolute value.} \label{fig:finD_trans_Vec}
  \end{center}
\end{figure}

Figure~\ref{fig:finD_trans_Vec} (a) 
displays the paths of the first quasinormal mode in the complex
frequency plane for different densities. The paths are
parametrized by the quark mass over temperature ratio.
When the quark mass is zero, \ie our embedding is flat, the quasinormal mode
is located near the point $2(1-\ii)$ and moves towards the real axis as the
quark mass increases. For small densities we see that the curves turn around
and tend towards smaller real values for larger quark mass. This behavior
disappears when we raise the density up to a critical density $\dt_c\approx
0.04$. 

%______________________________________________
\subsection{Scalar}
\label{sec:finiteD_trans_scalar}

In order to compute the spectral function for scalar modes we first
calculate the pullback of the metric to the D$7$-branes and expand in
the fluctuations of the embedding variables $\delta\Theta$ and
$\delta\phi$. $\delta\Theta$ corresponds to the scalar and
$\delta\phi$ to the pseudoscalar excitations. We consider 
time and $\rho$ dependence of the fluctuations only, 
since we stay at zero momentum. The induced metric then reads

\begin{eqnarray}\label{eq:metric_scalar}
\nonumber \dd s^2&=&\frac{1}{2}\frac{\rho^2}{R^2}\left(-\frac{f^2}{\ft}\dd t^2+\ft\dd x_i^2\right)+\frac{R^2}{\rho^2}\frac{1-\chi^2+\rho^2{\chi'}^2}{1-\chi^2}\dd\rho^2+R^2\sin(\Theta+\delta\Theta)\dd\Omega_3\\
&&-2 R^2\frac{\chi'}{\sqrt{1-\chi^2}}\del_a\delta\Theta \dd x^a\dd \rho+R^2\del_a\delta\Theta\del_b\delta\Theta \dd x^a\dd x^b\;.
\end{eqnarray}
This coincides with the result found in~\cite{Myers:2007we} at zero density.

To obtain a consistent solution at non-zero density, it is necessary to also
include fluctuations in the gauge field $\delta A_0$ since these couple to the
fluctuations $\delta \Theta$ of the induced metric of the
D$7$-brane~\cite{Mas:2008jz}. The coupling occurs since both fields
  transform as scalars under the group of rotations $SO(3)$. They cannot be
  distinguished by the different transformation under the $U(1)$ gauge
  symmetry anymore since this symmetry is broken at finite density. We may also think
  of the embedding scalar as being effectively charged and this explains its coupling
  to the gauge field fluctuations.
Then the action in this case differs from the
action in~\cite{Myers:2007we} by the non-vanishing gauge field and gauge field
fluctuation terms. The action can be found in
appendix~\ref{sec:LagrFiniteD}. The equation of motion is given by
\begin{equation}
  \label{eq:eomssclarafiniteD}
  0=\del_\rho[A\del_\rho(\delta\Theta)]+B\w^2(\delta\Theta)+C(\delta\Theta)
\end{equation}
where
\begin{equation}
  \begin{split}
  A&=\frac{\rho^5f\ft(1-\chi^2)^3}{(1-\chi^2+\rho^2\chi'^2)^{3/2}\sqrt{1-\frac{8\dt^2}{\rho^6\ft^3(1-\chi
^2)^3+8\dt^2}}}\,,\\
B&=\frac{\rho\ft^2(1-\chi^2)^2}{\sqrt{1-\chi^2+\rho^2\chi'^2}\sqrt{1-\frac{8\dt^2}{\rho^6\ft^3(1-\chi
^2)^3+8\dt^2}}}\,,\\
C&=\frac{3\rho^3f\ft(1-\chi^2)\sqrt{1-\frac{8\dt^2}{\rho^6\ft^3(1-\chi
^2)^3+8\dt^2}}}{\sqrt{1-\chi^2+\rho^2\chi'^2}}
\left[1-6\chi\left(\rho \frac{f}{\ft}\chi'+\chi\right)\right]\,,
  \end{split}
\end{equation}
with $f$ and $\tilde f$ defined in equation~\eqref{eq:AdSmetric_f}.
For a more detailed derivation of the Lagrangian we refer to appendix~\ref{sec:LagrFiniteD}.

Now we can, similar to the procedure with the transversal vector modes, split
the solution in a regulating and a regular part, and compute the asymptotic
solution for the latter. This allows us then to apply the shooting technique
(see~\ref{sec:appShooting}) to compute the quasinormal modes in the same
manner as above.

\paragraph{Results for scalars}

\begin{figure}
  \centering
  \includegraphics[width=0.5\textwidth]{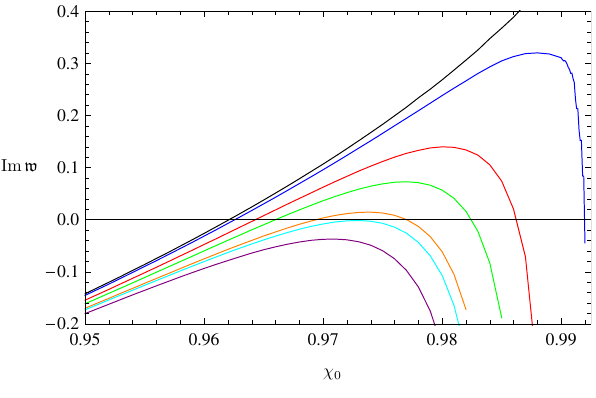}
   \caption{The scalar quasinormal modes with pure imaginary quasinormal frequencies
     for different densities $\tilde d$. For densities smaller than the critical
     density $\tilde d_c=0.00315$, the mode becomes tachyonic in some region of the
     parameter space. The several colors corresponds to different densities $\dt=$
    $0$ (black), $0.001$ (blue), $0.002$ (red),
    $0.0025$ green, 
    $0.003$ (orange), $0.00315$ (cyan), $0.0035$ (purple).}\label{fig:tachyonfiniteD}
\end{figure}

First we study the density 
dependence of the mode with purely imaginary quasinormal frequencies found
at zero density in fig.~\ref{fig:tachyonK0D0}. Our
numerical results are shown in fig.~\ref{fig:tachyonfiniteD}. In this figure
we observe that the critical parameter $\chi_0^{\text{crit}}$, at
which the instability occurs, \ie the quasinormal frequency enters the upper half plane,
increases with the density. Thus we write $\chi_0^{\text{crit}}(\dt)$.
Also we note that at finite densities the modes become stable again at some
larger value for $\chi_0$, which we denote by $\chi_{0,2}^{\text{crit}}(\dt)$.
If we increase the density further, we obtain a critical density $\dt_c=0.00315$
at which the mode is always stable. Therefore the system is unstable in the
parameter region
$\chi_0^{\text{crit}}(\dt)<\chi_0<\chi_{0,2}^{\text{crit}}(\dt)$ and
$\dt<\dt_c$. The numerical values can be obtained from
fig.~\ref{fig:tachyonfiniteD}. In section~\ref{sec:disc-finiteD} as well
  as in the introduction, we further
discuss this mode and relate the instability found to an instability already
known in thermodynamics.

\begin{figure}
  \centering
  \subfigure[]{\includegraphics[width=0.45\textwidth]{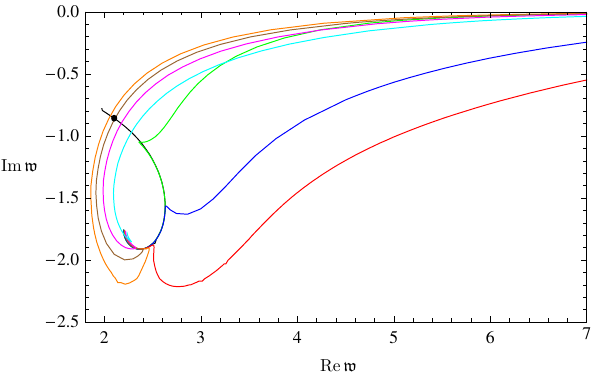}}
  \hfill
  \subfigure[]{\includegraphics[width=0.45\textwidth]{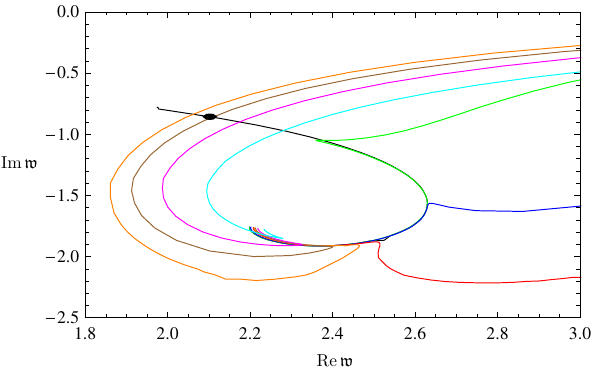}}
  \subfigure[]{\includegraphics[width=0.45\textwidth]{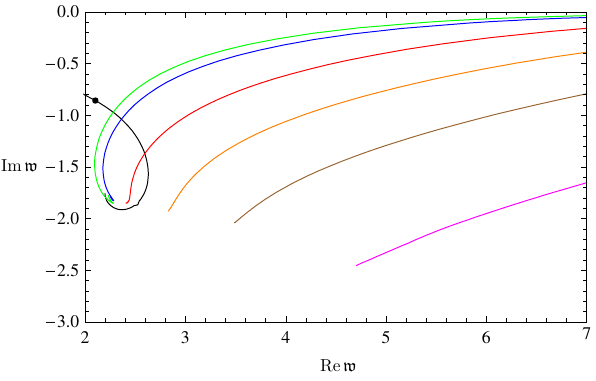}}
  \caption{Dependence of the first {\it scalar} quasinormal frequency on the density: In
    (a) and (b) the quasinormal frequencies for small densities $\dt\le 0.2$
    are shown while in (c) densities up to $\dt=5$ are considered. In (a) and
    (b) the different colors corresponds to distinct densities $\dt=$
    $0$ (black), $0.01$ (green), $0.05$ (blue),
   $0.1$ (red), $0.11$ (orange), $0.125$ (brown),
     $0.15$ (magenta), $0.2$ (cyan). In (c) the color coding is
   $\dt=$ $0$ (black), $0.2$ (green), $0.25$ (blue),
    $0.5$ (red), $1$ (orange), $2$ (brown),
      $5$ (magenta). In all three plots the black dot marks the
      critical value of of the quark mass/temperature parameter 
      $\chi_0$ where the instability occurs.}\label{fig:scalarfirstfiniteD}
\end{figure}

Let us now consider the behavior of the first quasinormal mode if we vary the
density $\dt$. In fig.~\ref{fig:scalarfirstfiniteD} we present our numerical
results. In this figure we observe that for each finite density we find that for
large enough $\chi_0$ the quasinormal frequencies behaves in a similar
way. The real
part increases while the imaginary part decreases as we increase $\chi_0$. For
smaller values of $\chi_0$ we observe three distinct movements of the quasinormal
frequencies. For small densities $\dt<0.1$ the quasinormal frequency first follows
the line of the quasinormal mode corresponding to zero density as we increase
the parameter $\chi_0$. After a critical value of $\chi_0$ is reached the
quasinormal frequency leaves this line as the real part of the quasinormal
frequency increases monotonically. For slightly larger densities
$0.11<\dt<0.2$, the frequencies also first moves along the line of the
quasinormal mode at zero densities but in contrast to the case discussed above
the real part of the frequencies now first decreases. For even
higher densities $\dt>0.2$, the frequency at $\chi_0=0$ strongly depends on the
densities. We find  from fig.~\ref{fig:scalarfirstfiniteD} (c) that both the
real and imaginary part of the quasinormal frequencies at $\chi_0=0$ increase
with $\dt$. As we increase $\chi_0$ we find the usual behavior, the real part
of the frequency increases while the imaginary part decreases.

%______________________________________________
\subsection{Schr\"odinger potential analysis}
\label{sec:finiteD_Schroedinger}

In this section we present the Schr\"odinger potential analysis which we
introduced in section \ref{sec:noDnoK} at finite baryon density. We use this
analysis to explain the qualitative movement of the quasinormal frequencies as we
change the baryon density.

For the vector fluctuations we observe in fig.~\ref{fig:finD_trans_Vec} (a) a turning point in the
movement of the quasinormal frequencies at low baryon density. The real part
of the quasinormal frequencies first increases and later decreases as we
increase the mass parameter $m$. This turning point disappears as the critical
density of $\dt_c=0.04$ is reached. After a critical quark mass is reached the real
part of the quasinormal frequencies always increases while the imaginary part
decreases.

\begin{figure}
  \centering
  \subfigure[]{\includegraphics[width=0.45\textwidth]{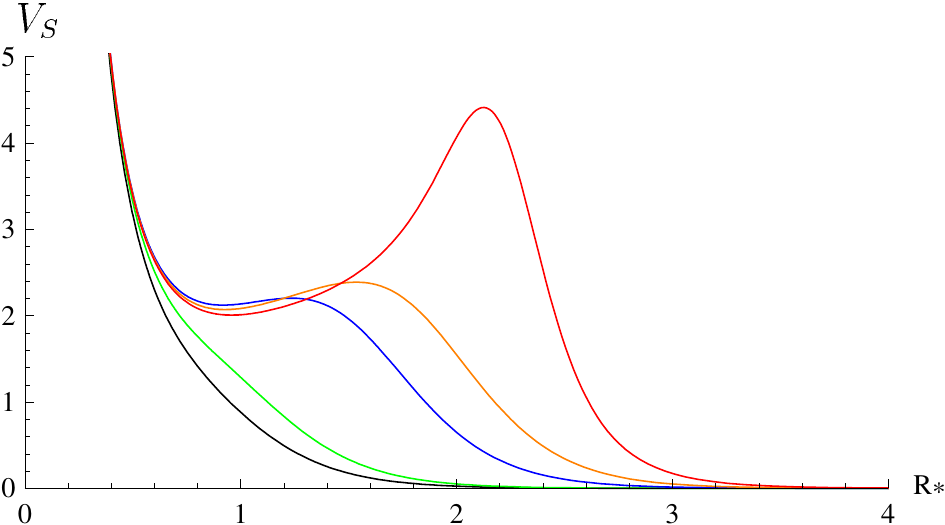}}
  \hfill
  \subfigure[]{\includegraphics[width=0.45\textwidth]{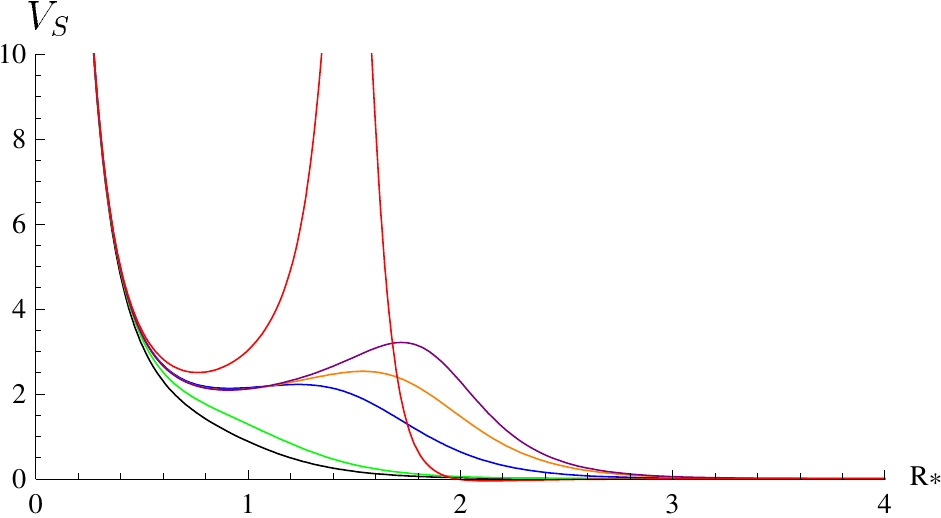}}\\
  \subfigure[]{\includegraphics[width=0.45\textwidth]{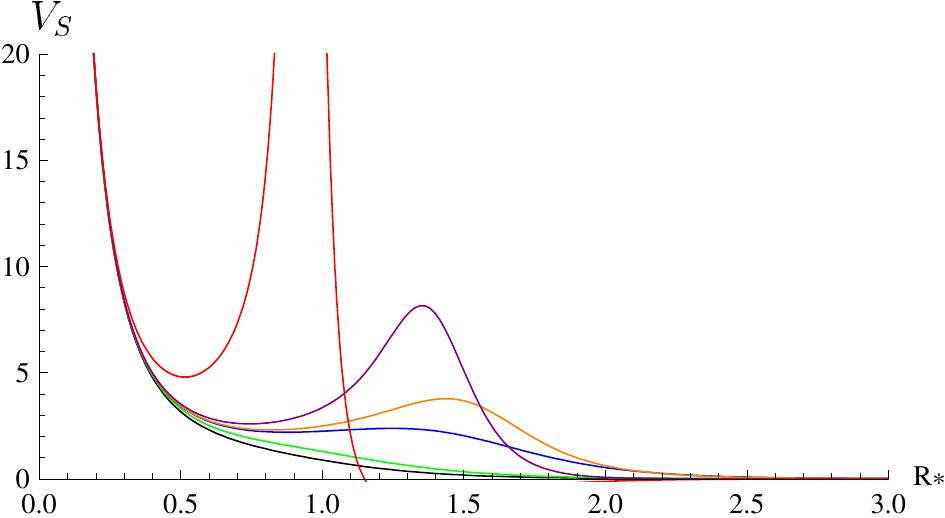}}
  \hfill
  \subfigure[]{\includegraphics[width=0.45\textwidth]{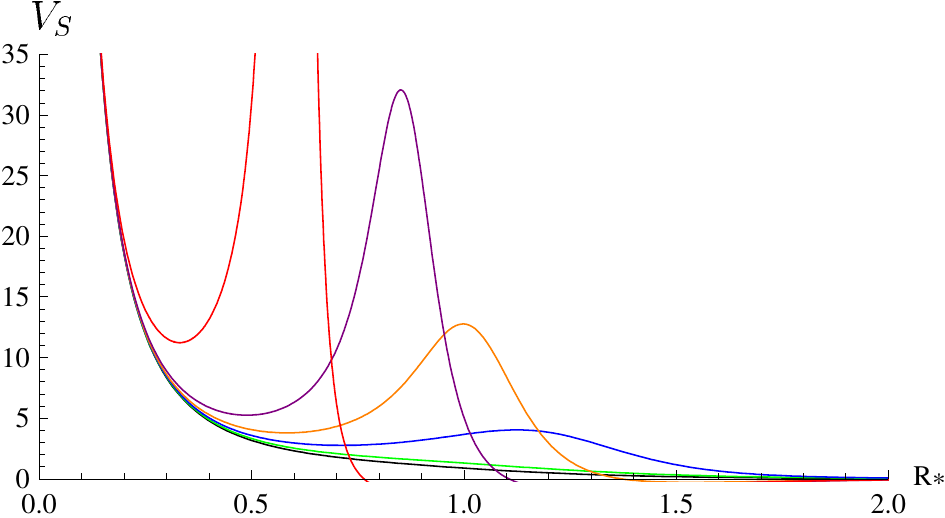}}
  \caption{Schr\"odinger potential of the vector fluctuations at finite baryon
  density (a) $\dt=0.003$, (b) $\dt=0.01$, (c) $\dt=0.03$ and (d) $\dt=0.1$.
  The different colors corresponds to distinct quark masses parametrized by $\chi_0=0$ (black), $0.5$ (green), $0.9$ (blue), $0.95$
    (orange), $0.97$ (purple), $0.99$ (red). See figure~\ref{fig:massplotD} for the relation between $\chi_0$,
    the temperature and the quark mass $M_q$.}
  \label{fig:potvectorfiniteD}
\end{figure}

Let us now consider the Schr\"odinger potential which correspond to these quasinormal modes in
fig.~\ref{fig:potvectorfiniteD}. A similar analysis was also done in
\cite{Myers:2008cj}. In addition to the infinite wall at $R*=0$ 
which corresponds to the AdS boundary an additional peak appears in the potential
as we increase the quark mass parametrized by $\chi_0$. For small densities
(see fig.~\ref{fig:potvectorfiniteD} (a)) this peak slowly grows out of the
step-shape potential already observed in section \ref{sec:noDnoK}. The step-shape potential is
also present at zero density and it is known from the analysis we presented there
that the corresponding quasinormal frequencies show the turning point behavior
discussed above. Thus the new feature of the finite density setup is
the appearance of the peak at high quark masses. As the peak grows, `bound'
states with positive energy can be formed, \ie the real part of the
corresponding quasinormal frequency is always bigger than its imaginary part.
Therefore we find quasiparticle excitations whose masses increase as we increase
the quark mass. 

If we increase the baryon density, the peak already appears at lower quark mass
and can thus destroy the step-shape potential (see
fig.~\ref{fig:potvectorfiniteD} (b) and (c)). For instance a step is still
observable in the orange and blue curve in fig.~\ref{fig:potvectorfiniteD} (b)
while in fig.~\ref{fig:potvectorfiniteD} (c) this step is gone. Since
we know that the step-shape potential is the reason for the turning point
potential, we also find in this analysis a critical baryon density at which
the turning point disappears. This critical density agrees with the value
found in fig.~\ref{fig:finD_trans_Vec} (a). For even larger densities (see
fig.~\ref{fig:potvectorfiniteD} (d)) the peak increases even faster, \ie the
real (imaginary) part of the corresponding quasinormal frequencies increase
(decrease) even faster. 

Let us now consider the scalar fluctuations at finite density which show several distinct
features: a tachyonic mode at small densities and three qualitative different
movements of the quasinormal frequencies.

\begin{figure}
  \centering
  \subfigure[]{\includegraphics[width=0.45\textwidth]{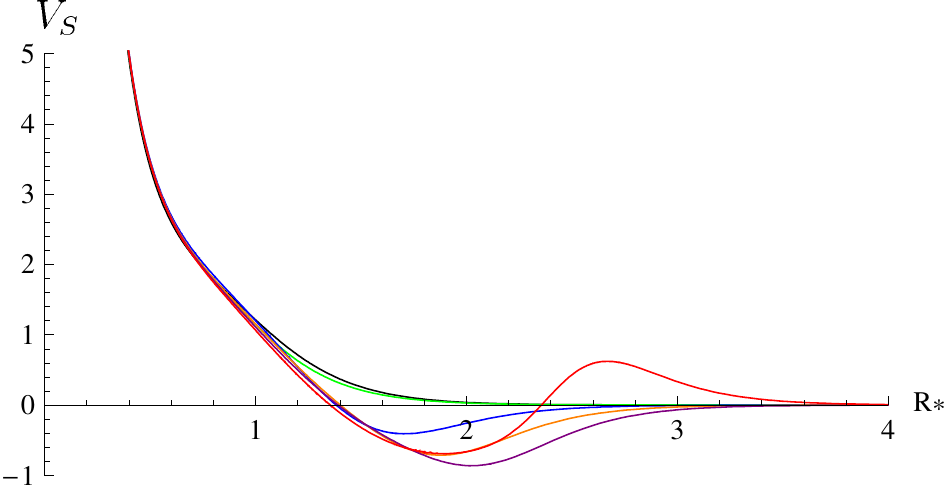}}
  \hfill
  \subfigure[]{\includegraphics[width=0.45\textwidth]{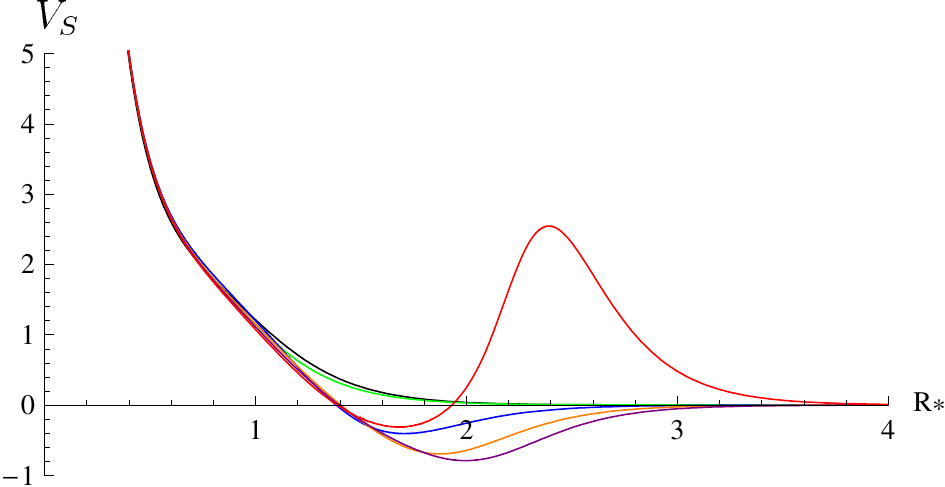}}
  \caption{Schr\"odinger potential of the scalar fluctuations at finite baryon density
    (a) $\dt=0.002$, (b) $\dt=0.00315$. The different colors
    corresponds to distinct quark masses parametrized by $\chi_0=0$ (black),
    $0.5$ (green), $0.9$ (blue), $0.95$ (orange), $0.97$ (purple), $0.99$ (red). 
    See figure~\ref{fig:massplotD} for the relation between $\chi_0$,
    the temperature and the quark mass $M_q$.}
  \label{fig:potscalarfiniteDtachyon}
\end{figure}

From fig.~\ref{fig:tachyonfiniteD} we know that there exists a parameter region
where a scalar mode becomes tachyonic. Especially interesting is that at
finite density the mode is stabilized as the quark mass is increased
and that there is a critical density $\dt=0.00315$ at which the mode is always
stable. In fig.~\ref{fig:potscalarfiniteDtachyon} we present the Schr\"odinger
potentials of the scalar fluctuations in the relevant density region. As in
the case of zero density a negative well appears in the potential. It also
first grows as we increase the quark mass (see
fig.~\ref{fig:potscalarfiniteDtachyon} (a)). However as we increase the quark
mass further, the well decreases while a peak forms in the potential. After a
critical quark mass is reached, this well cannot longer support a `bound' state with
negative energy, \ie the imaginary part of the quasinormal frequency becomes
negative again. At the critical density $\dt=0.00315$ (see
fig.~\ref{fig:potscalarfiniteDtachyon}), the potential also shows a negative
well. However this well just reaches a critical depth in order to support a zero energy
`bound' state which is due to the zero-point energy (cf. the three dimensional
  potential pot know from quantum mechanics).

\begin{figure}
  \centering
  \subfigure[]{\includegraphics[width=0.45\textwidth]{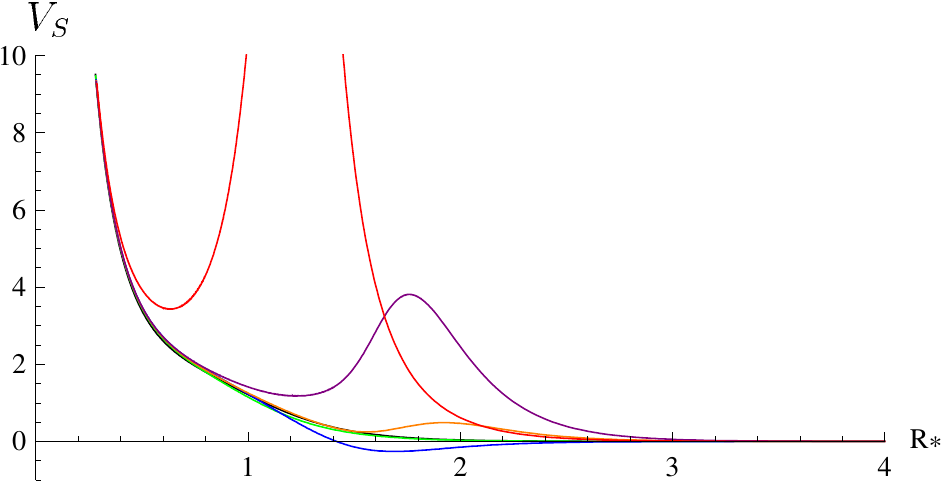}}
  \hfill
  \subfigure[]{\includegraphics[width=0.45\textwidth]{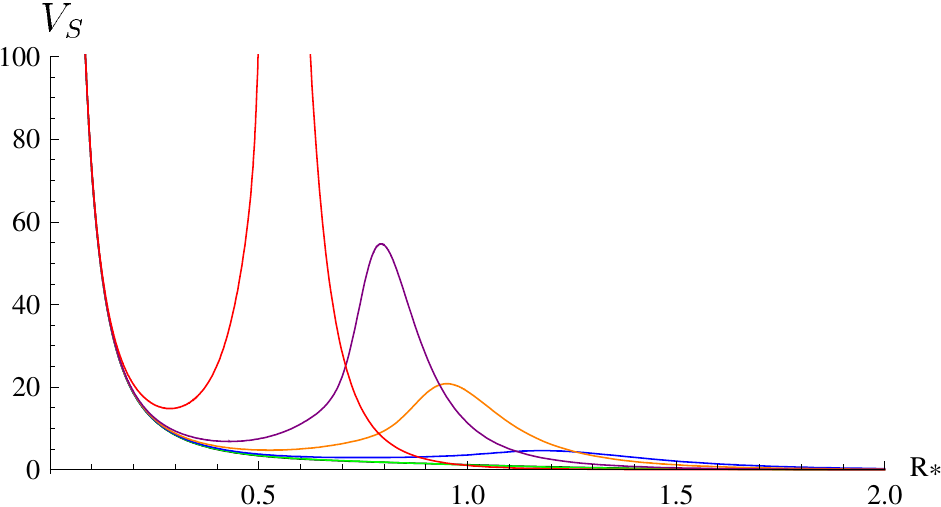}}\\
  \subfigure[]{\includegraphics[width=0.45\textwidth]{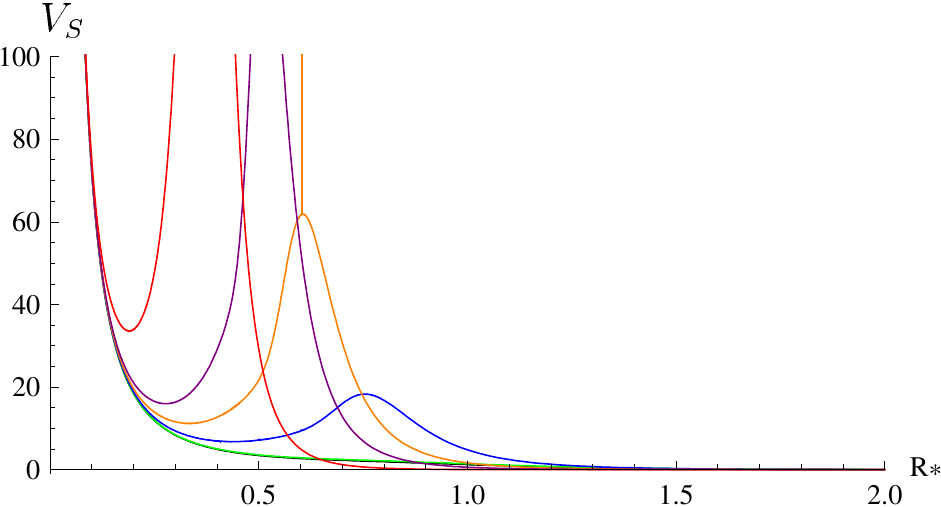}}
  \caption{Schr\"odinger potential of the scalar fluctuations at finite baryon density
    (a) $\dt=0.01$, (b) $\dt=0.15$, (c) $\dt=0.5$. The different colors
    corresponds to distinct quark masses parametrized by $\chi_0=0$ (black),
    $0.5$ (green), $0.9$ (blue), $0.95$ (orange), $0.97$ (purple), $0.99$ (red).
    See figure~\ref{fig:massplotD} for the relation between $\chi_0$,
    the temperature and the quark mass $M_q$.}
  \label{fig:potscalarfiniteD}
\end{figure}

Now we investigate the Schr\"odinger potentials which are relevant for the
movement of the first quasinormal frequency studied in
fig.~\ref{fig:scalarfirstfiniteD}. These potentials are plotted in
fig.~\ref{fig:potscalarfiniteD}. The first observation is that for small
densities $\dt\lesssim 0.15$ (see fig.~\ref{fig:potscalarfiniteD} (a) and (b))
the potentials at low quark mass agree with the potential at zero density.
Thus also the corresponding quasinormal frequencies must agree which we
already found in fig.~\ref{fig:scalarfirstfiniteD}. At larger densities (see
fig.~\ref{fig:potscalarfiniteD} (c)) even the potential at zero quark mass
differs from the one at the zero density such that the corresponding
quasinormal frequencies at zero quark mass depend on the density which is
consistent with the result found in fig.~\ref{fig:scalarfirstfiniteD} (c). As
for the vector fluctuations a peak appears in the potential at finite density
as we increase the quark mass. These peak can again support `bound' states 
which correspond to quasiparticle excitations. Thus the imaginary part of
the quasinormal frequencies decreases while the real part increases as we
increase the quark mass. This is the overall movement of the quasinormal
frequencies which is shown in fig.~\ref{fig:scalarfirstfiniteD}. 

\begin{figure}
  \centering
  \includegraphics[width=0.5\textwidth]{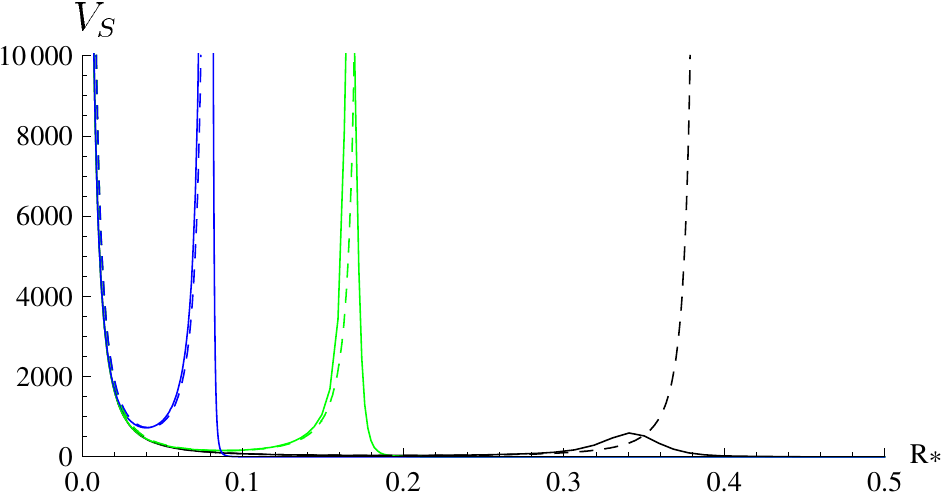}
  \caption{Comparison of the Schr\"odinger potential obtained from black hole
    embeddings (solid lines) and from supersymmetric Minkowski embeddings (dashed lines) at the
    same quark mass $m$. The different colors correspond to different quark
    masses parametrized by $\chi_0=0.99$ (black), $0.999$ (green), $0.9999$
    (blue) at a fixed density $\dt=0.5$.
    See figure~\ref{fig:massplotD} for the relation between $\chi_0$,
    the temperature and the quark mass $M_q$.}
  \label{fig:potentialbhvsminkfiniteD}
\end{figure}

\begin{figure}
  \centering
  \includegraphics[width=0.5\textwidth]{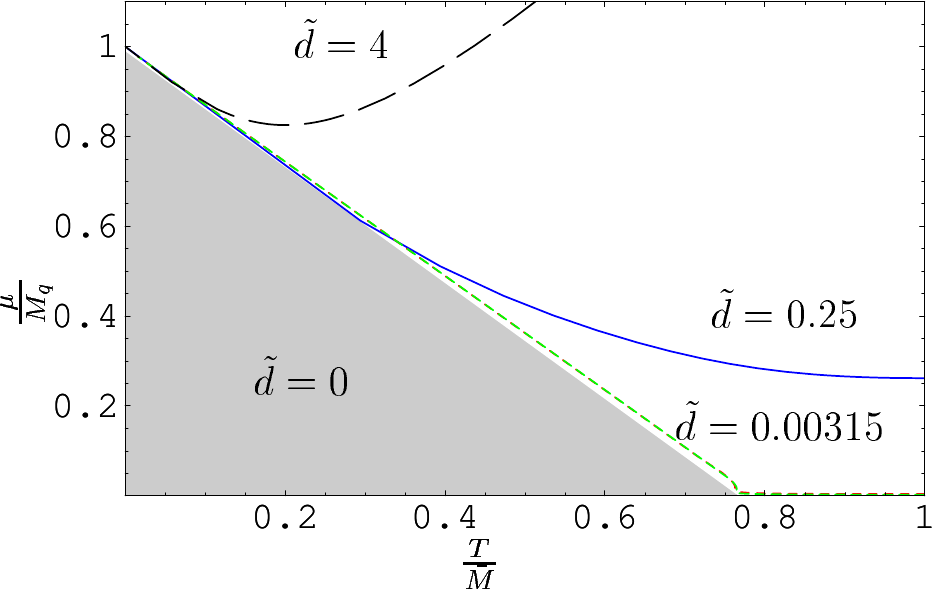}
  \caption{Sketch of the phase diagram: The chemical potential
$\mu$ divided by the quark mass $M_q$ is plotted versus the temperature
$T$ divided by $\bar M = 2 M_q/\sqrt{\lambda}$. Two different regions are displayed:
The shaded region with vanishing baryon density where Minkowski embeddings are preferred and the
region above the transition line with finite baryon density where the black hole embeddings are preferred.
Here we work in the second phase. The curves are lines of equal baryon density parametrized by $\dt$.}\label{fig:phasediagramsketch}
\end{figure}

In general we observe that at finite baryon density a peak appears in the
Schr\"odinger potential as we increase the quark mass. If the peak is high
enough, it separates the horizon of the black hole from the AdS boundary. Only
a small leak into the black hole is possible due to tunneling. The potential
thus approaches the one of a Minkowski embedding where the brane does not fall
into the horizon of the black hole and therefore can support stable normal
modes which are calculated in \cite{Kruczenski:2003be} in the supersymmetric
limit. In fig.~\ref{fig:potentialbhvsminkfiniteD} we explicitly confirm that
the Schr\"odinger potential obtained from the black hole embeddings converges
to the potential of the supersymmetric Minkowski embedding\footnote{The Schr\"odinger
  potential of a supersymmetric Minkowski embedding is given by \cite{Paredes:2008nf}
  \begin{equation*}
    V_S=m^2
    \left[1/4+3/8\left(\tan^2(m R*/\sqrt{2})+\cot^2(m R*/\sqrt{2})\right)\right]\,.
  \end{equation*}
}
with the same mass parameter $m$ as
the mass parameter goes to infinity. This convergence explains that also the
quasinormal frequencies approach this supersymmetric mass spectrum  if the
quark mass is big enough. This behavior was already found for the vector fluctuations in
\cite{Erdmenger:2007ja,Myers:2008cj}. We can also understand this convergence if we look
at the phase diagram (see fig.~\ref{fig:phasediagramsketch}). For large quark mass over
temperature ratios the equal density lines approach the shaded region where the Minkowski
embeddings are preferred. Since the phase transition is third or second order
for large quark mass over temperature ratios as it is shown in \cite{Faulkner:2008hm}, we also expect a smooth
transition from the spectrum of the quasinormal modes obtained from the black hole embeddings
to the spectrum of the normal modes obtained from the Minkowski embeddings.

\begin{figure}
  \centering
  \subfigure[]{\includegraphics[width=0.45\textwidth]{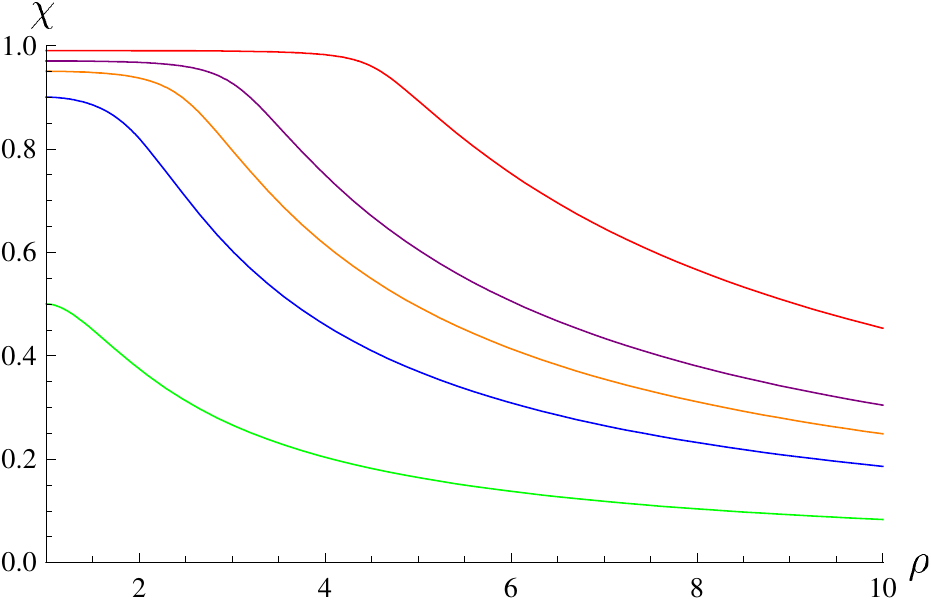}}
  \hfill
  \subfigure[]{\includegraphics[width=0.45\textwidth]{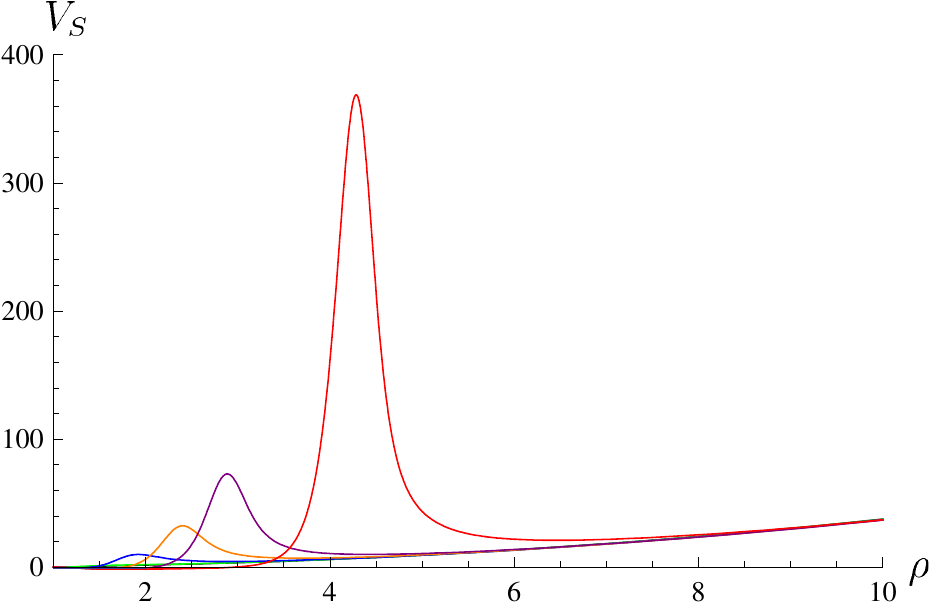}}
  \caption{Comparison of the location of the peak in the Schr\"odinger
    potential and the throat in the embedding of the D$7$-brane. The end of
    the throat is located where $\chi$ first changes its value. The different
    colors corresponds to different quark masses parametrized by $\chi_0=0$ (black),
    $0.5$ (green), $0.9$ (blue), $0.95$ (orange), $0.97$ (purple), $0.99$
    (red). See figure~\ref{fig:massplotD} for the relation between $\chi_0$,
    the temperature and the quark mass $M_q$. The baryon density is $\dt=0.25$.}
  \label{fig:cfpeaksthroat}
\end{figure}

Next we would like to understand the
appearance of the peak in terms of the D$7$-brane embedding. In
\cite{Kobayashi:2006sb} it was shown that the finite baryon density on
the brane is induced by fundamental strings which are stretched from the
horizon of the black hole to the D$7$-brane. At large quark masses these
strings form a throat close to the horizon. We confirm numerically that the end of
this throat and the peak in the Schr\"odinger potential are located at the same
value of the radial coordinate $\rho$ (see fig.~\ref{fig:cfpeaksthroat}). Thus
we can interpret this throat as a potential barrier for the fluctuations which
becomes bigger as the throat becomes deeper. 

%______________________________________________
\subsection{Analytic solution at high frequencies}
\begin{figure}
\begin{center}
  \includegraphics[angle=0,width=0.49\linewidth]{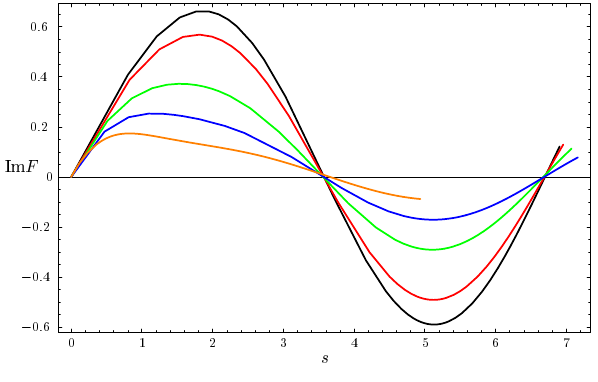}
  \hfill
  \includegraphics[angle=0,width=0.49\linewidth]{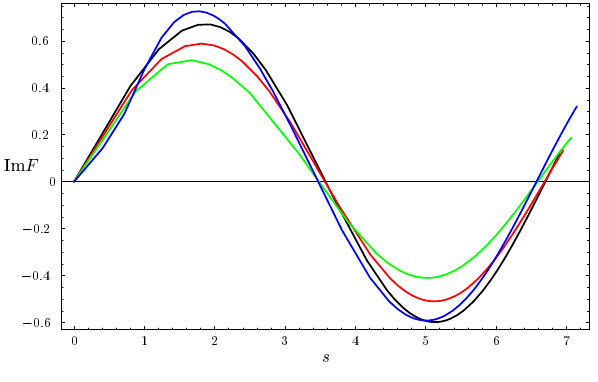}
                \caption{
                Imaginary part of the solution to the regular function~$F$ versus the
                proper radial
                coordinate~$s$~\cite{Kaminski:2008ai}.
                The left plot at vanishing baryon density~$\tilde d=0$ shows that the
                initially sinusoidal
                solution is deformed as the mass parameter~$\chi_0=0.01,\,0.5,\,0.8,\,0.9$
                is increased (see figure~\ref{fig:massplotD} for the relation between $\chi_0$,
    the temperature and the quark mass $M_q$). Furthermore, its amplitude
                decreases while the wave length increases. The right plot shows that
                introducing a finite
                baryon density~$\tilde d=0.2$ causes the solutions to change their behavior
                with increasing~$\chi_0$:
                While the first three curves for~$\chi_0=0.01,\,0.5,\, 0.8$ show the same
                qualitative behavior as
                those in the left plot, the blue curve for~$\chi_0=0.9$ clearly signals a
                qualitative change with its
                increased amplitude. Looking at the wave lengths in the lower plot we realize
                that already the
                green curve~($\chi_0=0.8$) shows a decreased wave length as well as the
                blue curve~($\chi_0=0.9$). The awkward oscillation pattern near $s=0$ stems
                from a second mode being superimposed on the one we are tracking here.
                }
                \label{fig:FOfS}
\end{center}
\end{figure}

\begin{figure}
  \includegraphics[angle=0,width=0.49\linewidth]{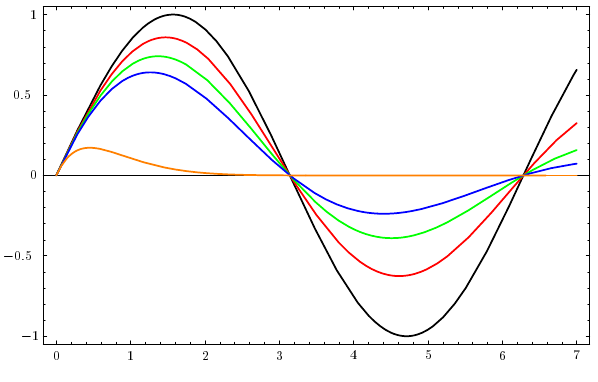}
         \hfill
        \includegraphics[angle=0,width=0.49\linewidth]{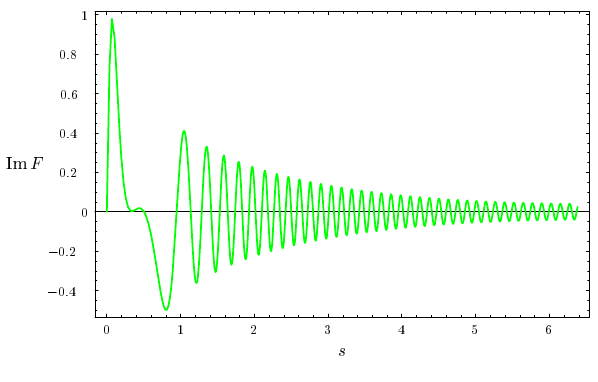}
         \caption{\label{fig:harmOsc}\label{fig:hiWSol}
         Left: Solutions to the damped harmonic oscillator
         equation~\eqref{eq:harmOscEOM}
         for increasing damping coefficient $\gamma=0,\, 0.1,\, 0.2,\, 0.3,\, 0.9$~(black,
         red, green, blue, orange). Here the eigenfrequency $\w$ is
         fixed to $1$ in analogy to the plots in figure~\ref{fig:FOfS}.
         Right: Hi-frequency ($\w=50$) numerical solution showing that the
         damping of the mode is strong near the horizon $s=0$ but
         decreases exponentially in the bulk towards the boundary $s\to\infty$.
         }
\end{figure}

Motivated by the numerical solution to the fluctuation equations of motion shown in figure~\ref{fig:FOfS}, we suspect
that this damped oscillating curve near the horizon can be approximated by a damped {\it quasi-harmonic} 
oscillator, i.e. we should be able to find an approximate equation of motion which is a generalization of the 
damped harmonic oscillator equation. By {\it quasi-harmonic} we mean that the oscillator is damped with the damping depending on the 
location of the mode in radial direction. From the observations in figure~\ref{fig:hiWSol} we 
have already concluded that the amplitude changes rapidly near the horizon and
change less in the bulk until the boundary is reached. Thus it is reasonable
to assume that the damping of the mode~$F$ already is strong near the horizon
and a near-horizon approximation can capture this effect to certain extent. 
In this spirit we take the near-horizon limit~$\varrho \sim 1$ and at the same time the high-frequency limit~$\w \gg 1$. 

\begin{figure}
  \includegraphics[width=0.9\linewidth]{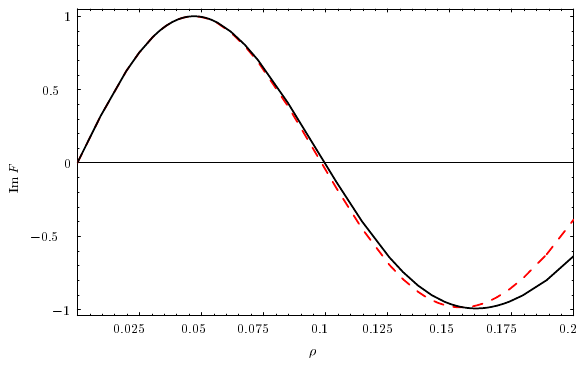}
                \caption{
                  Approximate analytic solution compared to the exact solution at~$\w = 70 \, ,\tilde d = 0\, ,\chi_0= 0.4$ (see figure~\ref{fig:massplotD} for the relation between $\chi_0$,
    the temperature and the quark mass $M_q$).
                }
                \label{fig:FApprox}
\end{figure}

\begin{figure}
          \includegraphics[width=0.49\linewidth]{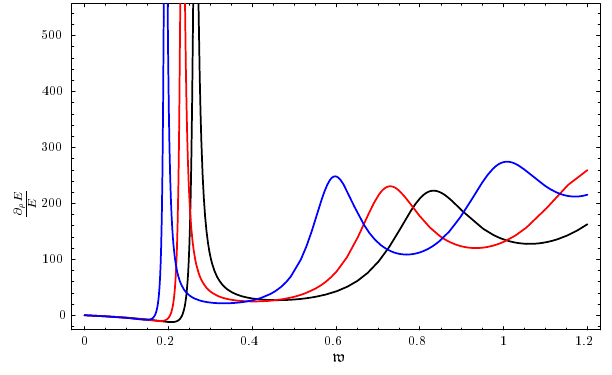}
        \hfill
        \includegraphics[width=0.49\linewidth]{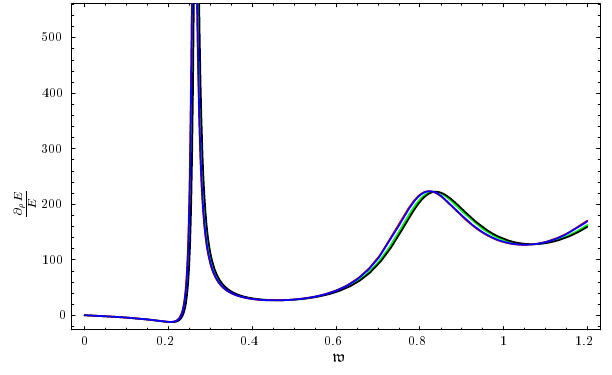}
                \caption{
                  Approximate spectral function fraction near the horizon computed with
                  the function~$E=(\rho-1)^\beta F(\rho)$ and~$F$ being the analytic approximation given
                  in equation~\eqref{eq:anaF}.
                }
                \label{fig:nearHorR}
\end{figure}

Applying these limits for the flat embedding~$\chi_0=0$ in the equation of motion~\eqref{eq:eom_vector_finDnoK}, 
we obtain the simplified equation of motion
\begin{equation}
y H'' +\left (- i \w -y \right ) H' + i\frac{\w}{2}\left( \frac{1}{\sqrt{7}} + 1 \right ) H = 0 \, ,
\end{equation}
where the variable is~$y= i\w\frac{\sqrt{7}}{4}(\varrho -1)$ and the regular 
function~$H(y)$ comes from the Ansatz~$E = (\varrho -1)^\beta F$ with the 
redefinition~$F=e^{-\sqrt{7}/4 \ii \w (\varrho-1)} H$.
This equation of motion has the form of {\it Kummers equation}, which
is solved by the confluent hypergeometric function of first~$H={}_1 F_1[-i\w(1/\sqrt{7}+1)/2,-i\w,\varrho-1]$ and second kind~$U$. 
Boundary conditions rule out the second kind solution which is non-regular at the horizon and therefore contradicts the 
assumptions put into the Ansatz~$E = (\varrho -1)^\beta F$. Since we are interested in how the solution changes
with decreasing~$m$, we need to choose~$\chi_0$ non-vanishing. Also with this complication we still get Kummers 
equation with changed parameters and the analytic solution for~$F$ is given
by

\begin{equation}
\label{eq:anaF}
\begin{split}
F= e^{-i \frac{\w}{2}(\varrho-1) \sqrt{\frac{7}{4} + \frac{4 \chi_2[\chi_0, \tilde d]^2}{1 - \chi_0^2}}} 
   {}_1F_1\Bigg [&-i \frac{\w}{2} \left (\frac{1}{2 \sqrt{\frac{7}{4} + \frac{4 \chi_2[\chi_0, \tilde d]^2}{1 - \chi_0^2}}} + 
                1\right ),\, -i \w,\\
              &i \w (\varrho-1) \sqrt{\frac{7}{4} +
                \frac{4 \chi_2[\chi_0, \tilde d]^2}{1 - \chi_0^2}}\Bigg] \, ,
\end{split}
\end{equation}
with the near horizon expansion of the embedding function~$\chi=\chi_0+\chi_2[\chi_0,\tilde d] (\varrho-1)^2 +\dots$ where
we determine recursively 
\begin{equation}
\chi_2[\chi_0, \tilde d] = 3 \chi_0 
 \frac{\chi_0^6 - 3 \chi_0^4 + 3 \chi_0^2 - 1}{4 (1 - 3 \chi_0^2 + 3 \chi_0^4 - \chi_0^6 + \tilde d^2)} \, .
\end{equation}
The approximate solution for~$F$ is shown in figure~\ref{fig:FApprox}. Furthermore we can calculate the 
fraction~$\partial_4 E/ E$ appearing in the spectral function near the horizon using this analytic solution. 
The result is displayed in figure~\ref{fig:nearHorR}. This near horizon limit is not the spectral function since 
we would have to evaluate it at the boundary which lies far beyond the validity of the near horizon 
approximation. Nevertheless, in the spirit of \cite{Iqbal:2008by} and
according to our initial assumptions that the effect of damping mainly
takes place near the horizon we further assume that the limit shown in
figure~\ref{fig:nearHorR} already contains the essential features of the
spectral function. Indeed the fraction shows distinct resonance peaks which
move to lower frequencies if we increase the mass parameter~$m$. The right
picture shows the same situation at a finite baryon density~$\tilde d=1$ and
we see that the peaks do not move to lower frequencies as much as before. Thus
also the vanishing of the turning point at large densities as observed before
is captured by this approximate solution. 

%______________________________________________
\subsection{Discussion: Turning point and Tachyon}
\label{sec:disc-finiteD}

We now discuss the turning point in the spectrum of the vector modes and its dependence on the density. For scalars the purely imaginary scalar quasinormal mode reaches into the upper half of the complex plane and thus yields a tachyonic excitation, which we also discuss here.

\subsubsection{Turning point}

In this section we discuss why the vector mesons at finite temperature get
smaller masses as the quark mass is increased. Further we investigate
why there is a turning behavior at finite density. In the latter case we also find
an analytic solution valid at high frequencies.

\paragraph{Line of argument}
Let us first summarize our results. The D3/D7-system at finite density
has two competing geometrical features. One is the formation of a {\it potential barrier}
near the horizon due to the charge located at the horizon. This causes the
resonances in the spectral function to become more stable and to move
to larger frequency when the quark mass is increased. The second
feature is the length of the brane \footnote{Since the AdS-boundary
is infinitely far away from the horizon, the length of the brane has to be renormalized by
subtracting for example the length of the brane at trivial embedding $\chi_0=0$.} 
 supporting an excitation from the horizon to the boundary. Increasing this length causes the vector meson resonances
to move to smaller frequency with increasing quark mass.
These two geometrical features compete at finite density while at zero
density the potential barrier is absent.
Below we will explain this in greater detail and give a field theoretic interpretation.

\begin{figure}
\includegraphics[width=0.9\textwidth]{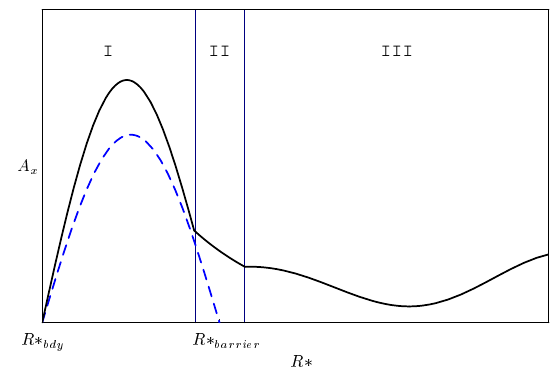}
\caption{ \label{fig:solSchematic}
Schematic outline of the solution $A_x(R*)$ to the vector fluctuation
equation of motion~\eqref{eq:eom_vector_finDnoK}. The horizon is located
at $R*=\infty$ where the wave function diverges, the AdS-boundary is $R*_{bdy}=0$. 
The solid black curve shows the
solution at finite temperature, density and quark mass, i.e. on a black
hole embedding. The dashed blue curve shows the
corresponding solution on a Minkowski embedding at the same
quark mass. The three regimes correspond to the nature of the potential
shown in figure~\ref{fig:potentialbhvsminkfiniteD}:
I) Minkowski-like, II) potential barrier, III) essentially vanishing potential.
}
\end{figure}

\paragraph{Guiding features of numerical solutions}
A detailed analysis of the numerical solutions~\cite{Kaminski:2008ai} shows
some interesting features. Most prominently there is a turning behavior, i.e. the
location of the vector resonance peaks first move towards lower real frequencies,
then turn around and move towards higher ones in order to asymptote to the
SUSY-spectrum (increasing the quark mass to infinity is equivalent to the
zero temperature case).
This turning point for the lowest of the
vector resonance peaks is only present at small densities $\tilde d\lesssim 1$.
For larger densities the peaks run towards higher frequencies with increasing
quark mass from the start, i.e. there is no turning point. As we see from the
vector quasinormal modes the imaginary part monotonically decreases
with increasing quark mass. At vanishing density the resonances only
move to lower frequencies with increasing quark mass in the regime of black hole embeddings.

Therefore we have two things to explain: 1) the turning of the real part of
the frequency and the decreasing imaginary part and 2) the motion
towards smaller frequencies at zero density or small densities and
quark masses ($\tilde d<1$, $\chi_0<0.5$). The solutions $E$ of the bulk
fluctuation equation of motion are composed of a singular
part and the regular part $F$. From these solutions
the spectral functions are computed. The regular part of the solutions
has interesting properties, in particular:

\begin{enumerate}
\item The proper distance $s$ on the brane
\begin{equation}
s=\int\limits_{\rho_H}^\rho \sqrt{G_{\rho\rho}}d\rho \, ,
\end{equation}
increases with increasing quark mass parameter $\chi_0$.
Integrating this expression from the horizon, to the boundary gives the proper length of 
the D7-brane. This also changes the wavelength of the fluctuations on the brane,
as we will explain below.

\item As we see from figure~\ref{fig:FOfS} the form of the numerical solutions
written in the proper radial coordinate $s$ shows a damped oscillation.
Thus we are motivated to map the fluctuation equation of motion
to a damped harmonic oscillator equation of motion.

\item At vanishing density $\tilde d=0$ a larger quark mass parameter
$\chi_0$ induces a stronger damping of the solutions $F(s)$ (see
figure~\ref{fig:FOfS}, left).

\item At finite density a larger quark mass parameter
$\chi_0$ first induces a stronger, then a weaker damping of the
solutions $F(s)$ (see figure~\ref{fig:FOfS}, right).

\item As seen from the high frequency ($\w=50$) solution in figure~\ref{fig:hiWSol}
most of the damping occurs close to the horizon while the fluctuation propagates
towards the boundary with exponential damping.

\item At $\tilde d=0$ for $\chi\to 1$ black hole embeddings asymptote to
$m=\rho\chi|_{\rho\to\rho_{bdy}}\approx 1.3$; i.e. this geometric construction
can not produce the $T\to 0$ limit in the black hole embeddings.
Furthermore, long before $\chi_0\to 1$ one has
to jump to Minkowski embeddings at the temperature where the instability
appears. In other words: at fixed $T$ you can not make the quarks heavier
than $M_q\sim1.3 T$. And there is no smooth transition
through the singularity at the limiting embedding $\chi_0=1$ to the Minkowski
embeddings.
\end{enumerate}

\paragraph{Resonances moving to higher frequencies towards the SUSY spectrum}
At finite density there are two effects present which compete. One effect is the
elongation of the brane. This is discussed below for the case of vanishing density.
A specialty at finite density is the appearance of a longer and longer spike
in the embedding reaching down to the horizon, connecting the Minkowski-like
bulk part of the black hole embeddings with the horizon.

The second and as it turns out also the more important effect at finite density
is the formation of a {\it potential barrier} near the horizon at $R*=\infty$, see
figure~\ref{fig:potentialbhvsminkfiniteD}. This barrier effectively cuts off the horizon from
the geometry. Only a small part of the wave function "leaks" into the region
behind the potential barrier where the black hole is located, that is region $III$ in
the schematic solution shown in figure~\ref{fig:solSchematic}.
In this region $III$ the Schr\"odinger potential asymptotically vanishes. Therefore
the solution shown in figure~\ref{fig:solSchematic} first drops to low
values in order to diverge only very close to the horizon~(not shown in the figure).
Region $II$ is the finite radial distance covered by
the potential barrier. Here the wave function qualitatively drops exponentially.
In region $I$ the main part of the solution is located between the boundary and the barrier.
Here the potential approaches a pot-like form more and more
similar to the corresponding potential generated by the Minkowski
embedding as the quark mass parameter is increased.
As this happens, the black hole fluctuation solution (solid black
curve in the schematic plot in figure~\ref{fig:solSchematic}) approach the
Minkowski solutions (dashed blue curve in the schematic plot in
figure~\ref{fig:solSchematic}).

As we see from figure~\ref{fig:cfpeaksthroat} the potential
barrier moves closer to the boundary as the mass
parameter $\chi_0$ is increased. Thus the potential gets more narrow
and the lowest possible excitation
is raised to a higher energy, i.e. the real part of the corresponding
quasinormal mode is increasing. Furthermore the barrier becomes
higher such that the corresponding excitation becomes more stable,
i.e. the imaginary part of the quasinormal frequency decreases.
In other words, less of the wave function leaks over the barrier
into the black hole.

\paragraph{Resonances moving to lower frequencies}
Let us for simplicity work at vanishing density first. All our considerations
will also apply to the finite density case.
The heuristic explanation for the left-movement of resonances has to do
with the proper length of the D7-brane. At $\chi_0=0$ we have a flat
embedding which is the minimal length the brane can have when measured in
$s$. Increasing the quark mass parameter $\chi_0$ the D7-brane becomes
"longer" in the sense that the proper length computed in the coordinate
$s$ increases. Therefore we can imagine a solution supported on the brane
to be "stretched" together with the brane, i.e. its effective wavelength
increases. Assuming a constant effective speed of light,
the effective frequency of this solution
has to decrease. In this sense $\chi_0$ here acts analogously to a damping coefficient
$\gamma$ in a damped harmonic oscillator.~\footnote{
This heuristic picture neglects the fact that the damping of the gravity
modes in general is a local effect, i.e. it depends on the radial
AdS-coordinate (through radially-dependent geometry). Tentatively we will
assume that we can average the
damping effects over the radial coordinate and express them globally
in an effective damping coefficient independent from the radial coordinate.}

We may quantify this intuition examining the solutions $F(s)$ shown
in figure~\ref{fig:FOfS}.
Indeed, comparing to the solutions (figure~\ref{fig:harmOsc}) of the
damped harmonic oscillator equation
\begin{equation}\label{eq:harmOscEOM}
0=\partial^2_t X(t)+2\w_0 \gamma \partial_t X(t)+{\w_0}^2 X(t)\, ,
\end{equation}
with damping coefficient $\gamma$, we find that increasing $\chi_0$
in figure~\ref{fig:FOfS} resembles the effect of increasing the damping
coefficient $\gamma$ in the analogous harmonic oscillator solutions,
figure~\ref{fig:harmOsc}.

This leads us to assume that the fluctuation equation of motion on the
D7-brane for real values of the frequency $\w\in\mathbf{R}$ can effectively be replaced by the equation of motion for a damped
harmonic oscillator with an effective eigenfrequency $\w_{eff}$
and with the replacements $\gamma\to\gamma(\chi_0)$, with
$\gamma(\chi_0)$ a monotonous function and
$\w_0\to\w(\chi_0=0)$. By $\w(\chi_0=0)$ we mean the effective
eigenfrequency corresponding to the solution mainly influenced by
the lowest of the quasinormal modes. $\gamma(\chi_0)$ can be seen as the
effective damping coefficient~\footnote{This identification of $\gamma(\chi_0)$ is
effective in the sense that the actual quantity appearing in the equation of
motion~\eqref{eq:eom_vector_finDnoK} is $\chi(\rho)$ which highly depends on the radial location.
So we understand the effective damping $\gamma(\chi_0)$ to be a constant
which averages damping effects over the whole AdS-radius.}.
So the effective solution at finite $\chi_0$ reads
\begin{equation}
F(s)\propto e^{-\gamma(\chi_0)\w_0 s} e^{i s \w_{eff}(\chi_0)}
\end{equation}
with the reduced eigenfrequency
\begin{equation} \label{eq:redW}
\w_{eff}(\chi_0)=\w_0 \sqrt{1-{\gamma(\chi_0)}^2}\, .
\end{equation}
Since $0\le\gamma(\chi_0)\le 1$ and $\gamma(\chi_0)$ is monotonous,
the frequency of the solution
which has the eigenfrequency $\w_0$ at $\chi_0$ is monotonously
decreasing with increasing $\chi_0$. So the quark mass parameter $\chi_0$ effectively
acts as a damping coefficient. We might even suspect that
the embedding $\chi(\rho)$ acts as a local damping coefficient depending
on the radial location. In this way the fact that the damping is strongest
at the horizon and vanishes towards the boundary (see figure~\ref{fig:hiWSol}, right)
is consistent with the fact that the embedding $\chi(\rho)$ assumes its
largest value near the horizon and quickly asymptotes to zero towards the
boundary.

Now let us (as a very crude approximation) choose $\w_0$ to be
the real part of the first vector quasinormal
mode on the D7-brane at vanishing $\chi_0$.
The quasinormal mode is already damped, i.e. it actually has no
real eigenfrequency. But let us nevertheless follow our recipe and replace
the complicated fluctuation equation by the simple damped harmonic oscillator
with an effective damping $\gamma(\chi_0)$ and effective reduced
eigenfrequency $\w(\chi_0)$ from equation~\eqref{eq:redW}.
The decreasing eigenfrequency $\w(\chi_0)$ with increasing $\chi_0$ effectively
explains the left-motion of the resonances in the corresponding spectral
functions computed from these solutions
$\text{Im} G^R\propto \partial_s F(s)/F(s)\left. \right |_{s\to s_{bdy}}$.
Here we have assumed that the lowest resonance peak in the
spectral function behaves in the same way (moving to lower frequencies)
as the effective eigenfrequency $\w_{eff}(\chi_0)$ of the solution $F(s)$
from which it is computed. This intuition we get from the fact that in
the exact computation both of these frequencies are mainly determined by
the behavior of the lowest quasinormal mode.

These considerations 
shall serve to give an intuition for the behavior of the resonances.
To be more precise the resonances are actually influenced by all the
quasinormal modes, i.e. by their location in the complex frequency
plane {\it and} by their residues.

\subsubsection{Killing the Tachyon}
The scalar tachyon appearing at zero density is stabilized by introducing baryon
charge. From a critical density $\tilde d_c=0.00315$ on the scalar does not become tachyonic
for any value of $\chi_0$. Thus the finite charge density $\tilde d$ cures the instability
and {\it stabilizes the black hole phase of the D3/D7 system} \footnote{Note that there exist Minkowski
embeddings with the same chemical potential and the same quark mass, but all states in
the  Minkowski phase have strictly vanishing density $\tilde d=0$~\cite{Mateos:2007vc,Kobayashi:2006sb}.}. We have described the scalar 
quasinormal mode signatures 
in detail in section \ref{sec:finiteD_trans_scalar}. The mechanisms explaining this effect 
are discussed with the help of Schr\"odinger potentials in section \ref{sec:finiteD_Schroedinger}.
The negative potential well supporting the tachyon is lifted with increasing charge density.
As discussed before the appearance of the tachyon is at non-zero density connected to the black hole to
black hole {\it first order phase transition} taking place at finite densities $0<\tilde d\le 0.00315$ 
between two distinct
black hole phases. In particular the tachyon appears on the unstable branch of the free
energy diagram of the phase transition. This branch connects two metastable branches
as shown in figure~\ref{fig:vdw_phtr}. At the
critical density $\tilde d_c=0.00315$ both the tachyon and the black hole to black hole 
phase transition disappear. It is not clear what is the physical ground state below $\tilde d_c$.
As was argued in \cite{Mateos:2007vc} the true ground state might be a mixed phase
for which the gravity description is not known so far.

%%%%%%%%%%%%%%%%%%%%%%%% C O N C L U S I O N S  
\section{Conclusions and Outlook} \label{sec:conclusions}
Our extensive study of the quasinormal modes of the D3/D7-brane system
has revealed some interesting relations between previously known and 
unknown phenomena. As a main result we have shown
that the system is completely stable above the critical density $\tilde d_c=0.00315$.
That means that the spectrum of scalar excitations does {\it not} contain
any tachyonic mode.
Also in this regime the spectrum of mesonic exictations in the field theory 
--corresponding directly to distinct quasinormal modes-- is de-singularized
at low temperature or large quark mass through the explicit breaking 
of a scaling symmetry near the limiting embedding. 
I.e. the different meson excitations behave in accordance with the 
mass formula derived for the supersymmetric case~\cite{Kruczenski:2003be}. 
Furthermore for the regime below the critical density, i.e. for $\tilde d<0.00315$ 
we have established the connection
between the black hole to black hole phase transition at finite charge 
density on one hand and the tachyonic scalar on the other.
Using the Schr\"odinger formulation of the problem we have explained
the movement of scalar and vector quasinormal modes in the complex
frequency plane in great detail.
A universal feature of all Schr\"odinger potentials at finite charge density 
is that they develop a barrier near the black hole horizon which hides
the horizon and the black hole from the boundary. In consequence the 
dissipation decreases with increasing density and the quasinormal modes
asymptote to a normal mode behavior. This behavior has also been observed
in~\cite{Horowitz:2009ij} in a setup where a black hole in $AdS_4$ 
develops scalar hair, i.e. there is also a charge density distributed near the horizon.\\
At vanishing baryon density but finite momentum we found a critical wavelength
at which the hydrodynamic approximation explicitly breaks down. Below
that wavelength the system at late times is no longer governed by hydrodynamic 
but by (propagating) collisionless modes only. 
A few unresolved issues remain for zero baryon density. In that case
we found a spiraling behavior of the first quasinormal mode's trajectory for
both scalars and vectors. The number $n$ of loops appears to be directly
related to what we coined "attractor" frequencies $\omega_n$ to which all
QNM-trajectories asymptote if their momentum $k_n$ lies in a certain 
momentum regime $k_{n-1}<k_n\le k_{n+1}$. A direct relation remains unrevealed.
In the same case we found that a scalar becomes tachyonic at low enough
temperature. This instability might give rise to a condensation process or
more generally to a phase transition as suggested by the hydrodynamic
behavior of the scalar mode where it just becomes tachyonic. It is interesting
to ask what the new phase could be and if it exists at all.\\
The analysis presented here can straightforwardly be extended to the D3/D7 setup
with finite isospin density or with spontaneously broken 
symmetry~\cite{Ammon:2009fe,Ammon:2008fc}.
In these setups there is a variety of modes to be studied, in the latter most prominently
the fluctuations of the order parameter phase which correspond to Goldstone modes
becoming new hydrodynamic modes in the phase with spontaneously
broken symmetry.

%%%%%%%%%%%%%%%%%%%%%%%% A C K N O W L E D G M E N T S
\section*{Acknowledgements}
We are grateful to J.~Mas, F.~Rust, J.~Shock, J.~Tarr{\'i}o for discussions.
The work of K.L. M.K and F.P.B has been supported by Plan Nacional de Alta Energ\'\i as FPA-2006-05485, 
Comunidad de Madrid HEPHACOS P-ESP-00346, Proyecto Intramural de CSIC 200840I257.
F.P.B has also been supported by the Programa de apoyo institucional para cursar estudios de doctorado de la Universidad Simon Bolivar.\\
The work of J.E. C.G. and P.K. was supported in part by {\it The Cluster of Excellence for
  Fundamental Physics - Origin and Structure of the Universe}. The work of M.K. was
 supported in part by the German Research Foundation {\it Deutsche Forschungsgemeinschaft~(DFG)}.

%%%%%%%%%%%%%%%%%%%%%%%% A P P E N D I X
\begin{appendix}
%______________________________________________
\section{Shooting method}\label{sec:appShooting}
\def\w{{\mathfrak{w}}}
\def\Im{{\mathrm{Im}\,}}
\def\Re{{\mathrm{Re}\,}}

We now discuss an improved shooting method for solving the equation of motion
for the fluctuations at complex frequencies. The problem arises when using the
standard method for solving the equations of motion by just numerically
integrating the differential equation. 

\begin{figure}[htbp]
        \includegraphics[width=7cm]{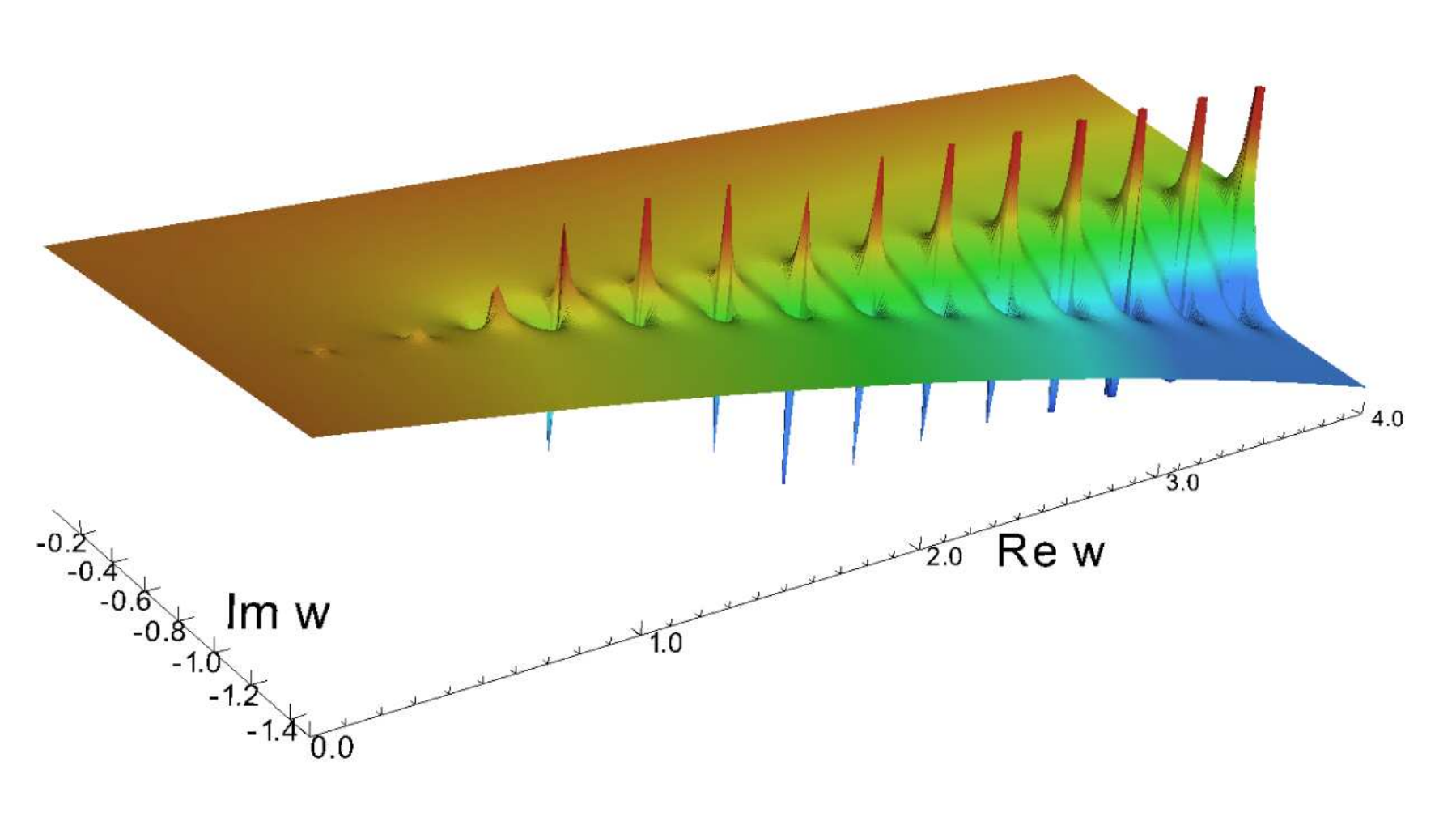}
        \caption{Breakdown of the standard numerical techniques at
          $\Im\,\w\lessapprox -1$. A series of spikes can be seen, being a
          firmly erroneous solution in view of the analytic
          result~\eqref{eq:ex_sol_QNM}. This problem will be solved by means
          of our improved method, discussed in the text.} 
        \label{fig:pole_nAdS_fail}
\end{figure}

As an example for the failure of this naive method we look at
figure~\ref{fig:pole_nAdS_fail}, which shows the spectral function for the
transverse vector modes for vanishing quark mass and vanishing density. Note
here that we use the dimensionless frequency $\w=\omega_{ph}/(2\pi
T)=\omega/2$ 
and momentum $\mathfrak{k}=k_{ph}/(2\pi T)=k/2$.
The line of poles found at $\Im\,\w =-1$ is definitely not correct since
in~\cite{Nunez:2003eq} the pole-structure of the very same configuration was
analytically determined to be  
\begin{equation}\label{eq:ex_sol_QNM}
\w=n\left(1-\ii\right)\;\textrm{ for }\; n\in\mathbb{N}_0\;.
\end{equation}

It turns out that the numerical errors hide the quasi-normal modes behind this
\glqq wall\grqq\, at $\Im\,\w =-1$. This wall could be misinterpreted as a
branch cut.

The problem is, that the equations of motion we investigated are not regular
at the horizon where the initial conditions are imposed. One therefore splits
the solution in a regular and regulating part:
$E(\rho)=\left(\rho-1\right)^{-\ii\w} F(\rho)$ by computing its Frobenius
expansion near the horizon. Nevertheless, we have to move the starting point
for the numerical integration slightly away from the real horizon to, say
$\rho=1.00001$ where we choose this value so close to $1$ that we can expect
only small deviations from the exact solution. 
However, for values of $\Im\,\w\leq -1$ the boundary condition yields

\begin{equation}\label{eq:problem_ic}
\left(\rho-1\right)^{-\ii\w}=\left(1\times 10^{-5}\right)^{\Im\,\w-\ii\Re\,\w}\,.
\end{equation}

This is problematic when $\Im\,\w$ is smaller than $-1$, the initial condition
is becoming large, leading to round off errors. These change the value of the
fluctuation at infinity dramatically and are responsible for the invalid
results of the spectral function in the specific region.

The cure is to shift the starting point for the numerical integration away
from the horizon, and thus dealing with not so small values in the basis
of~\eqref{eq:problem_ic}, resulting in better numerical initial conditions. Of
course, the error from starting the integration further away from the horizon
has to be compensated. This is simply done, by calculating the asymptotics of
the gauge field fluctuations in the vector case, or the embedding deviations
in the scalar case at the horizon to higher order.

When reimplementing the numerical integration with a starting value of \eg
$\rho=1.1$ and the new asymptotics, the numerical aberration is prevented
successfully and one can gain insight farther into the complex $\omega$-plane.

\begin{figure}[htbp]
        \includegraphics[width=7cm]{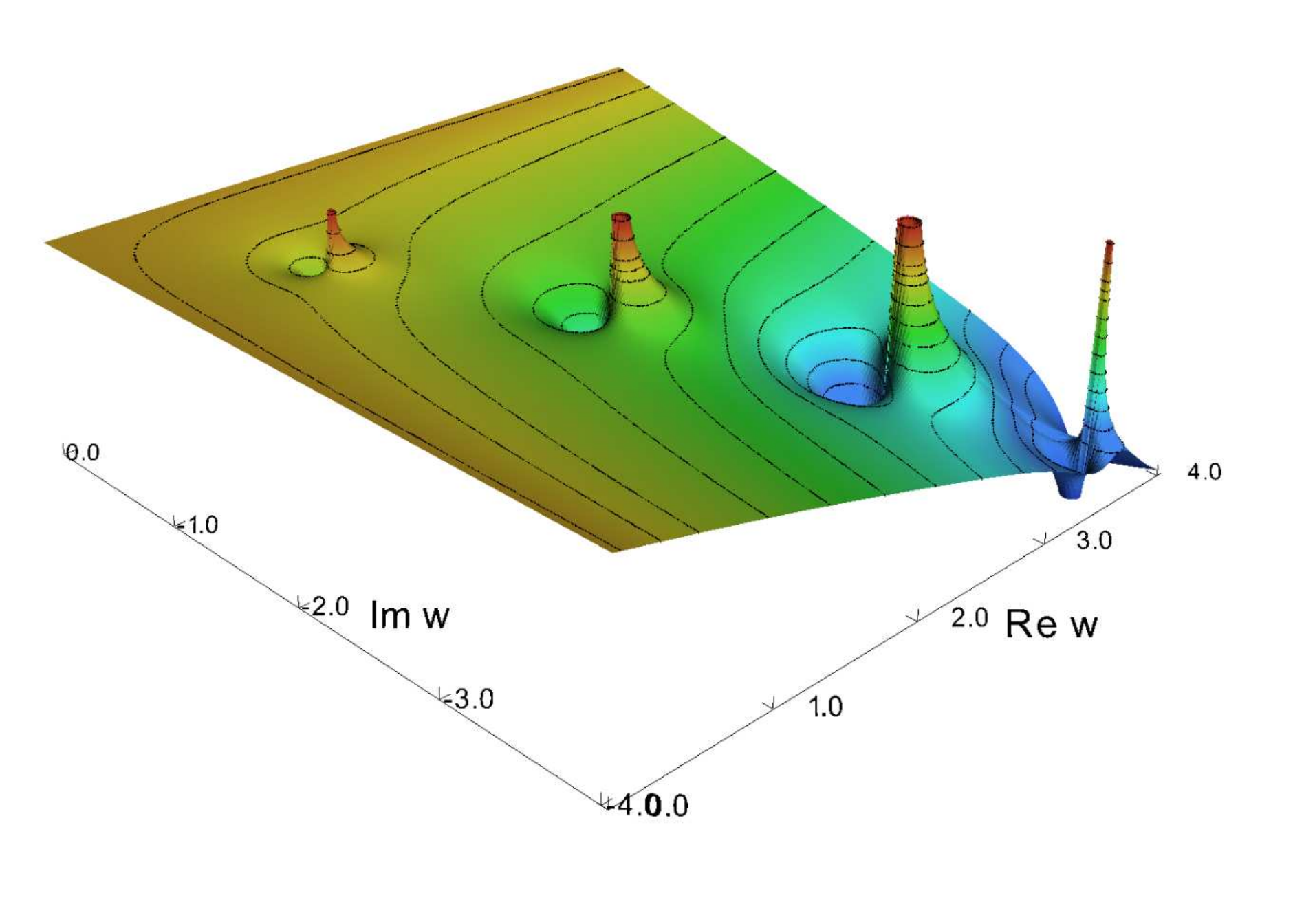}
        \caption{With the improved method nearly four poles can be resolved
          for vanishing density $\tilde d=0$ and massless quarks $m=0$. For
          the computation of the initial values the expansion in the
          fluctuation equation were evaluated up to eleventh order. As can be
          seen, the location of the poles fits the analytical
          solution~\eqref{eq:ex_sol_QNM} very well.} 
        \label{fig:poles_11c}
\end{figure}

A look at the results shows what can be achieved with this more sophisticated
method. A comparison between the surface plot of the old method
(figure~\ref{fig:pole_nAdS_fail}) and the improved one
(figure~\ref{fig:poles_11c}) reveals that the expected pole structure can now
be seen clearly. In fact it can be checked that the position of the poles
agrees with the analytical result~\eqref{eq:ex_sol_QNM} very well.

%______________________________________________
\section{Relaxation method}\label{sec:appRelax}

The determination of quasinormal frequencies can be understood as a particular case
of a  two-point boundary value problem. A common numerical method for solving such
problems is the shooting method, where one solves the differential equations with varying boundary conditions at
one boundary and searches for a solution that
approximates the wanted boundary condition at the other boundary within some
numerical error. We describe that method in appendix~\ref{sec:appShooting}. 

A different approach is provided by the relaxation method which allows to fix the
correct boundary values on both boundaries. Since this method is less frequently used
we give a short outline below\footnote{A more detailed exposition can be found in \cite{nrinc}.}.

The method is based on replacing the differential equations by a system of
finite-difference equations (FDE) on a discrete grid. Starting from an ansatz
solution 
obeying the correct boundary conditions, one varies the value of the dependent
variables at each point \textit{relaxing} to the configuration which provides
an approximate solution for the FDE within some given numerical error. In our
case, we convert our second order complex ODE into a set of four first order
equations by separating real and imaginary parts of the dependent variables. 

More generally we can consider a set of $\tilde N$ first order ODE's
\begin{equation} \label{eq:odesys}
\frac{\dd y_i(x)}{\dd x} =g_i(x,y_1,\ldots,y_{\tilde N};\lambda) ~,
\end{equation}
where each dependent variable $y_i(x)$ depends on the others and itself, on
the independent variable $x$ and possibly on additional parameters, like
$\lambda$ above. In our case this (complex) parameter is the  quasinormal
frequency. These extra parameters can be embedded into the problem by writing
trivial differential equations for them 
\begin{eqnarray}
\left\{ \begin{array}{rcl} y_{N+j} &\equiv& \lambda_j~,\\[-1em] \\
    \displaystyle \frac{\dd y_{N+j}}{\dd x} &=& 0~, ~~\hbox{since it is
      constant~.}\end{array} \right. 
\end{eqnarray}
We assume that there are $n$ additional parameters and include them from now on
into the set of dependent variables $y_1,\ldots,y_N$, $N=\tilde N +n$.

The solution to the problem involves $N\times M$ values, for the $N$ dependent
variables in a grid of $M$ points. We also have to fix $N$ boundary conditions
for the dependent variables.

The system is discretized as usual
\begin{equation}
x \to \frac{1}{2}(x_k +x_{k-1}) ~, ~~ y \to \frac{1}{2}(y_k +y_{k-1}) ~,
\end{equation}
for points in the bulk. One may arrange the whole set of $y_i$'s in a column
vector ${\mathbf y}_k =(y_1,\ldots,y_N)_k{}^T$, where the subscript $k$ refers
to evaluation at the point $x_k,~k=1,\ldots,M.$ With this matrix notation, the
system \eqref{eq:odesys} can be written as 
\begin{equation}
0 = \mathbf{E}_k \equiv \mathbf{y}_k -\mathbf{y}_{k-1} -(x_k-x_{k-1})
\,\mathbf{g}_k(x_k,x_{k-1},\mathbf{y}_k,\mathbf{y}_{k-1}) ~,~~k=2,\ldots,M ~, 
\end{equation}
where the $\mathbf{E}_k$ are the aforementioned FDE's. These are the equations
that we need to fulfil. Notice there are $N$ equations at $M-1$ points, so the
remaining $N$ equations are supplied by the boundary conditions. We will set
$n_1$ of them on the left at $x_1$, called $\mathbf{E}_1$, and the rest
$n_2=N-n_1$ at $x_M$, called $\mathbf{E}_{M+1}$. 

Now we take a trial solution that nearly solves the FDE's $\mathbf{E}_k$. By
shifting each solution $\mathbf{y}_k\to\mathbf{y}_k+\Delta\mathbf{y}_k$ and
Taylor expanding in the shift, one obtains the relation 
\begin{eqnarray}
0 =\mathbf{E}_k(\mathbf{y}+\Delta\mathbf{y}) &\simeq&
\mathbf{E}_k(\mathbf{y}_k,\mathbf{y}_{k-1}) +\sum_{n=1}^N
\frac{\partial\mathbf{E}_k}{\partial y_{n,k-1}} \,\Delta y_{n,k-1}
+\sum_{n=1}^N \frac{\partial\mathbf{E}_k}{\partial y_{n,k}} \,\Delta
y_{n,k}~,~~ \\ 
\Rightarrow ~-E_{j,k} &=& \sum_{n=1}^N \Big( S_{j,n} \,\Delta y_{n,k-1} \Big) +\sum_{n=N+1}^{2N} \Big( S_{j,n} \,\Delta y_{n-N,k} \Big) ~.
\end{eqnarray}
This allows to find the $\Delta\mathbf{y}_k$ that improve the solution. First we merge
the two differentials
\begin{equation}
S_{j,n} =\frac{\partial E_{j,k}}{\partial y_{n,k-1}} ~,~~ S_{j,n+N} =\frac{\partial E_{j,k}}{\partial y_{n,k}} ~,~~~ n=1,\ldots,N ~,
\end{equation}
in a $N\times 2N$ matrix, for each bulk grid position $x_k$. For the boundaries the expressions follow equally
\begin{eqnarray}
-E_{j,1} &=& \sum_{n=1}^N S_{j,n} \,\Delta y_{n,1} =\sum_{n=1}^N
\frac{\partial E_{j,1}}{\partial y_{n,1}} \,\Delta y_{n,1} ~, \quad
j=n_2+1,\ldots,N ~, \\ 
-E_{j,M+1} &=& \sum_{n=1}^N S_{j,n} \,\Delta y_{n,M} =\sum_{n=1}^N
\frac{\partial E_{j,M+1}}{\partial y_{n,M}} \,\Delta y_{n,M} ~, \quad
j=1,\ldots,n_2 ~, 
\end{eqnarray}
where $n$ runs in both from $1$ to $N$. The whole $(NM\times NM)$ matrix  $S$
possesses a block diagonal structure. In fact since for reasonable systems
$N<<M$ $S$ is a sparse matrix. This allows the usage of computer packages in
which the solution of linear systems 
with sparse matrices can be done in a very efficient way. The actual solution can now
be found by an iterative process until a desired accuracy is achieved. As
measure for the discrepancy of an approximate solution to the actual solution
we used 
\begin{equation}
\mathtt{err} = \frac{1}{MN} \sum_{k=1}^M \sum_{j=1}^N \left| \frac{\Delta y[j][k]}{\mathtt{scalevar}[j]} \right| < \mathtt{conv}~,
\end{equation}
where $\mathtt{scalevar}[j]$ is an associated scale for each of the dependent
variables (e.g. the value at the midpoint or so). The idea is that when that
averaged value of the shift to get a better solution is smaller than
\texttt{conv}, we accept the former values we had as the actual solution. In
our computations we set $\mathtt{conv}=10^{-6}$. 

In order to obtain the correct boundary conditions we have used the $z$
coordinate system where the horizon lies at $z=1$ and the boundary at $z=0$.
We split off the ingoing 
boundary condition on the horizon according to $\Phi(z) = (1-z)^{-i \omega/4} y(z)$ .
Demanding $y(1)=1$ and $y(0)=0$ gives four real boundary conditions. Since in total we
have however six dependent variables, counting also the real and imaginary part of the
quasinormal frequency we need two more boundary conditions. We found it convenient to
expand the function $y(z)$ in a Taylor series at the horizon and compute also $y'(1)$
which provides the additional two real boundary conditions. The typical gridsize we used 
was consisted of 5000 points. Finally we note that we have implemented to outlined algorithm in 
GNU/Octave \cite{eaton:2002}.

%______________________________________________
%------------------------------------
\section{Schr\"odinger potentials}
\label{sec:appSchroedingerPot}
This appendix shows how to compute the effective potentials, i.e. the
Schr\"odinger potentials for the scalar and vector fluctuations on the
D7-probe-brane. In order to do so the linearized fluctuation equations
of motion have to be rewritten in terms of a new radial coordinate $R*$.
This procedure has been described before (e.g. in \cite{Myers:2007we},
\cite{Paredes:2008nf},
\dots), nevertheless we include it here for completeness.
For convenience we stick to the notation made use of in \cite{Myers:2007we},
and we compute all Schr\"odinger potentials in the $\rho$-coordinates
introduced in section \ref{sec:setup}.

We are not interested in the higher angular excitations on the $S^3$, so we
separate the fluctuations according to $\phi(\rho, S^3)=
%(\rho-1)^{-i\frak{w}}
y(\rho) \Phi(S^3)$.
Let us consider exclusively fluctuations without angular momentum
on the $S^3$, i.e. $\Phi(S^3)=1$.
All the linearized vector and scalar fluctuation equations of motion are second
order ordinary differential equations and can be rewritten in the form
\begin{equation}\label{eq:genFlucEq}
-\frac{H_0}{H_1}\partial_\rho\left[H_1\partial_\rho y(\rho)\right]+
\left[\frak{k}^2 H_2 +H_{\theta} \right] y(\rho) = \frak{w}^2 y(\rho),
\end{equation}
where $H_0,\, H_1,\, H_2$ and $H_{\theta}$ are in general functions
of $\rho$ and depend on the particular field fluctuation considered.
$H_\theta$ only appears in the scalar fluctuations.
For {\it transverse vector} fluctuations we have
\begin{eqnarray}\label{eq:transVecH}
H_0=-\frac{g^{\rho\rho}}{g^{tt}}, & H_1 = \sqrt{-g} g^{\rho\rho} g^{xx} ,\\
H_2= -\frac{g^{zz}}{g^{tt}}, & H_{\theta} = 0 \, .
\end{eqnarray}
For {\it longitudinal vector} fluctuations we have
\begin{eqnarray}\label{eq:longVecH}
H_0=-\frac{g^{\rho\rho}}{g^{tt}}, & H_1 = \sqrt{-g} g^{\rho\rho} g^{zz} ,\\
H_2= -\frac{g^{zz}}{g^{tt}}, & H_{\theta} = 0\, .
\end{eqnarray}
For {\it scalar} fluctuations the equations of motion do not take
such a simple general form in terms of metric components since the metric
itself contains scalar fluctuations. So we choose to write explicitly
\begin{eqnarray}\label{eq:scalarH}
H_0=-\frac{f^2\rho^4(1-\chi^2)}{8{\tilde{f}}(1-\chi^2+\rho^2{\chi'}^2)}, &
H_1 =  \frac{f \tilde f \rho^5 (1-\chi^2)^3}{(1-\chi^2+\rho^2{\chi'}^2)^{3/2}}
\sqrt{1+\frac{8{\tilde d}^2}{\tilde{f}^3 \rho^6 (1-\chi^2)^3}},\\
H_{\theta} =& \frac{3f^2 {\tilde f}^2\rho8 (1-\chi^2)^2 \left[ {\tilde
      f}^3\rho^6(1-\chi^2)^3+8{\tilde d}^2 (1-6{\chi}^2)-144{\tilde d}^2
    f^3{\tilde f}\rho^9\chi\chi'(1-\chi^2)^2 \right]}{8 \left[ {\tilde
      f}^3\rho^6 (1-\chi^2)^3+8{\tilde d}^2 \right]^2}\, , \nonumber
\end{eqnarray}
where we do not include $H_2$ since the above coefficients are computed
in the case of vanishing momentum $\frak{k}=0$ but at finite density $\tilde d$.
At finite momentum and at vanishing density we have $H_2=f2/{\tilde f}2$ in
addition to the $\tilde d\to0$ limits of $H_0,\, H_1$ and $H_\theta$ from
\eqref{eq:scalarH}. At finite momentum {\it and} density the scalar
fluctuations couple to the vector fluctuations and we will not address
this complication in this work.

In any case we can substitute $y(\rho)=h \psi$ in equation
\eqref{eq:genFlucEq} to obtain
\begin{equation}
-H_0\psi''-H_0\left( 2\frac{h'}{h}+\frac{{H_1}'}{H_1} \right)\psi'
+\left[ {\frak{k}}^2 H_2+H_\theta-H_0\left( \frac{h''}{h}+\frac{{H_1}'}{H_1}\frac{h'}{h} \right) \right] \psi = {\frak{w}}^2 \psi\, .
\end{equation}
Introducing the new radial coordinate 
\begin{equation}
  \label{eq:defR*}
  R*=\int\limits_\rho^\infty d\tilde\rho/\sqrt{H_0(\tilde \rho)}\,,
\end{equation}
we can rewrite the first term $-H_0 \psi''=-{\partial_{R*}}^2\psi+ {H_0}'\psi'/2$.
The special choice of $h={H_0}^{1/4}/{H_1}^{1/2}$ eliminates all the
terms containing $\psi'$. Thus the fluctuation equation of motion finally
assumes Schr\"odinger form
\begin{equation}
-{\partial_{R*}}^2\psi+V_S \psi = E \psi \,
\end{equation}
with the effective energy $E={\frak{w}}^2$ and the Schr\"odinger potential
given by
\begin{equation}
V_S=-H_0\left( \frac{h''}{h}+\frac{H_1'}{H_1}\frac{h'}{h} \right)+{\frak{k}}^2 H_2+H_\theta\, .
\end{equation}
From this general formula the specific scalar, longitudinal and transversal vector
Schr\"odinger potentials may be obtained by substituting for $H_0,\, H_1,\,
H_2,\, H_\theta,\, h$ using the corresponding values from
equations \eqref{eq:scalarH}, \eqref{eq:longVecH} and \eqref{eq:transVecH},
respectively.

%______________________________________________
\section{Finite density but vanishing momentum} \label{sec:LagrFiniteD}
\def\Im{{\mathrm{Im}\,}}
\def\Re{{\mathrm{Re}\,}}
\def\L{{\mathcal{L}}}
\def\pd{{\partial}}
\def\wn{{\mathfrak{w}}}

To abbreviate the longish Lagrangian we introduced the following notation

\begin{equation}
a=1-\chi^2,\quad b=a+\rho^2{\chi'}^2,\quad c=\frac{8\tilde d^2}{\rho^6\tilde
  f^3 a^3+8\tilde d^2}, \text{ so }\quad 1-c=\frac{\rho^6{\tilde f}^3
  a^3}{\rho^6{\tilde f}^3 a^3+8\tilde d^2}\;. 
\end{equation}
In terms of these expressions the Lagrangian, expanded to second order, reads

\begin{alignat}{2}
\nonumber \frac{\L}{\left(-N_f
    T_{D7}\sqrt{h_3}\frac{r_H^4}{4}\right)}=&\;\L_0+\L_1+\L_2-\rho^3 f\tilde f
a\sqrt{b}\sqrt{1-c}F^{04}(\delta F_{40})-\frac{3}{2}\rho^3 f\tilde
f\frac{a-bc}{\sqrt{b}\sqrt{1-c}}(\delta\theta)^2\\ 
\nonumber &-\frac{R^4}{r_H^2}\rho\frac{{\tilde
    f}^2}{f}\frac{a^2}{\sqrt{b}\sqrt{1-c}}(\pd_t\delta\theta)^2+\frac{1}{2}\rho^5
f\tilde f\frac{a^2(a-bc)}{b^{3/2}(1-c)^{3/2}}(\pd_\rho\delta\theta)^2\\ 
\nonumber &+\frac{R^2}{2}\sum_i \G_0^{ii}\chi^2(\pd_i \delta\phi)^2
+\rho^5\tilde
f^{3/2}\chi'\frac{\sqrt{2}}{r_H}\frac{a^2\sqrt{c}}{b\,(1-c)^{3/2}}(\delta
F_{40})(\pd_\rho\delta\theta)\\ 
\nonumber &-\rho^3{\tilde
  f}^{3/2}\chi\frac{3\sqrt{2}}{r_H}\frac{a\sqrt{c}}{\sqrt{1-c}}(\delta
F_{40})(\delta\theta) +\frac{1}{4}\rho^3 f{\tilde f}
a\sqrt{b}\sqrt{1-c}\sum_{ik}\G_0^{ii}\G_0^{kk}(\delta F_{ik})^2\\ 
&-\rho^3 f\tilde f a\sqrt{b}\sqrt{1-c}F^{04}\sum_i
\G_0^{ii}\left(-\frac{(1+\delta_{4i})R^2\chi'}{\sqrt{a}}\right)(\delta
F_{i0})(\pd_i\delta\theta) 
\end{alignat}
where $\L_0$ is simply the Lagrangian without any fluctuations but nonvanishing density

\begin{equation}
\L_0=\rho^3 f\tilde fa\sqrt{b}\sqrt{1-c}
\end{equation}
and the boundary terms $\L_1$ and $\L_2$ are given by

\begin{equation}
\L_1=\pd_\rho\left[-\frac{\rho^5 f\tilde f a^{3/2}\chi'}{\sqrt{b}\sqrt{1-c}}(\delta \theta)\right]
\quad\text{ and }\quad
\L_2=\pd_\rho\left[-\frac{3}{2} \frac{ \rho^5 f\tilde f \chi\chi'  a}{\sqrt{b}\sqrt{1-c}}(\delta\theta)^2\right]\;.
\end{equation}
The symbols denoted by $\G$ with upper indices are the components (diagonal
part) of the inverse background tensor $\G=G+2\pi\alpha'F$. The most important
ones are 

\begin{equation}
\G_0^{00}=-2\frac{R^2}{r_H^2}\frac{\tilde f}{f^2\rho^2(1-c)}
\quad\text{ and }\quad
\G_0^{44}=\frac{\rho^2 a}{R^2 b(1-c)}\;.
\end{equation}
For the off-diagonal part of the inverse background tensor $\G^{04}\equiv
F^{04}=-F^{40}$ holds, because of the antisymmetry of the field strength
tensor $F$, and has the value 

\begin{equation}
F^{04}=-\frac{\sqrt{2\tilde fac}}{r_H f\sqrt{b}(1-c)}\;.
\end{equation}

Finally, $\delta F_{ij}=\pd_i\delta A_j -\pd_j\delta A_i$ is the field strength for the vector fluctuation.\\

%______________________________________________
\section{Results for second quasinormal modes}  
\label{app:secondQNMs} 
In this appendix we collect results for the location of the second quasinormal modes
(scalar, transverse and longitudinal vectors) at finite momentum but vanishing density.     
%.............................................
\subsection{Transverse vector fluctuations} 
\begin{figure}[h!]
 \begin{center}  
$$
\begin{array}{cc}
\includegraphics[scale=0.7]{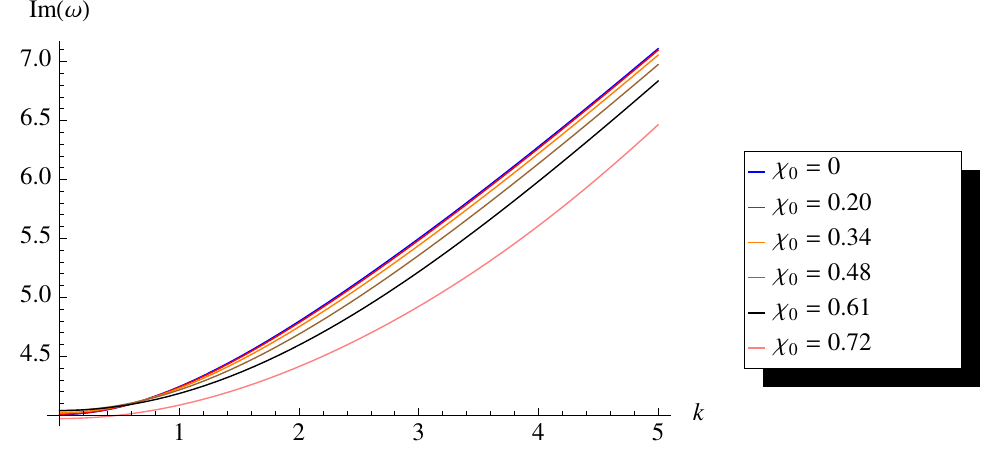} & \includegraphics[scale=0.7]{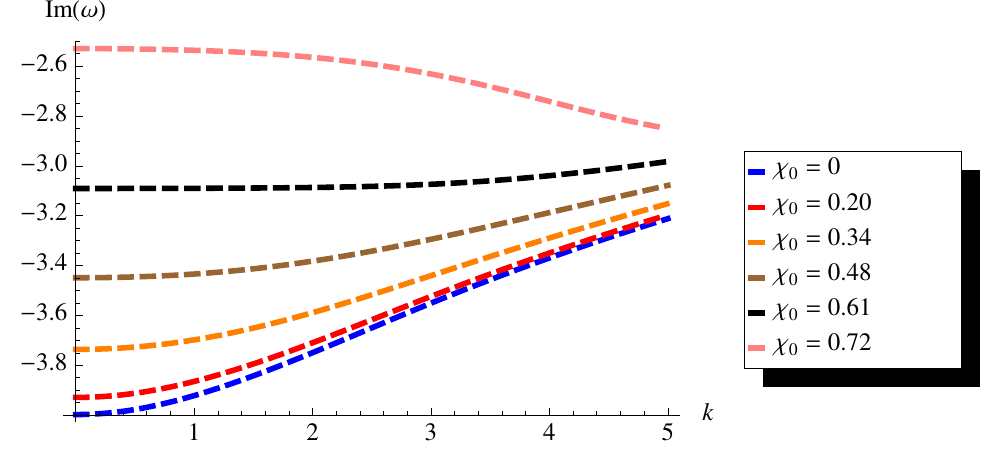}
\end{array}
$$
\caption{\label{fig:dispTransVec2ndQNM}
Dispersion relation for the second transverse vector quasinormal mode.}
\end{center}
\end{figure}

Figure \ref{fig:dispTransVec2ndQNM} shows the dispersion relation of
the second transverse vector QNM at different values
for the mass parameter $\theta_0$.
\begin{figure}[h!]
 \begin{center}
$$
\begin{array}{cc}
\includegraphics[scale=0.7]{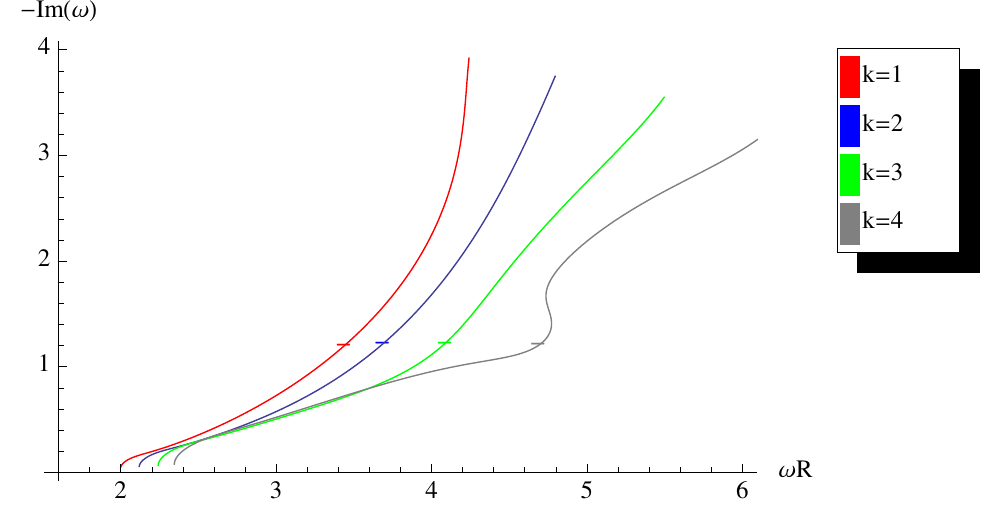} & \includegraphics[scale=0.7]{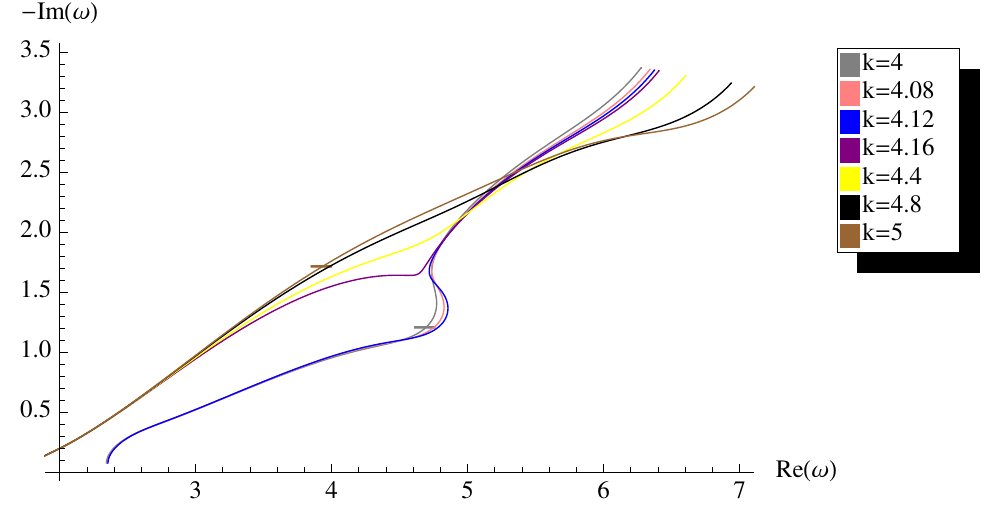}
\end{array}
$$
\caption{\label{fig:transVecK2ndQNM}
Second quasinormal mode of transverse vector fluctuations for
distinct values of the spatial momentum $k$.}
\end{center}
\end{figure}

 \newpage
%.............................................
\subsection{Longitudinal vector fluctuations} 
\begin{figure}[h!]
 \begin{center}
$$
\begin{array}{cc}
\includegraphics[scale=0.7]{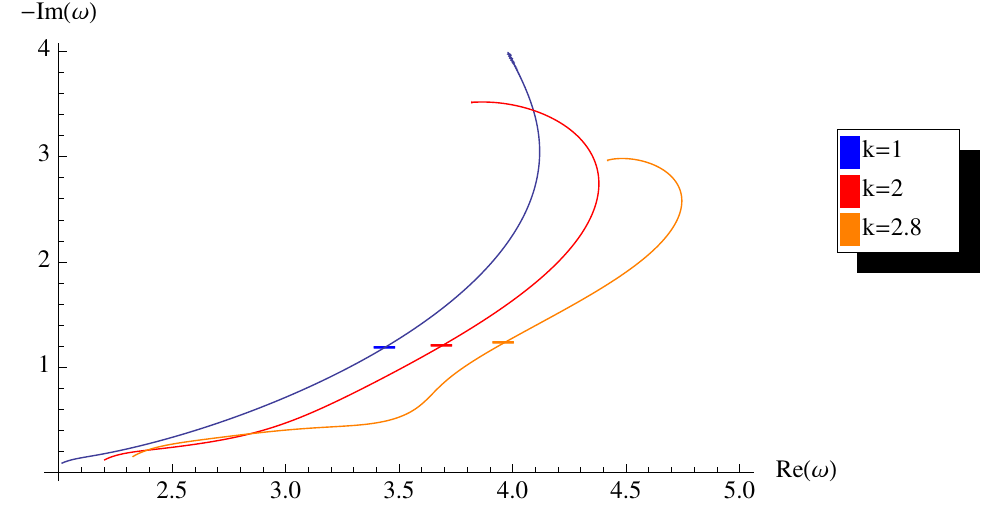} & \includegraphics[scale=0.7]{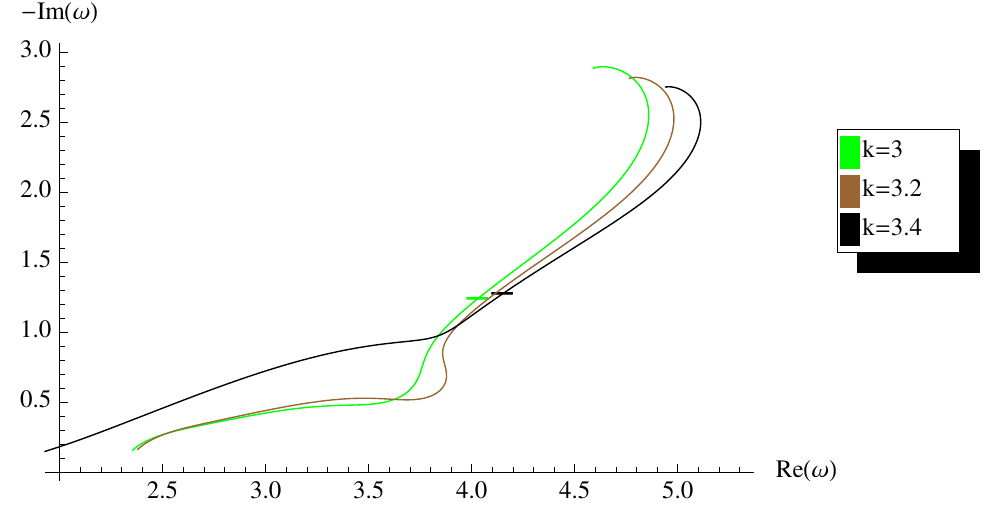}\\
\includegraphics[scale=0.7]{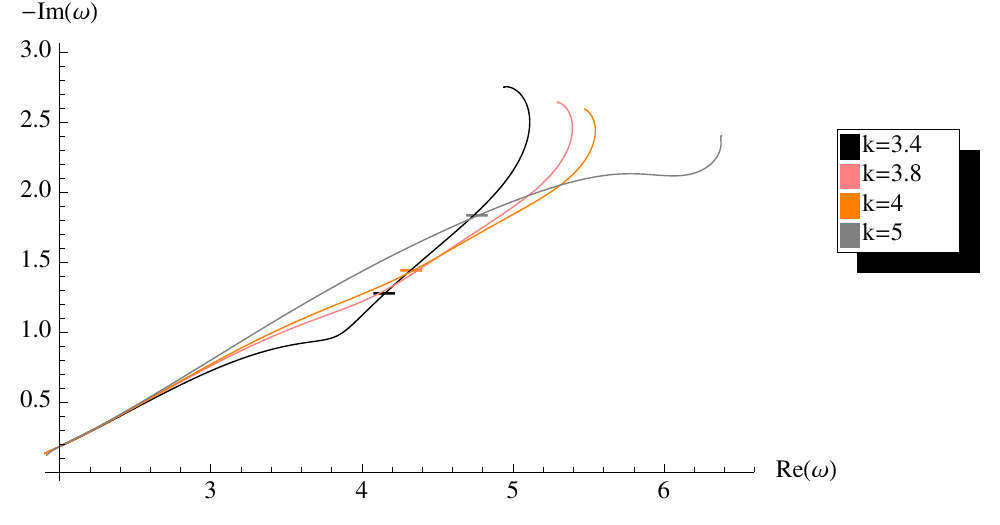} &
\end{array}
$$
\caption{\label{fig:longVecK2ndQNM}
Second quasinormal mode for longitudinal vector fluctuations at distinct
spatial momenta $k$.}
\end{center}
\end{figure}
\begin{figure}[h!]
 \begin{center}
$$
\begin{array}{cc}
\includegraphics[scale=0.7]{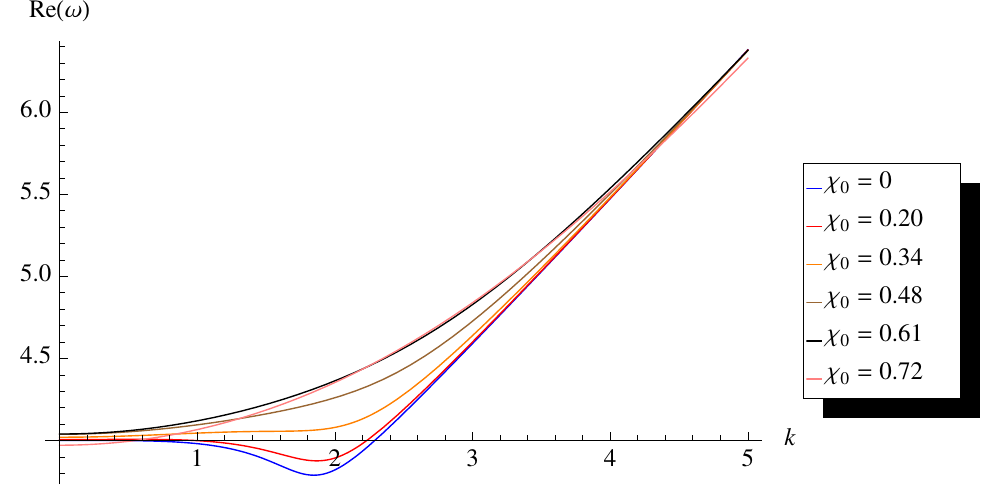} & \includegraphics[scale=0.7]{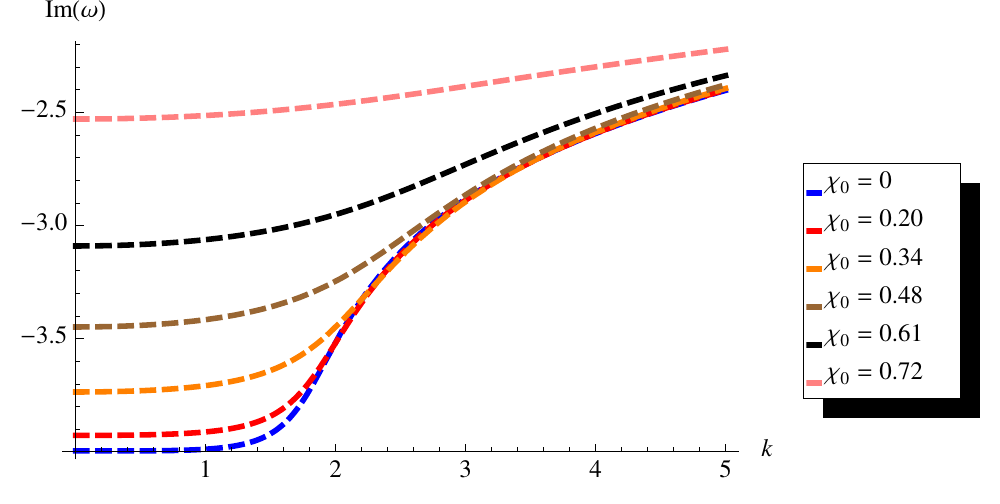}
\end{array}
$$
\caption{\label{fig:dispLongVec2ndQNM}
Dispersion relation for the second longitudinal vector quasinormal mode
fluctuations at distinct spatial momenta $k$.}
\end{center}
\end{figure}
\newpage
%.............................................
\subsection{Scalar fluctuations} 

\begin{figure}[h!]
 \begin{center}
\includegraphics[scale=0.7]{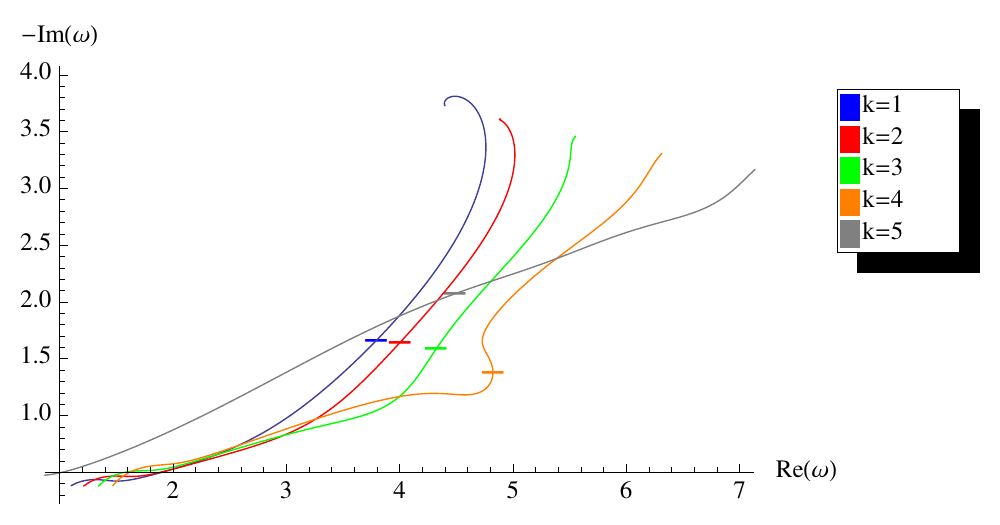}
\caption{\label{fig:scalarK2ndQNM}
Second scalar quasinormal mode at distinct $k$.}
\end{center}
\end{figure}
\begin{figure}[h!]
 \begin{center}
$$
\begin{array}{cc}
\includegraphics[scale=0.7]{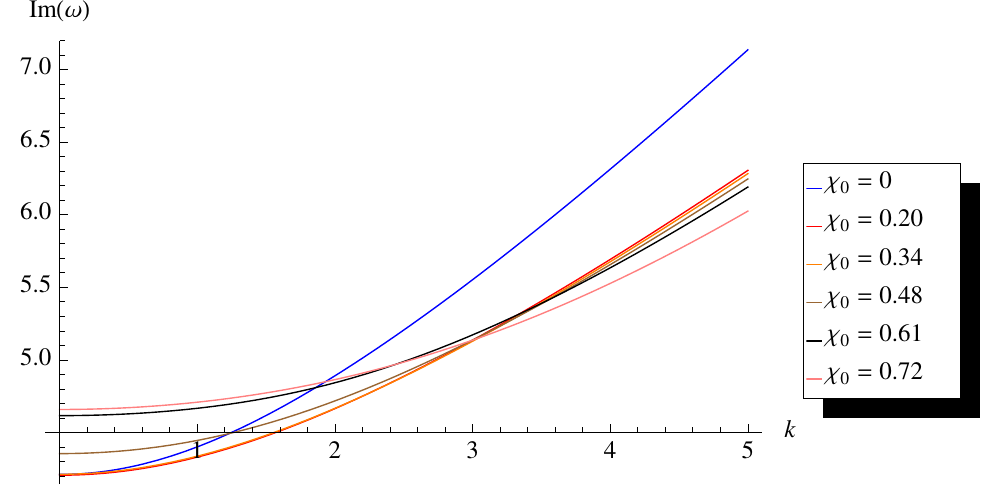} & \includegraphics[scale=0.7]{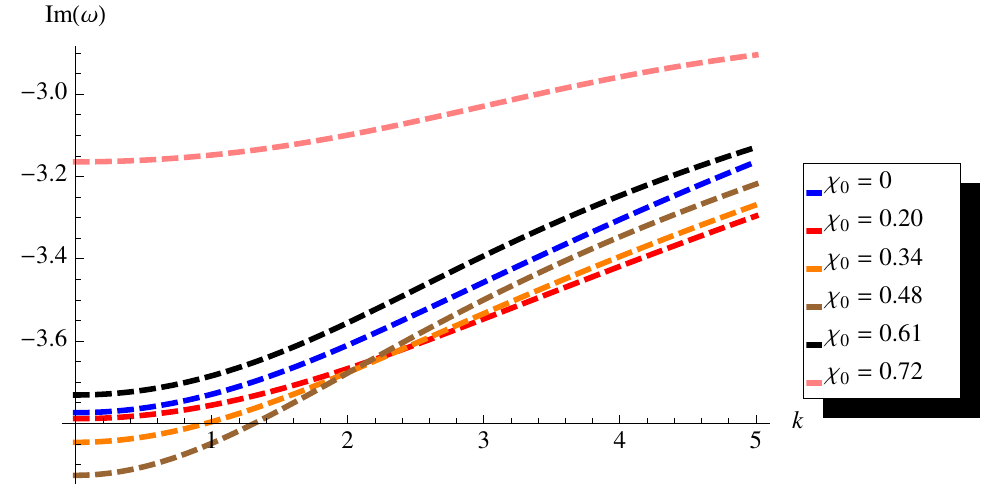}
\end{array}
$$
\caption{\label{fig:dispScalarK2ndQNM}
Dispersion relation for the second scalar quasinormal mode for
distinct mass or temperature parameter $\theta_0$.}
\end{center}
\end{figure}

\end{appendix}

%%%%%%%%%%%%%%%%%%%%%%%% R E F E R E N C E S
\providecommand{\href}[2]{#2}\begingroup\raggedright\endgroup

%\bibliographystyle{JHEP}     % bibtex style file
%\bibliography{papers,papersunpub,books}

\end{document}